\documentclass[12pt,a4paper]{article}            
 \usepackage[skins,theorems]{tcolorbox}
\tcbset{highlight math style={enhanced,
  colframe=red,colback=white,arc=0pt,boxrule=1pt}}
  \usepackage{tikz}
  \usepackage{tikz-3dplot}
 \usetikzlibrary{calc}
 \usetikzlibrary{decorations} %
 \usepackage[UKenglish]{babel}
 \usepackage[toc,page]{appendix}
 \usepackage{amsmath}
 \usepackage{amssymb}
 \usepackage{mathrsfs}
 \usepackage{graphicx}
 \usepackage{hhline}
 \usepackage[bf]{caption}
\usepackage{cite}
\usepackage[vcentermath]{youngtab}
\usepackage{geometry}
\usepackage{slashed}
\usepackage{color}
\usepackage{stackrel}
\usepackage{tikz-cd} 
\usepackage{cancel} 
\usepackage{multirow} 
\usepackage{physics}
\usepackage{longtable}
\usepackage{float}
\usepackage[aligntableaux=center]{ytableau}
\RequirePackage{pgfkeys}

\usepackage{tikz}
\usetikzlibrary{shadows}
\usepackage[framemethod=tikz]{mdframed}

\usetikzlibrary{shapes.geometric,calc,arrows.meta,positioning,decorations.pathreplacing}

\tikzset{
	flavor/.style={regular polygon,regular polygon sides=4, inner sep=-3,draw,minimum size=38pt},
    smallflavor/.style={regular polygon,regular polygon sides=4, inner sep=-3,draw,minimum size=20pt},
    emptyflavor/.style={regular polygon,regular polygon sides=4, inner sep=-3,draw,minimum size=20pt,dotted},
	gauge/.style={circle, draw, minimum size=30pt,inner sep=1},
    smallgauge/.style={circle, draw, minimum size=16pt,inner sep=1},
    largegauge/.style={circle, draw, minimum size=60pt,inner sep=1},
    emptygauge/.style={circle, draw, minimum size=16pt,inner sep=1,dotted},
}

\allowdisplaybreaks

\global\mdfdefinestyle{myboxstyle}{%
  shadow=true,
  linecolor=black,
  shadowcolor=black,
  shadowsize=6pt,
  nobreak=false,
  innertopmargin=10pt,
  innerbottommargin=10pt,
  leftmargin=5pt,
  rightmargin=5pt,
  needspace=1cm,
  skipabove=10pt,
  skipbelow=15pt,
  middlelinewidth=1pt,
  afterlastframe={\vspace{5pt}},
  aftersingleframe={\vspace{5pt}},
  tikzsetting={%
draw=black,
very thick} }

\newmdenv[style=myboxstyle]{whitebox} \newmdenv[style=myboxstyle,backgroundcolor=black!20]{graybox}

\definecolor{bluish}{rgb}{0.0, 0.2, 0.4}
\definecolor{darkred}{rgb}{0.7, 0.0, 0.0}
\definecolor{darkgreen}{rgb}{0.0, 0.5, 0.0}

\newmdenv[style=myboxstyle,nobreak=true]{blockwhitebox}
\newmdenv[style=myboxstyle,backgroundcolor=black!20,nobreak=true]{blockgraybox}
\newmdenv[nobreak=true,hidealllines=true]{blockbox}

\usepackage{jheppub}
\usepackage{cleveref}

\newcommand{\bqa}{\begin{eqnarray}}
\newcommand{\eqa}{\end{eqnarray}}

\newcommand{\nn}{\nonumber}

\def\et24{\eta^{24}}
\def\oet24{\frac1{\eta^{24}}}
\def\IH{\mathbb{H}}
\def\IE{\mathbb{E}}
\def\IC{\mathbb{C}}

\hypersetup{
    pdftitle={},
    pdfauthor={},
    pdfsubject={}
}
\numberwithin{equation}{section}
\numberwithin{table}{section}

\makeatletter

\newcommand{\be}{\begin{equation}}
\newcommand{\ee}{\end{equation}}
\newcommand{\beq}{\begin{equation}}
\newcommand{\eeq}{\end{equation}}
\newcommand{\ba}{\begin{aligned}}
\newcommand{\ea}{\end{aligned}}
\newcommand{\su}{\mathfrak{su}}
\newcommand{\so}{\mathfrak{so}}
\newcommand{\bea}{\begin{eqnarray}}
\newcommand{\eea}{\end{eqnarray}}

\newcommand{\cO}{\mathcal{O}}
\newcommand{\cT}{\mathcal{T}}

\newcommand{\cC}{\mathcal{C}}

\newcommand{\cN}{\mathcal{N}}

\newcommand{\cH}{\mathcal{H}}

\newcommand{\cI}{\mathcal{I}}
\newcommand{\cJ}{\mathcal{J}}
\newcommand{\cR}{\mathcal{R}}
\newcommand{\cS}{\mathcal{S}}

\newcommand{\cV}{\mathcal{V}}

\newcommand{\cM}{\mathcal M}
\newcommand{\cQ}{\mathcal Q}

\newcommand\bi{\begin{itemize}}
\newcommand\ei{\end{itemize}}

\def\Tr{\mathop{\mathrm{Tr}}\nolimits}

\def\unit{{1\kern-.65ex {\rm l}}}
\def\1{{1\kern-.65ex {\rm l}}}

\def\IZ{\mathbb{Z}}

\def\CT{{\cal T}}

\def\IR{{\mathbb{R}}}

\makeatother

\begin{document}
	
%\red{ XXX ------------------------------------------------------\\
%EDITS FOR V2:\\
%	 -  corrected \eqref{eq:leading_order} \\
%	 -  notation \eqref{eq:cS1}-\eqref{eq:cS4}\\
%	 -  notation 	\eqref{eq:SU2toZn_projection}, and 	\eqref{eq:Zn-tensor}-\eqref{eq:equiv_hs_sum}%
%	 }

\title{M5 branes on ADE singularities: BPS spectrum and partition functions}

\author{Daniele Ceppi}
\author{and Guglielmo Lockhart}

\affiliation{Bethe Center for Theoretical Physics, Universit\"at Bonn, D-53115, Germany}

\emailAdd{dceppi@uni-bonn.de}
\emailAdd{glockhar@uni-bonn.de}

\abstract{
The dynamics of a stack of M5 branes probing a transverse multi-centered Taub--NUT space are described by a class of 6d $\mathcal{N}=(1,0)$ superconformal field theories known as the M-string orbifold SCFTs. We determine the equivariant partition functions for this class of theories on a geometric background of type $T^2\times\mathbb{C}^2/\Gamma$, where $\Gamma \in\{\cC_N,\cQ_N, \cT,\cO,\cI\}$ is an arbitrary finite subgroup of $SU(2)$. The partition functions are built out of contributions from BPS strings as well as BPS particles that arise upon putting the 6d theory on a circle. We find that BPS particle contributions can be expressed in terms of $\Gamma$\emph{-covariant Hilbert series} which count holomorphic sections of vector bundles on the orbifold singularity with monodromy specified by an irreducible representation of $\Gamma$. The BPS string contributions, on the other hand, are given by the elliptic genera of 2d $\mathcal{N}=(0,4)$ $\Gamma$\emph{-dressed quiver gauge theories}, obtained by stacking Kronheimer--Nakajima quivers of type $\Gamma$ between interfaces that support current algebras for the McKay dual affine Lie algebra $\widehat{\mathfrak{g}}$. We obtain explicit expressions for the elliptic genera of arbitrary BPS string configurations corresponding to fractional instanton strings on $\mathbb{C}^2/\Gamma$, and for the case of \emph{star-shaped} quivers of type $\Gamma\in\{\cQ_4,\cT,\cO,\cI\}$ we give a prescription  to compute the elliptic genera by gluing 2d analogues of Gaiotto and Witten's $T[SU(N)]$ theories.
}
\maketitle

\section{Introduction}
One of the overarching themes in the study of supersymmetric quantum field theory over the last few decades has been the realization that many of the structures that govern the physics of lower dimensional theories can be understood most naturally as having a six-dimensional origin, which follows from the existence of the still-mysterious 6d $\mathcal{N}=(2,0)$ worldvolume theory of M5 branes. The imprint of this can be directly seen in the supersymmetric partition functions of the lower dimensional theories. Celebrated examples of this include the Vafa--Witten partition function of $\mathcal{N}=4$ SYM theory \cite{Vafa:1994tf}, which can be interpreted as the torus partition function of an auxiliary theory in $6-4 = 2$ dimensions \cite{Minahan:1998vr}, and the AGT correspondence relating Nekrasov's partition function \cite{Nekrasov:2002qd} of $\mathcal{N}=2$ theories in the Omega background to CFT correlators on the Gaiotto curve  \cite{Alday:2009aq}.

A second direct route by which one can connect four-dimensional theories with eight supercharges to six-dimensional theories, this time with $\mathcal{N}=(1,0)$ supersymmetry, is to view the former as torus compactifications of the latter. Generalizations of the Nekrasov partition function appear naturally in five~\cite{Nekrasov:2002qd} and six \cite{Haghighat:2013gba} dimensions, where one can consider equivariant partition functions respectively on the product of $\mathbb{C}^2$ and a circle or a torus. In 5d the partition function has an interpretation as a supersymmetric index counting BPS particles. In the 6d case, in addition to contributions from the KK modes of the 6d fields, the partition function picks up contributions from BPS strings wrapped on the torus, which carry instanton charge with respect to the 6d gauge symmetry as a consequence of the Green--Schwarz--Sagnotti--West \cite{Green:1984bx,Sagnotti:1992qw} mechanism.

A natural generalization of Nekrasov's partition function is obtained by  replacing the $\mathbb{C}^2$ component of the spacetime geometry with other toric four-manifolds~\cite{Gasparim:2009sns,Bonelli:2012ny,Fucito:2004ry,Fucito:2006kn,Bonelli:2011jx,Ito:2013kpa,Alfimov:2013cqa,Dey:2013fea,Bruzzo:2013daa,Mekareeya:2015bla,Kim:2019uqw}. The simplest example of this are the asymptotically locally Euclidean (ALE) spaces, which can be constructed as (resolutions) of orbifold singularities $X_\Gamma = \mathbb{C}^2/\Gamma$~\cite{Kronheimer:1989zs}, where $\Gamma$ is one of the finite subgroups of $SU(2)$ listed in Table \ref{tab:fin}.
\begin{table}[t]
    \centering
    \begin{tabular}{c|r|c}
$\Gamma$& Description & $\mathfrak{g}$\\
\hline\hline
$\cC_N$ & Cyclic group of order $N$ & $\su(N)$\\
$\cQ_{2N}$ & Binary dihedral group of order $4N$ & $\mathfrak{so}(2N+4)$\\
$\cT$ & Binary tetrahedral group of order $24$& $\mathfrak{e}_6$\\
$\cO$ & Binary octahedral group of order $48$& $\mathfrak{e}_7$\\
$\cI$ & Binary icosahedral group of order $120$& $\mathfrak{e}_8$\\
    \end{tabular}
    \caption{The finite subgroups of $SU(2)$ and their McKay dual.}
    \label{tab:fin}
\end{table}
The best studied example is by far the one of abelian orbifolds by $\mathbb{Z}_N\simeq \cC_N$. In this case, the partition function can be localized to the fixed points of the equivariant action and decomposes into $N$ copies of the $\mathbb{C}^2$ partition function according to Nekrasov's master formula~\cite{Nekrasov:2003vi}.  No analogue of the master formula is available for non-abelian choices of orbifold group, and these cases remain much less explored, although results are available in the context of pure 5d $\mathcal{N}=1$  $U(W)$ SYM on $S^1 \times \mathbb{C}^2/\Gamma$ \cite{Mekareeya:2015bla}.  In addition to their relevance to physics, geometric engineering provides a separate motivation for studying the partition functions on orbifolds of $\mathbb{C}^2$. Namely, they are expected to be generating functions of \emph{higher-rank BPS invariants} of ADE type on Calabi--Yau threefolds as discussed in \cite{DelZotto:2021gzy,DelZotto:2023ryf}, although the details remain to be spelled out.

Six-dimensional theories turn out to be an illuminating context in which one can disentangle the various ingredients that contribute to the equivariant partition functions on ALE spaces and understand the way they interact with each other. In this paper we focus on a class of 6d $\mathcal{N}=(1,0)$ superconformal field theories on the tensor branch which encode the dynamics of a stack of $r$ M5 branes on a transverse Taub--NUT space of charge $W$, and determine their partition functions on all backgrounds of type $T^2\ltimes \mathbb{C}^2/\Gamma$, generalizing the results of \cite{DelZotto:2023ryf,DelZotto:2023rct} for abelian orbifolds. The class of theories we consider includes in particular the 6d M-string SCFT \cite{Haghighat:2013gba}, which upon compactification gives rise to $\cN=2^*$ $U(r)$ SYM theory in 4d.
We arrive first of all at a better understanding of the contributions from the KK modes of the 6d fields, which we find can be expressed in terms of $\Gamma$-covariant Hilbert series which count holomorphic sections of vector bundles on $\mathbb{C}^2/\Gamma$ with prescribed monodromy at asymptotic infinity. We also obtain a detailed description of the 2d $\mathcal{N}=(0,4)$ relative QFTs that describe the BPS strings. We find that these can be described in terms of collections of Kronheimer--Nakajima \cite{kronheimer1990yang} quiver gauge theories interacting through interfaces. A main focus of this paper is to construct these theories and provide expressions for their elliptic genera in the form of integrals over gauge group holonomies. The elliptic genera determine the contributions of the BPS strings to the 6d partition function. The computation of the elliptic genera is carried out explicitly in a number of examples corresponding to BPS strings probing non-abelian orbifolds of $\mathbb{C}^2$.

The added bonus of working in a six-dimensional setup is that it makes the connection between equivariant partition functions and the Vafa--Witten partition function transparent. The existence of relations between them is expected based on observations made in the context of 4d $\mathcal{N}=2^*$ $SU(2)$ gauge theory \cite{Bruzzo:2013daa,Bruzzo:2014jza,Manschot:2021qqe}. Going up to six-dimensions helps to demystify this connection: the 2d degrees of freedom that contribute to the Vafa--Witten partition function (which in this setting correspond to integrable highest weight representations for the McKay dual of $\Gamma$) are localized on the same torus as the BPS strings which contribute to the equivariant partition function, and the two types of degrees of freedom interact with each other in a way which is needed to ensure the freedom from gauge anomalies of the worldsheet theory of the BPS strings.\\

The remainder of the paper is organized as follows: In Section \ref{sec:6dSCFT} we review basic properties of the 6d SCFTs $\cT^{6d}_{r,W}$ and discuss their partition function on $T^2\times\mathbb{C}^2$. In Section \ref{sec:Z6d}, after reviewing relevant aspects of the ALE geometries and introducing a notion of $\Gamma$-covariant Hilbert series for them, we discuss superselection sectors of the 6d SCFTs on ALE backgrounds and give an expression for their supersymmetric partition function on $T^2\times \mathbb{C}^2/\Gamma$ in terms of contributions from BPS particles, chiral algebras, and BPS strings. In Section \ref{sec:strings} we find a convenient description for the degrees of freedom in terms of $\Gamma$\emph{-dressed quivers}. For the case of star-shaped Dynkin diagrams, we discuss a gluing formalism for constructing the $\Gamma$-dressed quivers out of a 2d version of the $T[SU(N)]$ theories of Gaiotto and Witten. In Section \ref{sec:ell} we provide formulas for the elliptic genus of the BPS strings. In Section \ref{sec:stringsIR} we determine some basic properties of the nonlinear sigma models that describe the IR physics of the strings. In Section \ref{sec:examples} we discuss a number of concrete examples of BPS string configurations for various choices of $\Gamma$ and work out their elliptic genera. Finally, in Section \ref{sec:concl} we present our conclusions and discuss directions for future research. Further technical results are contained in the appendices, including explicit expressions for the $\Gamma$-covariant Hilbert series of ALE spaces of arbitrary type in Appendix \ref{app:Hilbert}.

\paragraph{Notation:}Throughout the paper we mark by a boldface symbol quantities that carry an upper index: $\boldsymbol{b}=(b^{(0)},\dots,b^{(r)})$, and by a vector quantities that carry a lower index, for example: $\vec{v}=(v_0,\dots,v_{\text{rk }\mathfrak{g}})$. Quantities that depend on a second lower index are underlined: $\underline{u_j} = (u_{j,1},\dots,u_{j,v_j})$.
\section{Review of the 6d SCFTs}
\label{sec:6dSCFT}
In this review section we recall basic properties of the \emph{M-string orbifold} SCFTs~\cite{Haghighat:2013tka}, a class of 6d superconformal field theories that belong to a Higgsing chain terminating on the $\mathcal{N}=(2,0)$ SCFTs that describe a stack of $r$ parallel M5 branes. We will denote this class of theories by
\be
\mathcal{T}^{6d}_{r,W},
\ee
where $r,W$ are a pair of nonnegative integers. The case $W=0$ corresponds to the $\cN=(2,0)$ SCFT, while $W = 1$ corresponds to the \emph{M-string} SCFT~\cite{Haghighat:2013gba}.\\

Recall first of all that the $\mathcal{N}=(2,0)$ SCFT describing a stack of $r$ parallel M5 branes, upon compactification on a circle of radius $R_{6d}$, gives rise to maximally supersymmetric $U(r)$ SYM theory in five dimensions, where the 5d gauge coupling gets identified with the radius of the circle:
\be
g_{5d}^2 = 8\pi R_{6d}.
\ee
We are interested in studying the tensor branch of the 6d theory, which describes the situation in which the M5 branes are separated along one common direction, say $x_6$; upon compactification to 5d, this corresponds to going on the Coulomb branch, which breaks the non-abelian gauge group $U(r)$ to $U(1)^r$. The resulting abelian gauge fields can be viewed as arising from $r$ anti-symmetric two-form fields $B_{\mu\nu}^{(1)},\dots,B_{\mu\nu}^{(r)}$ living on the individual M5 branes.

The R-symmetry, which at the superconformal point is $Sp(2)\simeq SO(5)$, is broken on the tensor branch to $SO(4)\sim SU(2)_I\times SU(2)_m$. In terms of $\mathcal{N}=(1,0)$ SUSY, $SU(2)_I$ plays the role of the R-symmetry, while $SU(2)_m$ appears as a flavor symmetry. Upon compactification to 5d, we can turn on a Wilson line for the Cartan of $SU(2)_m$:
\be
\mu = \int_{S^1_{6d}} A_{SU(2)_m}.
\ee
From the five-dimensional point of view, turning on the Wilson line corresponds to giving a mass to the adjoint hypermultiplet; the resulting theory is known as the $\mathcal{N}=1^*$ $U(r)$ SYM theory.\\

\begin{table}[t]
    \centering
    \begin{tabular}{c|cc|cccc|c|cccc}
    & \multicolumn{2}{c|}{$T^2$} 
         & \multicolumn{4}{c|}{$\mathbb{C}^2/\Gamma$} 
         & \multicolumn{1}{c|}{$\IR$} 
         & \multicolumn{4}{c}{$TN_W$} \\
         &  0 & 1 & 2 & 3 & 4 & 5 & 6 & 7 & 8 & 9 &10 \\ \hline
        $r$ $M5$& $\cross$ &$\cross$  & $\cross$ & $\cross$  & $\cross$  & $\cross$  & & & &&     \\
	$M2$& $\cross$  &$\cross$ & & & & & $\cross$ & & &&     
    \end{tabular}
    \caption{ The M-theory setup corresponding to the M-string orbifold SCFT $\cT^{6d}_{r,W}$. BPS strings arise from stretched M2 branes shown in the table.}
    \label{tab:Mset}
\end{table}
The 5d $\mathcal{N}=1^*$ theory can alternatively be obtained as the compactification of a 6d  $\mathcal{N}=(1,0)$ UV fixed point, the rank-$r$ M-string SCFT $\mathcal{T}^{6d}_{r,1}$. This theory is realized within M-theory by placing the stack of $r$ M5 branes at the origin of a single-charge Taub--NUT space $TN_1$, extended along directions $x_7,\dots,x_{10}$. The M5 branes extend along directions $x_0,\dots,x_5$, and we take the first two coordinates to parametrize a torus $T^2$ while the remaining four are now reserved for $\IC^2$ but will be later taken to parametrize $\IC^2/\Gamma$. The tensor branch again corresponds to spacing the M5 branes along $x_6$.  On the tensor branch one finds two-dimensional BPS strings charged under the two-form fields, which are realized by M2 branes suspended between neighboring M5 branes and extended along directions $x_0,x_1$. By compactifying along the circle fiber of $TN_1$ we reach a Type IIA description where the Taub--NUT is replaced by a D6 brane extended along directions $x_0,\dots,x_6$, while the M5 and M2 branes are replaced by NS5 and D2 branes respectively. The generalization to the M-string orbifold SCFT $\mathcal{T}^{6d}_{r,W}$ is obtained simply by replacing the single-centered Taub--NUT space with its $W$-centered generalization $TN_W$ which locally around the origin has a singularity of type $\mathbb{C}^2/\mathbb{Z}_W$. The M-theory setup is summarized in Table \ref{tab:Mset}. The presence of a transverse singularity leads to a 6d gauge symmetry
\be
\mathfrak{g}^{6d}=\prod_{a=1}^{r-1} \mathfrak{u}(W)^{(a)}
\label{eq:GC2}
\ee
and flavor symmetry 
\be
\mathfrak{f}^{6d}=\mathfrak{u}(W)^{(0)}\times\mathfrak{u}(W)^{(r)}.
\label{eq:FC2}
\ee
It is a well known fact \cite{Berkooz:1996iz,Douglas:1996sw,Hanany:1997gh} that the the St\"uckelberg mechanism leads to the photons corresponding to the abelian factors in Equations \eqref{eq:GC2} and \eqref{eq:FC2} acquiring a mass. The exception is the diagonal factor
\be
\mathfrak{u}(1)^{diag}
=
\text{diag}(
\mathfrak{u}(1)^{(0)}\times
\mathfrak{u}(1)^{(1)}\times
\dots\times
\mathfrak{u}(1)^{(r)}
),
\ee
which remains unbroken.
\begin{figure}[t!]
	\centering
	\scalebox{0.8}{\begin{tikzpicture}[line width=0.8]
      \coordinate (center) at (0,0);
         \draw (center) ++(-60:5) arc[start angle=-60, end angle=180, radius=5];
         \draw[dashed] (center) ++(180:5) arc[start angle=180, end angle=300, radius=5];
        \node[fill=white,circle, draw, minimum size=60pt,inner sep=1] (g0) at (0,5) {$U(r)^{(0)}$};
        \node[fill=white,circle, draw, minimum size=60pt,inner sep=1]  (gW-1) at (-4.33,2.5) {$U(r)^{(W-1)}$};
        \node[fill=white,circle, draw, minimum size=60pt,inner sep=1](g2) at (4.33,2.5) {$U(r)^{(1)}$};
        \node[fill=white,circle, draw, minimum size=60pt,inner sep=1]  (g3) at (4.33,-2.5) {$U(r)^{(2)}$};
	\end{tikzpicture}}
    \caption{The 5d $\mathcal{N}=1$ quiver gauge theory corresponding to the M-string orbifold SCFT $\mathcal{T}^{6d}_{r,W}$.}
    \label{fig:5dorbi}
\end{figure}
\\

In the Type IIA frame, the $\mathfrak{u}(W)$ gauge degrees of freedom are supported on stacks of D6 branes which are extended along directions $x_0,\dots,x_5$ and along the tensor branch direction $x_6$, and are suspended between neighboring NS5 branes. The Type IIA brane configuration is summarized in the following table, where we also indicate the orientation of the D2 branes giving rise to BPS strings:
{\begin{center}
\begin{tabular}{c|cc|cccc|c|ccc}
&0&1&2&3&4&5&6&7&8&9
\\
\hline
$r $ NS5 & $\cross$ & $\cross$  & $\cross$  & $\cross$  &$\cross$  &$\cross$  & & & & 
\\
$W $ D6 & $\cross$  & $\cross$ &  $\cross$ & $\cross$ & $\cross$ &$\cross$ & $\cross$ & & & 
\\
$\boldsymbol{v} $ D2 &  $\cross$  & $\cross$ &  & & & & $\cross$  & & &
\end{tabular}
\end{center}
}
The process of Higgsing can be understood as progressively removing D6 branes from the locus of the NS5 branes and moving them off to infinity; in the M-theory picture this corresponds to lowering the degree of the singularity at the origin by a resolution and reducing the number of centers of Taub--NUT. It is also worth noting that the $\mathcal{T}^{6d}_{r,W}$ theories also admit a dual description as 5d $\mathcal{N}=1$ quiver gauge theories with $W$ unitary gauge nodes. These quiver gauge theories are depicted in Figure~\ref{fig:5dorbi}~\cite{Haghighat:2013tka}.

\subsection{The $T^2\ltimes\mathbb{C}^2$ partition function}
\label{sec:T2C2} 
Upon compactifying the 6d theory $\cT_{r,W}^{6d}$ on a circle, one obtains a 5d KK theory~\cite{Jefferson:2018irk} which is dual to the one portrayed in Figure \ref{fig:5dorbi}. Turning on Wilson lines $\vec{s}^{(a)}=(s^{(a)}_0,\dots,s^{(a)}_{W-1})$ for the Cartan of $\mathfrak{u}(W)^{(a)},$ for $a = 0,1,\dots,r$ before compactification corresponds to deforming this 5d theory to the Coulomb branch.  As a consequence of the Stückelberg mechanism, the fugacities turn out to be related to each other~\cite{Haghighat:2013tka}:
\be
\sum_{A=1}^{W} s^{(a+1)}_A = \sum_{A=1}^{W} s^{(a)}_A + W \mathfrak{m},\qquad a = 0,\dots,r-1,
\label{eq:st}
\ee
where the shift parameter $\mathfrak{m}$ gets identified with the parameter $\mu$ in the $W=1$ case corresponding to the M-string SCFT. A natural quantity to compute from the five-dimensional perspective is the K-theoretic Nekrasov partition function \cite{Nekrasov:2002qd}, for which one takes the spacetime to be the Omega-deformed $S^1\ltimes\mathbb{C}^2_{\epsilon_1,\epsilon_2}$, where we adopt the usual convention that $\epsilon_\pm = \tfrac{\epsilon_1\pm\epsilon_2}{2}$ denote respectively the deformation parameters for $SU(2)_R$ and $SU(2)_L$, where $SU(2)_R\times SU(2)_L\sim SO(4)$ parametrize rotations of $\mathbb{C}^2$. In particular $SU(2)_R$ is identified with the group of hyperkähler rotations of $\mathbb{C}^2$. In the case corresponding to the M-string SCFT $\mathcal{T}^{6d}_{r,1}$, the 5d K-theoretic partition function is computed by summing over superselection sectors labeled by $U(r)$ instanton charge $k$, giving rise to an expansion of the form
\be
Z_{5d\, U(r)}(\boldsymbol{\phi},\mu,\epsilon_+,\epsilon_-,\tau)
=
Z^{pert}(\boldsymbol{\phi},\mu,\epsilon_+,\epsilon_-)
\sum_{k \geq 0} 
q^k Z_{5d\, U(r)}^{k \text{ inst}}(\boldsymbol{\phi},\mu,\epsilon_+,\epsilon_-),
\ee
where we denote by $\boldsymbol{\phi} = (\phi^{(1)},\dots, \phi^{(r)}) $ the vevs of the $U(r)$ vector multiplet scalars, whose differences $\varphi^{(a)}=\phi^{(a+1)}-\phi^{(a)}$ parametrize the Coulomb branch. The same partition function admits a dual interpretation in terms of the 6d theory as a sum over contributions of bound states of BPS strings~\cite{Haghighat:2013gba}:
\be
Z_{5d\, U(r)}
=
Z_{\mathcal{T}^{6d}_{r,1}}
=
q^{\frac{r}{24}}
\chi_{\mathcal{H}}(\tau)^r
Z_{\mathcal{T}^{6d}_{r,1}}^{\text{BPS particles}}(\mu,\epsilon_+,\epsilon_-,\tau)
\sum_{\boldsymbol{{\kappa}}\in\mathbb{Z}^{r-1}_{\geq 0}}
e^{-\boldsymbol{\varphi}\cdot\boldsymbol{{\kappa}}}
\mathbb{E}_{\boldsymbol{{\kappa}}}(\mu,\epsilon_+,\epsilon_-,\tau)
.
\label{eq:Z6dC2}
\ee
Here, the tuple $\boldsymbol{{\kappa}} = (\kappa^{(1)},\dots,\kappa^{(r)})$ labels a bound state of $\kappa^{(1)}$ M2 branes suspended between the first and second M5 brane, $\kappa^{(2)}$ suspended between the second and third, and so on. Moreover,
\be
\chi_{\mathcal{H}}(\tau)=\frac{1}{\eta(\tau)}
\ee
is the character of the Heisenberg algebra, which captures the degrees of freedom associated to displacing an M5 brane along the direction $x_6$, and
\be\label{eq:elliptic_genus_on_C2}
\mathbb{E}_{\boldsymbol{\kappa}}(\mu,\epsilon_+,\epsilon_-,\tau) = \Tr (-1)^F q^{L_0-\frac{c_L}{24}}\overline{q}^{\overline{L}_0-\frac{c_{R}}{24}}m^{J_{SU(2)_m}}x^{J_{SU(2)_L}}t^{J_{SU(2)_I}-J_{SU(2)_R}}
\ee
denotes the flavored elliptic genus of the worldsheet theory corresponding to the bound state of BPS strings. Our notation here is that $q = e^{2\pi i\tau}$, $m=e^{2\pi i \mu}$, $t = e^{2\pi i\epsilon_+}$, $x = e^{2\pi i \epsilon_-}$, $\tau$ is the complex structure of $T^2$, $F$ is the fermion number, $J_G$ are Cartan currents of the various global symmetries of the string, and the trace is taken with periodic boundary conditions for the fermions.\\

The prefactor $Z^{\text{BPS particles}}$ in Equation \eqref{eq:Z6dC2} is a product of contributions from 5d BPS particles arising from the 6d $\mathcal{N}=(1,0)$ tensor and hypermultiplets and their KK modes along the 6d circle. Its explicit expression is given by:
\be
Z^{\text{BPS particles}}(\mu,\epsilon_+,\epsilon_-,\tau)
=
PE
\left[
r\left(m-t + \frac{q}{1-q}\frac{(1-m t)(1-m/t)}{m}\right)t \mathcal{H}_{\mathbb{C}^2}(t,x)
\right]
\label{eq:zbpsc2}
\ee
where the plethystic exponential $PE[\dots]$ is given by
\be
PE[f(x_1,\dots,x_n)] = \exp\left(\sum_{k=1}^\infty \frac{f(x_1^k,\dots,x_n^k)}{k}\right),
\ee
and 
\be
\mathcal{H}_{\mathbb{C}^2}(t,x) = \frac{1}{(1-t x)(1-t x^{-1})}
\ee
is the Hilbert series of $\mathbb{C}^2$, whose $t$-expansion coefficients enumerate the holomorphic functions on $\mathbb{C}^2$ with given charge under the Cartan of $SU(2)_R$.\\

Generalizing now to the M-string orbifold SCFT, the 6d partition function can be written as:
\be
Z_{ \mathcal{T}^{6d}_{r,W}}
=
q^{\frac{r}{24}}\chi_{\mathcal{H}}(\tau)^r
Z^{\text{BPS particles}}_{ \mathcal{T}^{6d}_{r,W}}
(\vec{\boldsymbol{s}},\epsilon_+,\epsilon_-,\tau)
\left(
\sum_{\boldsymbol{{\kappa}}\in\mathbb{Z}^{r-1}_{\geq 0}}
e^{-\boldsymbol{\varphi}\cdot\boldsymbol{{\kappa}}}
\mathbb{E}_{\boldsymbol{{\kappa}}}(\vec{\boldsymbol{s}},\epsilon_+,\epsilon_-,\tau)
\right),
\label{eq:Z6dC2ZW}
\ee
which depends on the Wilson lines
\be
\vec{\boldsymbol{s}} = (\vec{s}^{(0)},\vec{s}^{(1)},\dots,\vec{s}^{(r)})
\ee
for the 6d $\mathfrak{u}(W)^{r+1}$ gauge and flavor symmetry. The partition function also depends implicitly on the parameter $\mathfrak{m}$ through the relation \eqref{eq:st}. We choose the $\va*{s}$ to have generic values consistent with the St\"{u}ckelberg constraint, so that no two Wilson lines for a given gauge factor are identical, and also $s_A^{(a-1)}\neq s_B^{(a)}$ for all $A\neq B$. In particular $\mathfrak{u}(W)^{(a)}$ is broken to its Cartan $\mathfrak{u}(1)^{(a)}_1\times\dots \times \mathfrak{u}(1)^{(a)}_{W}$, and we fix a Weyl chamber such that $s_A^{(a)}< s_B^{(a)}$ for $A <B$.  We also introduce the notation $M_{A}^{(a)} =e^{2\pi i s_A^{(a)}}$ for the corresponding exponentiated fugacities. The partition function \eqref{eq:Z6dC2ZW} includes a sum over instanton sectors, \emph{i.e.}\ over bound states of BPS strings carrying instanton charges $\boldsymbol{{\kappa}}$ under the 6d gauge algebra; their elliptic genera, which encode the contributions to the partition function, were determined in \cite{Haghighat:2013tka}. The BPS particle factor in the partition function, $Z^{\text{BPS particles}}_{ \mathcal{T}^{6d}_{r,W}}$, receives contributions from the KK modes of the tensor, vector, and hypermultiplets of the theory along the 6d circle. Keeping in mind the embedding of the SCFT into M-theory/Type IIA, we can write this term as a product of contributions from the individual M5/NS5 branes:
\be
Z^{\text{BPS particles}}_{ \mathcal{T}^{6d}_{r,W}} = \prod_{a=1}^{r} Z_{M5/TN_W}^{(a)}.
\ee
We can express the factor associated to a given fivebrane in terms of the plethystic exponential as:
\bea
\nonumber
&&
Z_{M5/TN_W}^{(a)}
=
PE\bigg[
-\frac{t\mathcal{H}_{\mathbb{C}^2}(t,x)}{1-q}
\bigg(
\sum_{\substack{A,B=1\\A\leq B}}^{W}\left(t \frac{M^{(a)}_B}{M^{(a)}_A}+qt^{-1}\frac{M^{(a)}_A}{M^{(a)}_B}\right)
\\
\nonumber
&&
+
\sum_{\substack{A,B=1\\A < B}}^{W}\left(t^{-1} \frac{M^{(a-1)}_B}{M^{(a-1)}_A}+qt\frac{M^{(a-1)}_A}{M^{(a-1)}_B}\right)
-\sum_{(A,B)\in{\mathfrak{S}^{(a)}}}\left(\frac{M^{(a)}_B}{M^{(a-1)}_A}+
q \frac{M^{(a-1)}_A}{M^{(a)}_B}\right)
\\
&&
-\sum_{(A,B)\in{\overline{\mathfrak{S}}^{(a)}}}\left(\frac{M^{(a-1)}_A}{M^{(a)}_B}+q \frac{M^{(a)}_B}{M^{(a-1)}_A}\right)
\bigg)
\bigg].
\label{eq:five}
\eea
In this expression, the set $\mathfrak{S}^{(a)}$ contains the pairs of indices $(A,B)$ with $A=1,\dots,W$, $B=1,\dots,W$ such that $s_{A}^{(a-1)}\leq s_B^{(a)}$, and $\overline{\mathfrak{S}^{(a)}}$ denotes its complement in the set of all $(A,B)$. This guarantees that \eqref{eq:five} only receives contributions from BPS hypermultiplets within the given chamber specified by the choice of $\va*{s}$. \\

Consistent with the 6d/5d duality discussed above, the 6d partition function~\eqref{eq:Z6dC2ZW} coincides with the 5d Nekrasov partition function for the quiver gauge theory of Figure \ref{fig:5dorbi}, upon performing a nontrivial mapping between the parameters of the dual 6d and 5d theories, which can be seen most easily by realizing the dual pair of theories in terms of $(p,q)$-webs \cite{Haghighat:2013tka}. For further details on the equivariant partition functions on $T^2\ltimes\mathbb{C}^2$ and their computation, we refer the reader to the articles~\cite{Haghighat:2013tka,Hohenegger:2013ala,DelZotto:2023ryf}.

\section{The  partition functions on $T^2 \ltimes \mathbb{C}^2/\Gamma$}
\label{sec:Z6d}

In this section we define a supersymmetric partition function for the 6d M-string SCFT and its orbifold theories $\mathcal{T}^{6d}_{r,W}$ on the equivariant background $T^2 \ltimes \mathbb{C}^2/\Gamma$ for arbitrary $\Gamma $. We begin in Section \ref{sec:ALE} with a discussion of relevant aspects of ALE spaces; in Section \ref{sec:gammaH} we introduce the notion of $\Gamma$-covariant Hilbert series, which encode the dependence of the partition function on spacetime degrees of freedom; in Section \ref{sec:sel} we discuss the data required to specify superselection sectors of the 6d theories which enter the definition of the partition function; finally, in Section \ref{sec:contr} we discuss the various BPS objects that contribute to the partition function, whose complete expression we present in Equations \eqref{eq:PF} and \eqref{eq:PFM}.

\subsection{ALE spaces}
\label{sec:ALE}

It is a well known result that four-dimensional, asymptotically locally Euclidean hyperkähler manifolds admit a classification in terms of $ADE$ Dynkin diagrams~\cite{kronheimer1993construction}, which arises in constructing them as resolutions $\widetilde{\mathbb{C}^2/\Gamma}$ of orbifold singularities of  $\mathbb{C}^2$ by a discrete subgroup $\Gamma\in SU(2)_L $. The possible choices of subgroup, as well as the (affine) Dynkin diagrams associated to them via the McKay correspondence, are displayed in Figure \ref{fig:ADEc}. The irreducible representations of the discrete group $\Gamma$ are in one-to-one correspondence with the nodes of the associated affine Dynkin diagram. In particular the number of irreps is given by $\text{rk }\mathfrak{g}+1$, where $\mathfrak{g}$ is the McKay dual Lie algebra of $\Gamma$.  We will denote the irreducible representations of the discrete group as $(\rho_0,\dots,\rho_{\text{rk }\mathfrak{g}})$. The dimension of an irreducible representation is given by the comark of the corresponding Dynkin diagram node:
\be
\text{dim}(\rho_j) = a_j.
\ee
We will sometimes also denote these irreps in terms of their dimensions as $\underline{a_j}$, and employ superscripts as needed to distinguish between representations of identical dimensions, as displayed in Figure \ref{fig:ADEc}. The order of $\Gamma$ is given in terms of the comarks as
\begin{equation}\label{eq:orderGamma}
	|\Gamma|= \sum_{j=0}^{\mathrm{rk}\;\mathfrak{g}} a_j^2. 
\end{equation}

The second homology of $H_2(\widetilde{\mathbb{C}^2/\Gamma},\mathbb{Z})$ is generated by a collection of genus zero curves $\Sigma_1,\dots,\Sigma_{\text{rk }\mathfrak{g}}$ whose intersection  matrix is given by:
\be
\Sigma_j\cdot\Sigma_k = -(C^{\mathfrak{g}})_{j k}
\ee
where $C^{\mathfrak{g}}$ is the Cartan matrix associated to the Lie algebra $\mathfrak{g}$. For $\Gamma = \mathcal{C}_N\simeq \mathbb{Z}_N$ (the cyclic group of order $N$) and $\mathcal{Q}_N$ (the binary dihedral group of order $2N$ for $N$ even), the ALE spaces can be obtained as deformations of corresponding asymptotically locally flat (ALF) hyperkähler spaces $ALF^\Gamma$. These can be viewed as circle fibrations over $\mathbb{R}^3$, where the radius of the circle fiber attains an asymptotic value $R_\infty$ on $\partial\mathbb{R}^3$. The ALE limit is obtained by taking $R_\infty\to\infty$. In particular, for $\Gamma = \mathcal{C}_N$ the space $ALF^\Gamma$ is the $N$-centered Taub--NUT space $TN_N$. On the other hand, there is no ALF space corresponding to orbifolds by the binary tetrahedral group $\mathcal{T},$ the binary octahedral group $\mathcal{O}$, or the binary icosahedral group $\mathcal{I}$.\\

\begin{figure}

		\begin{tabular}{c|c|cc}
			$\Gamma$ & $\mathfrak{g}$ & &Dynkin diagram\\ \hline\hline
			%%%%%%%%%%%%%%%
			%                       A
			%%%%%%%%%%%%%%%
			$\mathcal{C}_N$&		$\mathfrak{su}(N)$ & $A_{N-1}:$&
			\raisebox{-0.5\height}{\scalebox{0.65}{\begin{tikzpicture}[line width=0.8]
						%Global
						\def\shift{0.55};
						%A-type
						\def\yA{0};
						\def\aA{1.2};
						\node at (-1,0) {$\phantom{ZZZZZ}$};
						\node[smallgauge]   (a0) at (3.75*\aA,\yA+2) {};
						\node at (3.75*\aA,\yA+2-\shift) {$0$};
						\node at (3.75*\aA,\yA+2+\shift) {$(\underline{1},\vb*{1})$};
						\node[smallgauge]   (a1) at  (0,\yA) {};
						\node at (0,\yA-\shift) {$1$};
						\node at (0,\yA+\shift) {$(\underline{1}^{(1)},\vb*{N})$};
						\node[smallgauge]   (a2) at (3*\aA,\yA) {};
						\node at (3*\aA,\yA-\shift) {$2$};
						\node at (3*\aA,\yA+\shift) {$(\underline{1}^{(2)},\tfrac{\vb*{N(N+1)}}{2})$};
						\node[smallgauge]   (an-1) at (7.5*\aA,\yA) {};
						\node at (7.5*\aA,\yA-\shift) {$N-1$};
						\node at (8.4*\aA-0.5,\yA+\shift) {$(\underline{1}^{(N-1)},\tfrac{\vb*{N(N-1)}}{2})$};
						\draw   (a0)--(a1);
						\draw  (a1)--(a2);
						\draw  (a2)--(4*\aA,0);
						\draw[dashed] (4*\aA,0)--(6.5*\aA,0);
						\draw   (6.5*\aA,0)--(an-1);
						\draw   (an-1)--(a0);
			\end{tikzpicture}}}\\%
			%%%%%%%%%%%%%%%
			%                       D
			%%%%%%%%%%%%%%%
			$\mathcal{Q}_{N} $  &		$\mathfrak{so}(N+4)$ & $D_{\frac{N}{2}+2}:$& \raisebox{-0.5\height}{\scalebox{0.65}{\begin{tikzpicture}[line width=0.8]
						\def\shift{0.55};
						\def\aD{1.2}
						\def\yD{0}
						\node at (-1,0) {$\phantom{ZZZZZ}$};
						\node[smallgauge]   (d0) at (0,\yD-1.5) {};
						\node at (0,\yD-1.5-\shift) {$0$};
						\node at (0,\yD-1.5+\shift) {$(\underline{1},\vb*{1})$};
						\node[smallgauge]   (d1) at (0,\yD+1.5) {};
						\node at (0,\yD+1.5-\shift) {$1$};
						\node at (0,\yD+1.5+\shift) {$(\underline{1}^{v},\vb*{N+4})$};
						\node[smallgauge]   (d2) at (2*\aD,\yD) {};
						\node at (2*\aD,\yD-\shift) {$2$};
						\node at (3.1*\aD+0.2,\yD+\shift) {$(\underline{2}^{(1)},\vb*{(\tfrac{N}{2}+2)(N+3)})$};
						\node[smallgauge]   (dn-2) at (7.5*\aD,\yD) {};
						\node at (7.5*\aD,\yD-1.2*\shift) {$\frac{N}{2}$};
						\node at (6.7*\aD,\yD+1.2*\shift) {$(\underline{2}^{(\tfrac{N}{2}-1)},\vb*{\vb*{\smqty(N+4\\N/2)}})$};
						\node[smallgauge]   (dn) at (9.5*\aD,\yD-1.5) {};
						\node at (9.5*\aD,\yD-1.5-\shift) {$\frac{N}{2}+2$};
						\node at (9.8*\aD+0.3,\yD-1.5+\shift) {$(\underline{1}^s,\vb*{(2^{\frac{N+2}{2}})}^s)$};
						\node[smallgauge]   (dn-1) at (9.5*\aD,\yD+1.5) {};
						\node at (9.5*\aD,\yD+1.5-\shift) {$\frac{N}{2}+1$};
						\node at (9.8*\aD,\yD+1.5+\shift) {$(\underline{1}^c,\vb*{(2^{\frac{N+2}{2}})}^c)$};
						\draw (d0)--(d2);
						\draw (d1)--(d2);
						\draw (d2)--(3*\aD,\yD);
						\draw[dashed] (3*\aD,\yD)--(6.5*\aD,\yD);
						\draw (6.5*\aD,\yD)--(dn-2);
						\draw (dn-2)--(dn-1);
						\draw (dn-2)--(dn);
			\end{tikzpicture}}}\\ 
			%%%%%%%%%%%%%%%
			%                       E6
			%%%%%%%%%%%%%%%
			$\mathcal{T}$&		$\mathfrak{e}_6$ & $E_6:$ & \raisebox{-0.5\height}{\scalebox{0.65}{\begin{tikzpicture}[line width=0.8]
						\def\shift{0.55};
						\def\aEsix{1.3}
						\def\ayEsix{0.8}
						\def\yEsix{0}
						\node[smallgauge]   (e6-1) at (0,\yEsix) {};
						\node at (0,\yEsix-\shift) {$1$};
						\node at (0,\yEsix+\shift) {$(\underline{1}',\vb*{27})$};
						\node[smallgauge]   (e6-2) at (2*\aEsix,\yEsix) {};
						\node at (2*\aEsix,\yEsix-\shift) {$2$};
						\node at (2*\aEsix,\yEsix+\shift) {$(\underline{2}',\vb*{\overline{35}})$};
						\node[smallgauge]   (e6-3) at (4*\aEsix,\yEsix) {};
						\node at (4*\aEsix,\yEsix-\shift) {$3$};
						\node at (4*\aEsix,\yEsix-2*\shift) {$(\underline{3},\vb*{6925})$};
						\node[smallgauge]   (e6-4) at (6*\aEsix,\yEsix) {};
						\node at (6*\aEsix,\yEsix-\shift) {$4$};
						\node at (6*\aEsix,\yEsix+\shift) {$(\underline{2}'',\vb*{35})$};
						\node[smallgauge]   (e6-5) at (8*\aEsix,\yEsix) {};
						\node at (8*\aEsix,\yEsix-\shift) {$5$};
						\node at (8*\aEsix,\yEsix+\shift) {$(\underline{1}'',\vb*{27}')$};
						\node[smallgauge]   (e6-6) at (4*\aEsix,\yEsix+2*\ayEsix) {};
						\node at (4*\aEsix-\shift,\yEsix+2*\ayEsix) {$6$};
						\node at (4*\aEsix+1.8*\shift,\yEsix+2*\ayEsix) {$(\underline{2},\vb*{78})$};
						\node[smallgauge]   (e6-0) at (4*\aEsix,\yEsix+4*\ayEsix) {};
						\node at (4*\aEsix-\shift,\yEsix+4*\ayEsix) {$0$};
						\node at (4*\aEsix+1.6*\shift,\yEsix+4*\ayEsix) {$(\underline{1},\vb*{1})$};
						\draw (e6-1)--(e6-2);
						\draw (e6-2)--(e6-3);
						\draw (e6-3)--(e6-4);
						\draw (e6-3)--(e6-6);
						\draw (e6-4)--(e6-5);
						\draw (e6-6)--(e6-0);
			\end{tikzpicture}}}\\%
			%%%%%%%%%%%%%%%
			%                       E7
			%%%%%%%%%%%%%%%
			$\mathcal{O} $  &		$\mathfrak{e}_7$ & $E_7:$ &  \raisebox{-0.5\height}{ \scalebox{0.65}{\begin{tikzpicture}[line width=0.8]
			\def\shift{0.55};
			\def\aEsev{0.85}
			\def\yEsev{0}
			\node[smallgauge]   (e7-0) at (0,\yEsev) {};
			\node at (0,\yEsev-\shift) {$0$};
			\node at (0,\yEsev+\shift) {$(\underline{1},\vb*{1})$};
			\node[smallgauge]   (e7-1) at (2*\aEsev,\yEsev) {};
			\node at (2*\aEsev,\yEsev-\shift) {$1$};
			\node at (2*\aEsev,\yEsev-2*\shift) {$(\underline{2},\vb*{133})$};
			\node[smallgauge]   (e7-2) at (4*\aEsev,\yEsev) {};
			\node at (4*\aEsev,\yEsev-\shift) {$2$};
			\node at (4*\aEsev,\yEsev+\shift) {$(\underline{3},\vb*{8645})$};
			\node[smallgauge]   (e7-3) at (6*\aEsev,\yEsev) {};
			\node at (6*\aEsev,\yEsev-\shift) {$3$};
			\node at (6*\aEsev,\yEsev-2*\shift) {$(\underline{4},\vb*{365750})$};
			\node[smallgauge]   (e7-4) at (8*\aEsev,\yEsev) {};
			\node at (8*\aEsev,\yEsev-\shift) {$4$};
			\node at (8*\aEsev,\yEsev+\shift) {$(\underline{3}',\vb*{27664})$};
			\node[smallgauge]   (e7-5) at (10*\aEsev,\yEsev) {};
			\node at (10*\aEsev,\yEsev-\shift) {$5$};
			\node at (10*\aEsev,\yEsev-2*\shift) {$(\underline{2}',\vb*{1539})$};
			\node[smallgauge]   (e7-6) at (12*\aEsev,\yEsev) {};
			\node at (12*\aEsev,\yEsev-\shift) {$6$};
			\node at (12*\aEsev,\yEsev+\shift) {$(\underline{1}',\vb*{56})$};
			\node[smallgauge]   (e7-7) at (6*\aEsev,\yEsev+2) {};
			\node at (6*\aEsev-\shift,\yEsev+2) {$7$};
			\node at (6*\aEsev+2.2*\shift,\yEsev+2) {$(\underline{2}'',\vb*{912})$};
			\draw (e7-0)--(e7-1);
			\draw (e7-1)--(e7-2);
			\draw (e7-2)--(e7-3);
			\draw (e7-3)--(e7-4);
			\draw (e7-4)--(e7-5);
			\draw (e7-5)--(e7-6);
			\draw (e7-3)--(e7-7);
\end{tikzpicture}}}\\
			%%%%%%%%%%%%%%%%
			%						E8
			%%%%%%%%%%%%%%%%
			$\mathcal{I} $  &		$\mathfrak{e}_8$ &$E_8:$ & \raisebox{-0.5\height}{ \scalebox{0.65}{\begin{tikzpicture}[line width=0.8]
						\def\shift{0.55};
						\def\aEott{0.95};
						\def\yEott{-17.5};
						\node[smallgauge]   (e8-0) at (0,\yEott) {};
						\node at (0,\yEott-\shift) {$0$};
						\node at (0,\yEott+\shift) {$(\underline{1},\vb*{1})$};
						\node[smallgauge]   (e8-1) at (2*\aEott,\yEott) {};
						\node at (2*\aEott,\yEott-\shift) {$1$};
						\node at (2*\aEott,\yEott+\shift) {$(\underline{2},\vb*{248})$};
						\node[smallgauge]   (e8-2) at (4*\aEott,\yEott) {};
						\node at (4*\aEott,\yEott-\shift) {$2$};
						\node at (4*\aEott,\yEott+\shift) {$(\underline{3},\vb*{30380})$};
						\node[smallgauge]   (e8-3) at (6*\aEott,\yEott) {};
						\node at (6*\aEott,\yEott-\shift) {$3$};
						\node at (6*\aEott,\yEott-2*\shift) {$(\underline{4},\vb*{2450240})$};
						\node[smallgauge]   (e8-4) at (8*\aEott,\yEott) {};
						\node at (8*\aEott,\yEott-\shift) {$4$};146325270
						\node at (8*\aEott,\yEott+\shift) {$(\underline{5},\vb*{146325270})$};
						\node[smallgauge]   (e8-5) at (10*\aEott,\yEott) {};
						\node at (10*\aEott,\yEott-\shift) {$5$};
						\node at (10*\aEott,\yEott-2*\shift) {$(\underline{6},\vb*{6899079269})$};
						\node[smallgauge]   (e8-6) at (12*\aEott,\yEott) {};
						\node at (12*\aEott,\yEott-\shift) {$6$};
						\node at (12*\aEott,\yEott+\shift) {$(\underline{4}',\vb*{6696000})$};
						\node[smallgauge]   (e8-7) at (14*\aEott,\yEott) {};
						\node at (14*\aEott,\yEott-\shift) {$7$};
						\node at (14*\aEott,\yEott-2*\shift) {$(\underline{2}',\vb*{3875})$};
						\node[smallgauge]   (e8-8) at (10*\aEott,\yEott+2) {};
						\node at (10*\aEott-\shift,\yEott+2) {$8$};
						\node at (10*\aEott+2.8*\shift,\yEott+2) {$(\underline{3}',\vb*{147250})$};
						\draw (e8-0)--(e8-1);
						\draw (e8-1)--(e8-2);
						\draw (e8-2)--(e8-3);
						\draw (e8-3)--(e8-4);
						\draw (e8-4)--(e8-5);
						\draw (e8-5)--(e8-6);
						\draw (e8-6)--(e8-7);
						\draw (e8-5)--(e8-8);
			\end{tikzpicture}}}
		\end{tabular}

\caption{ Correspondence between discrete subgroups $\Gamma$ of $SU(2)$, simply-laced Lie algebras $\mathfrak{g}$, and affine Dynkin diagrams. The label $(\underline{r_j},\boldsymbol{R}_j)$ of the $j$-th node in the Dynkin diagram indicates the corresponding irreducible representations of $\Gamma$ and $\mathfrak{g}$.}
\label{fig:ADEc}
\end{figure}

We will also need to use some basic facts about vector bundles on ALE spaces~\cite{KronheimerNakajima,gocho1992einstein}. For a given $\widetilde{\mathbb{C}^2/\Gamma}$, there exists a canonical set of vector bundles 
\be\label{eq:canonical_vector_bundles}
\mathcal{R}_1, \dots, \mathcal{R}_{\text{rk }\mathfrak{g}}
\ee
in one-to-one correspondence with the nodes of the Dynkin diagram of $\Gamma$, whose rank is given by the comark of the corresponding node. The vector bundles are equipped with an anti-self-dual connection, and their first Chern classes $c_1(\mathcal{R}_j)$ form a basis of $H^2(\widetilde{\mathbb{C}^2/\Gamma})$. Moreover, the boundary of the ALE space has a nontrivial first homotopy group $\pi_1(\partial ALE^\Gamma)\simeq \Gamma$, and upon parallel transport along paths at asymptotic infinity the fiber of the bundle $\mathcal{R}_j$ transforms as the $a_j$-dimensional representation $\rho_j$ of $\Gamma$. A generic rank $W$ complex vector bundle $\mathcal{V}$ with anti-selfdual connection can be decomposed in terms of this basis as:
\be
\mathcal{V} = \bigoplus_{j=0}^{\text{rk }\mathfrak{g}} w_j \mathcal{R}_j,\qquad \sum_{j=0}^{\text{rk }\mathfrak{g}} a_j w_j = W.
\ee
Its topological classes are specified in terms of a pair of $(\text{rk }\mathfrak{g}+1)$-tuples $(w_0,\dots,w_{\text{rk }\mathfrak{g}})$ and $(v_0,\dots,v_{\text{rk }\mathfrak{g}})$ of nonnegative integers. Specifically, its first Chern class is given by:
\be
c_1(\mathcal{V}) = \sum_{j=1}^{\text{rk }\mathfrak{g}}u_j c_1(\mathcal{R}_j),
\ee
where
\be
\qquad {u}_j = {w_j} - (C^{\widehat{\mathfrak{g}}}\cdot\vec{v})_j\in \mathbb{Z} \qquad j=1,\dots,\text{rk }\mathfrak{g}
\label{eq:uw}
\ee
and $C^{\widehat{\mathfrak{g}}}$ is the affine Cartan matrix of $\mathfrak{g}$. The second Chern character is likewise given in terms of the $ch_2(\cR_j)$, which are generally fractional and can be inferred from~\cite{KronheimerNakajima}:
\begin{equation}
    N_j=\int_{\widetilde{\IC^2/\Gamma}}ch_2(\cR_j) =\frac{1}{|\Gamma|}\sum_{k=1}^{\text{rk }\mathfrak{g}}\qty(C^{\mathfrak{g}})^{-1}_{jk} a_k.
\end{equation}
From this one obtains:
\be
    N_{\mathcal{V}}=\int_{\widetilde{\IC^2/\Gamma}}ch_2(\mathcal{V})=\sum_{j=1}^{\text{rk }\mathfrak{g}} u_j N_j + \frac{\sum_{j=0}^{\text{rk }\mathfrak{g}} a_j v_j}{|\Gamma|}.
    \label{eq:ch2V}
\ee
\newline
In the Type IIA setup the K\"ahler parameters get complexified by the addition of the $B$-field:
\be
\int_{\Sigma_j}\eta \to \int_{\Sigma_j}\eta +i \int_{\Sigma_j}B^{NS},
\ee
where we denote by $\eta$ the K\"ahler form on $ALE^\Gamma$. The supersymmetric partition function is expected to be invariant under K\"ahler deformations of the ALE space~\cite{DelZotto:2021gzy,DelZotto:2023rct}, and in the remainder of the paper we will be considering the orbifold limit where the real component is switched off but the B-field is switched on.\\

In this paper we work equivariantly with respect to the isometries of the orbifold spaces $\mathbb{C}^2/\Gamma$. As the action of $\Gamma$ is embedded in $SU(2)_L\subset SU(2)_L\times SU(2)_R$, the orbifold space inherits the $SU(2)_R$ isometry from $\mathbb{C}^2$. On the other hand, the action of $\Gamma$ on $\mathbb{C}^2$ does not commute with $SU(2)_L$ except for $\Gamma=\mathcal{C}_N$, in which case it leaves the Cartan subgroup $U(1)_L$ unbroken. Therefore, for the arbitrary orbifold space $\mathbb{C}^2/\Gamma$ one can turn on an equivariant parameter $\epsilon_+$ for $SU(2)_R$, and for $\Gamma = \mathcal{C}_N$ we can turn on a second equivariant parameter $\epsilon_-$ for $U(1)_L$.

\subsection{$\Gamma$-covariant Hilbert series}
\label{sec:gammaH}
The Hilbert series of $\mathbb{C}^2$ is defined as the generating function of number of holomorphic functions over $\mathbb{C}^2$ of given degree. It is given by:
\be
\mathcal{H}(t) = \frac{1}{(1-t)^2} = \sum_{n=0}^\infty t^n {(n+1)}.
\ee
Let us parametrize $\mathbb{C}^2$ by two complex variables $(z_1,z_2)$, which are rotated by the isometry group $SU(2)_R\times SU(2)_L$. The Cartan of $SU(2)_R$ and $SU(2)_L$ act respectively as $(z_1,z_2)\to (t z_1,t z_2)$ and $(z_1,z_2)\to (x z_1,x^{-1} z_2)$. The space of polynomials in $(z_1,z_2)$ of fixed degree $n$ forms an irreducible $(n+1)$-dimensional representation $\boldsymbol{n+1}$ of $SU(2)_L$, and we can define an equivariant Hilbert series that encodes the $SU(2)_L$ representation content:
\be
\mathcal{H}(t,x)
=
\frac{1}{(1- t x)(1- t x^{-1})}
=
\sum_{n=0}^\infty t^n \boldsymbol{(n+1)},
\ee
where by abuse of notation we denote by $\boldsymbol{n}= \frac{x^{-n}-x^{n}}{x^{-1}-x}$ the character of the $n$-dimensional irreducible representation of $SU(2)$. \\

The embedding of a discrete group $\Gamma$ into $SU(2)_L$ gives rise to a branching of irreducible representations of $SU(2)_L$ into irreducible representations $\rho_0,\dots,\rho_{\text{rk }\mathfrak{g}}$ of $\Gamma$. This allows us to decompose the Hilbert series of $\mathbb{C}^2$ into a set of $\Gamma$\emph{-covariant} Hilbert series which count holomorphic functions that transform under a given irreducible representation of $\Gamma$:
\footnote{ Hilbert series for covariants are well known objects in representation theory and invariant theory \cite{Konstant1984,springer_poincare_1987,broer1994hilbert,kostant2006coxeter}, which recently have also found applications in particle physics \cite{Grinstein:2023njq}.}
\be
\sum_{n=0}^\infty t^n \boldsymbol{(n+1)}
=
\sum_{j=0}^{\text{rk }\mathfrak{g}}
\rho_j \mathcal{H}^\Gamma_{\rho_j}(t)
.
\ee
Evaluated on the identity element of $\Gamma$ this implies the identity
\be
\mathcal{H}(t)
=
\sum_{j=0}^{\text{rk }\mathfrak{g}}
a_j \mathcal{H}^\Gamma_{\rho_j}(t).
\ee

In particular, the Hilbert series $\mathcal{H}^{\Gamma}_{\rho_0}$ for the trivial representation $\rho_0$ of $\Gamma$ coincides with the standard Hilbert series which counts holomorphic functions on $\mathbb{C}^2/\Gamma$, for which expressions are well known, see \emph{e.g.}\ \cite{Benvenuti:2006qr}. On the other hand, we can interpret the $\Gamma$-covariant Hilbert series for the remaining representations as counting holomorphic sections of the vector bundles $\mathcal{R}_j$ over $\mathbb{C}^2/\Gamma$ which transform with the corresponding monodromy. Explicit expressions for the corresponding Hilbert series are provided in Appendix \ref{app:Hilbert}.

\subsection{Superselection sectors}
\label{sec:sel}
Let us now turn to the discussion of superselection sectors for the partition functions on $\mathbb{C}^2/\Gamma$. The data required to specify a superselection sector consist both of discrete and of continuous parameters. We first discuss the discrete parameters. Since the asymptotic boundary of $\mathbb{C}^2/\Gamma$ has nontrivial first homotopy group,
\be
\pi_1(\partial (\mathbb{C}^2/\Gamma)) = \Gamma,
\ee
we can allow for the possibility of gauge field configurations with nontrivial monodromy at infinity for each of the gauge and background gauge fields of the theory on the tensor branch. Specifically, for a $\mathfrak{u}(R)$ gauge field  a choice of flat connection at infinity determines a choice of an element
\be
\rho\in\mathrm{Hom}\qty(\pi_1(\partial (\mathbb{C}^2/\Gamma)),\mathfrak{u}(R)),
\ee
that is, an $R$-dimensional representation of $\Gamma$.
In the present context, the theory $\mathcal{T}^{6d}_{r,W}$ depends on the following monodromy data:
\begin{itemize}
\item For each factor $\mathfrak{u}(W)^{(a)}$, $a=0,\dots, r$, of the 6d gauge and flavor symmetry, a choice of representation
\be
\rho^{(a)}= \sum_{j=0}^{\mathrm{rk}\;\mathfrak{g}} w^{(a)}_j \rho_j, \qquad \sum_{j=0}^{\mathrm{rk}\;\mathfrak{g}} w^{(a)}_j a_j = W.
\label{eq:mondec}
\ee
This has the effect of breaking the gauge symmetry into factors which commute with the action of $\Gamma$:
\be
\mathfrak{u}(W)^{(a)}\to \mathfrak{u}(w^{(a)}_0)\times \dots\times \mathfrak{u}(w^{(a)}_{\text{rk }\mathfrak{g}}),
\label{eq:breaking}
\ee
Notice that generally this results in a reduction of rank of the gauge symmetry, except for the case $\Gamma = \mathcal{C}_N$ in which all irreps of $\Gamma$ are one-dimensional and $\sum_j w_j^{(a)}=W$.
\item Each of the $r$ two-form fields admits a dual 5d interpretation as an abelian gauge field $A^{KK,(a)}_\mu$, for which we must also specify a choice of one-dimensional representation of $\Gamma$, that is, a choice of:
\be
\rho^{KK,(a)}
\in
\begin{cases}
\{\rho_0,\dots,\rho_{N-1}\} \qquad & \Gamma = \mathcal{C}_N\\
\{\rho_0,\rho_1,\rho_{N/2+1},\rho_{N/2+2}\}  & \Gamma = \mathcal{Q}_N\\
\{\rho_0,\rho_1,\rho_5\} \qquad & \Gamma = \mathcal{T}\\
\{\rho_0,\rho_6\} \qquad & \Gamma = \mathcal{O}\\
\{\rho_0\} \qquad & \Gamma = \mathcal{I}\\
\end{cases}
,
\ee
which we equivalently can denote in terms of the integrable highest weight representation (i.h.w.r.)\ $\omega^{KK,(a)}$ associated to the corresponding node of the ADE Dynkin diagram for $\Gamma$.
\end{itemize}
 Additionally, for the abelian global symmetry group factor $\mathfrak{u}(1)^{diag}$ one also needs to specify a choice of first Chern class $\vec u^{diag} \in H_2(\widetilde{\mathbb{C}^2/\Gamma})$; due to the St\"{u}ckelberg mechanism, this coincides with the first Chern class for each of the $\mathfrak{u}(W)^{(a)}$ factors:
\be
u^{diag}_j = u^{(a)}_j \qquad\forall a=0,\dots,r.
\label{eq:c1diag}
\ee
In particular, it coincides with the first Chern class of the flavor symmetry group $\mathfrak{u}(W)^{(0)}$, which is given by Equation \eqref{eq:uw}:
\be
u^{(0)}_j = w^{(0)}_j - (C^{\widehat{\mathfrak{g}}}\cdot\vec{v}^{(0)})_j = w^{(0)}_j.
\label{eq:uw0}
\ee
We have chosen to set $\vec{v}^{(0)}=\vec{v}^{(r)} = 0$; nonzero values would correspond to computing the partition function in the presence of defect BPS strings of infinite tension. While this is a natural and straightforward extension of our work, in the present paper we restrict our attention to the computation of the partition function without such insertions. As a consequence of \eqref{eq:uw0}, the first Chern class parameter $\vec{u}^{diag}$ is identified with the flavor symmetry group monodromy $\vec{w}^{(0)}$ and therefore does not give rise to additional superselection sector data. We also remark that the requirement that the parameters $v_j^{(a)}$ in Equation \eqref{eq:uw0} be nonnegative integers places restrictions on the possible values that can be taken by the $w^{(a)}_j$ for $a = 1,\dots,r-1$, as discussed for the $\Gamma=\mathcal{C}_N$ case in \cite{DelZotto:2023ryf}. In the case of M-strings in particular the only solution is that, for all $a$,
\be
\vec{w}^{(a)}=\vec{w}^{(0)}\qquad\text{and}\qquad \vec{v}^{(a)} = \kappa^{(a)}(a_0,a_1,\dots,a_{\text{rk }\mathfrak{g}}),
\ee
where $\kappa^{(a)}\in\mathbb{Z}_{\geq 0}$.\\

Let us now turn to the continuous data. First of all, we must specify a choice of tensor branch parameters in terms of the vevs of the tensor multiplet scalars:
\footnote{ The center of mass parameter $\sum_{a=1}^{r} \phi^{(a)}$ decouples.}
\be
\varphi^{(a)} = \phi^{(a+1)}-\phi^{(a)}.
\ee
 Moreover, one can also turn on Wilson lines for the $\mathfrak{u}(W)^{(a)}$. These must be compatible with the monodromy of the vector field, \emph{i.e.}\ given the decomposition \eqref{eq:mondec}, we are allowed to turn on Wilson lines for a block-diagonal set of components of the gauge field:
\bea
\nonumber
&&\left\{\int_{S^1_{6d}} A^{\mathfrak{u}(W)^{(a)}}\right\}_{A,B=1,\dots,W}
\\
&&=
\text{diag}
\left(
s^{(a)}_{0,1} \rho^{(a)}_0,
\dots,
s^{(a)}_{0,w^{(a)}_0}\rho^{(a)}_0,
\cdots,
s^{(a)}_{\text{rk }\mathfrak{g},1} \rho^{(a)}_{\text{rk }\mathfrak{g}},
\dots,
s^{(a)}_{\text{rk }\mathfrak{g},w^{(a)}_{\text{rk }\mathfrak{g}}}\rho^{(a)}_{\text{rk }\mathfrak{g}}
\right),
\eea
as well as chemical potentials $\vec\xi$ which couple to the first Chern classes of the $A^{KK,(a)}_\mu$ gauge fields \cite{DelZotto:2023ryf}.\\

A final remark is that we expect the partition function to transform covariantly under the outer automorphism group $\mathcal{O}(\widehat{\mathfrak{g}})$ of the corresponding affine algebra $\widehat{\mathfrak{g}}$, whose generators act as follows on the labels of the Dynkin diagram:
{\begin{center}
\begin{tabular}{c|c|ccc}
$\Gamma$ & $\mathcal{O}(\widehat{\mathfrak{g}})$& &$o$\\
\hline\hline
$\mathcal{C}_N$& $\mathbb{Z}_N$ &$(0,1,\dots,n-2,n-1)$& $\mapsto$ & $(1,2,\dots,n-1,0)$\\
\hline
\multirow{2}*{$\mathcal{Q}_{4n}$}& \multirow{2}*{$\mathbb{Z}_2\times\mathbb{Z}_2$} &$(0,1,2,\dots,2n+1,2n+2)$& $\mapsto$ & $(1,0,2,\dots,2n+2,2n+1)$\\
&&$(0,1,2,\dots,2n+1,2n+2)$& $\mapsto$ & $(2n+2,2n+1,2n,\dots,1,0)$\\
\hline
$\mathcal{Q}_{4n+2}$& $\mathbb{Z}_4$ &$(0,1,2,\dots,2n+2,2n+3)$& $\mapsto$ & $(2n+2,2n+3,2n+1,\dots,1,0)$\\
\hline
$\mathcal{T}$& $\mathbb{Z}_3$ &$(0,1,2,3,4,5,6)$& $\mapsto$ & $(1,5,4,3,6,0,2)$\\
\hline
$\mathcal{O}$& $\mathbb{Z}_2$ &$(0,1,2,3,4,5,6,7)$& $\mapsto$ & $(6,5,4,3,2,1,0,7)$\\
\hline
$\mathcal{I}$& $\emptyset$ & & --
\end{tabular}
\end{center}
}
\noindent The group $\mathcal{O}(\widehat{\mathfrak{g}})$ permutes the superselection sector data as follows:
\bea
\boldsymbol{w}_j
&\mapsto&
\boldsymbol{w}_{o(j)},
\\
\boldsymbol{s}_{j,K},
&\mapsto&
\boldsymbol{s}_{o(j),K}
\\
\boldsymbol{\omega}^{KK}_j
&\mapsto&
\boldsymbol{\omega}^{KK}_{o(j)},
\\
\xi_j
&\mapsto&
\begin{cases}
\xi_{o(j)} & o(j) \neq 0\\
-\sum_{k=1}^{\text{rk }\mathfrak{g}}a_k\xi_k & o(j)=0
\end{cases}
,
\label{eq:OG}
\eea
while the remaining parameters remain unchanged. In particular, in the case $W=1$ of M-strings, we can always use the action of $\mathcal{O}(\widehat{\mathfrak{g}})$ to set $w_j^{(a)}= \delta_{j0}$ for all $a$. This generalizes an observation made in \cite{DelZotto:2023ryf} for the case $\Gamma=\mathcal{C}_N$, where the invariance under gauge transformations of the NS-NS B-field in Type IIA on $\mathbb{C}^2/\Gamma$ \cite{Witten:2009xu} was ultimately found to be responsible for covariance. We will verify in the following sections that the various constituents of the partition function indeed transform covariantly with respect to the action of $\mathcal{O}(\widehat{\mathfrak{g}})$ for all choices of $\Gamma$.

\subsection{BPS contributions}
\label{sec:contr}
Within a specific superselection sector, we can assemble a supersymmetric partition function by summing over BPS configurations, graded by the topological charges for the various gauge fields. Among these we must include the 6d $\mathfrak{u}(W)^{r-1}$ vector fields as well as the $r$ two-form fields $B_{\mu\nu}^{(1)},\dots,B_{\mu\nu}^{(r)}$. For the former set of fields, the relevant topological charges are the instanton numbers $\boldsymbol{\kappa}=(\kappa^{(1)},\dots,\kappa^{(r-1)})$, which couple to the tensor branch parameters $\boldsymbol{\varphi}$; the first Chern classes, on the other hand, are fixed by virtue of Equation \eqref{eq:c1diag}. For the latter set of fields, it is again convenient to resort to compactification to 5d and trade off the two-form fields for ordinary abelian vector fields; their topological charges are the instanton numbers $\boldsymbol{n}^{KK}$ which couple to $\tau$, as well as the first Chern classes $\vec{u}^{KK,(a)}$, and are to be summed over.

\noindent Based on the $\Gamma= \mathcal{C}_N$ example, which was studied in detail in \cite{DelZotto:2023rct,DelZotto:2023ryf}, we expect the following BPS contributions to the partition function:
\begin{itemize}
\item[--] BPS particles arising from the KK modes along the 6d circle of the tensor, vector, and hypermultiplets;
\item[--] A 2d sector $(\mathcal{H}\times\widehat{\mathfrak{g}}_1)^r$ localized on the $T^2$, consisting of a product of Heisenberg algebras and affine current algebras. Each of the $r$ factors arises from a distinct M5 brane; this is the obvious generalization of the $(\mathcal{H}\times \widehat{\mathfrak{su}}(N)_1)^r$ degrees of freedom found in \cite{DelZotto:2023rct} in the $\Gamma=\cC_N$ case;
\item[--] BPS strings wrapped on $T^2$, carrying instanton charge under the 6d gauge algebra.
\end{itemize}
Let us discuss these contributions in turn, before presenting the complete expression for the partition function at the end of this section.\\\newline

\noindent\textbf{BPS particles contribution.} The contribution from BPS particles on $\mathbb{C}^2$ can be obtained by projecting the modes of BPS particles on $\mathbb{C}^2$ onto the $\Gamma$-invariant subspace. To determine this contribution it is convenient to refer to the Type IIA frame that was reviewed in Section \ref{sec:6dSCFT}. The particles that contribute are localized at the $r$ NS5 interfaces, and arise from fundamental strings ending on D6 branes.\footnote{ With the exception of a massive photon which decouples; its contribution to the partition function is replaced by an identical contribution arising from the two-form field on the NS5 brane.} Let us focus on the $a$-th interface; strings ending on neighboring stacks of branes give rise to hypermultiplets in the bifundamental representation of $\mathfrak{u}(W)^{(a-1)}\times \mathfrak{u}(W)^{(a)}$, while string ending on the same stack of branes lead to vector multiplets in the adjoint representations of $\mathfrak{u}(W)^{(a)}$ as well as $\mathfrak{u}(W)^{(a-1)}$, which we will also think of as products of the fundamental and anti-fundamental representation. The choice of monodromy discussed in Section \ref{sec:sel} determines a branching of the fundamental and anti-fundamental representation of $\mathfrak{u}(W)^{(a)}$ in terms of irreducible representations of $\Gamma$ as in Equation \eqref{eq:mondec}:
\be
\square \to \rho^{(a)} = \bigoplus_{j=0}^{\text{rk }\mathfrak{g}}w^{(a)}_j\rho_j,
\qquad
\overline\square \to \overline\rho^{(a)} = \bigoplus_{j=0}^{\text{rk }\mathfrak{g}}w^{(a)}_j\overline\rho_j,
\ee
where $\overline\rho_j$ denotes the conjugate representation to $\rho_j$. Therefore, a bifundamental representation under $\mathfrak{u}(W)^{(a)}\times \mathfrak{u}(W)^{(a')}$ branches into products of $\mathfrak{u}(w^{(a)}_j)\times \mathfrak{u}(w^{(a')}_{j'})$ bifundamentals, which we denote schematically as $(\square^{(a)}_j,\overline\square^{(a')}_{j'})$:
\be
(\square^{(a)},\overline\square^{(a')}) \simeq \bigoplus_{j,j'=0}^{\text{rk }\mathfrak{g}} (\square^{(a)}_j,\overline\square^{(a')}_{j'}) \otimes(\rho_j\otimes \overline\rho_{j'}) 
\simeq
\bigoplus_{j,j',\ell=0}^{\text{rk }\mathfrak{g}} c_{j,j'}^\ell(\square^{(a)}_j,\overline\square^{(a')}_{j'})\otimes \overline\rho_\ell,
\ee
where we used the decomposition $\rho_j\otimes \overline\rho_{j'} = \oplus_{\ell=0}^{\text{rk }\mathfrak{g}} c_{j,j'}^\ell \overline\rho_\ell$. Additionally, to determine the BPS particle contribution to the partition function we must take into account the embedding of $\Gamma$ into the spacetime $SU(2)_L$ and project onto BPS states which are invariant under $\Gamma$. Using the fact that
\be
c_{j,j'}^{0}=\delta_{j,j'},
\ee
we see that for BPS particles that transform as $(\square^{(a)}_j,\overline\square^{(a')}_{j'})\otimes \overline\rho_\ell$ we must project the $SU(2)_L$ spin content onto the $\rho_\ell$ representation. That is, their contribution to the BPS partition function is given in terms of the $\Gamma$-covariant Hilbert series $\mathcal{H}^{\Gamma}_{\rho_\ell}(t)$ discussed in Section \ref{sec:gammaH}. Combining the contributions from vector and hypermultiplets, we obtain the following contribution from the $(a)$-th interface to the partition function\footnote{ For $\Gamma =\mathcal{C}_N$, the equivariant Hilbert series also depends on $x$, as in Appendix \ref{app:Hilbert}.}:
\bea
\nonumber
&&
Z_{M5/TN_W}^{(a)}
=
PE
\bigg[
-\frac{t}{1-q}
\sum_{j,j',\ell=0}^{\text{rk }\mathfrak{g}}
\mathcal{H}^{\Gamma}_{\rho_\ell}(t)
c_{j,j'}^\ell
\times
\\
\nonumber
&&
\bigg(
\sum_{A=1}^{w^{(a)}_j}
\sum_{B=1}^{w^{(a)}_{j'}}
f^+_{vm}(s^{(a)}_{j,A},s^{(a)}_{j',B},\epsilon_+,\tau)
+
\sum_{A=1}^{w^{(a-1)}_j}
\sum_{B=1}^{w^{(a-1)}_{j'}}
f^-_{vm}(s^{(a-1)}_{j,A},s^{(a-1)}_{j',B},\epsilon_+,\tau)
\\
&&
+
\sum_{A=1}^{w^{(a-1)}_j}
\sum_{B=1}^{w^{(a)}_{j'}}
f_{hm}(s^{(a-1)}_{j,A},s^{(a)}_{j',B},\epsilon_+,\tau)
\bigg)
\bigg],
\eea
where
\bea
f^+_{vm}(s^{(a)}_{j,A},s^{(a)}_{j',B},\epsilon_+,\tau)
&=&
\begin{cases}
t \frac{M^{(a)}_{j',B}}{M^{(a)}_{j,A}}
+
q t^{-1} \frac{M^{(a)}_{j,A}}{M^{(a)}_{j',B}}
&
s^{(a)}_{j',B} \geq s^{(a)}_{j,A}
\\
0
&
s^{(a)}_{j',B} > s^{(a)}_{j,A}
\end{cases}
,
\\
f^-_{vm}(s^{(a)}_{j,A},s^{(a)}_{j',B},\epsilon_+,\tau)
&=&
\begin{cases}
t^{-1} \frac{M^{(a-1)}_{j',B}}{M^{(a-1)}_{j,A}}
+
q t \frac{M^{(a-1)}_{j',B}}{M^{(a-1)}_{j,A}}
&
s^{(a-1)}_{j',B} > s^{(a-1)}_{j,A}
\\
0
&
s^{(a-1)}_{j',B} \leq s^{(a-1)}_{j,A}
\end{cases}
,
\eea
and
\be
f_{hm}(s^{(a)}_{j,A},s^{(a)}_{j',B},\epsilon_+,\tau)
=
\begin{cases}
-\frac{M^{(a)}_{j',B}}{M^{(a-1)}_{j,A}}
-
q \frac{M^{(a-1)}_{j,A}}{M^{(a)}_{j',B}}
&
s^{(a)}_{j',B} > s^{(a-1)}_{j,A}
\\
-\frac{M^{(a-1)}_{j,A}}{M^{(a)}_{j',B}}
-
q \frac{M^{(a)}_{j',B}}{M^{(a-1)}_{j,A}}
&
s^{(a)}_{j',B} < s^{(a-1)}_{j,A}
\end{cases}
.
\ee
Here we have defined $ M^{(a)}_{j,A} = e^{2\pi i s^{(a)}_{j,A}}$, and we make a generic choice of Wilson line parameters such that $s^{(a)}_{j,A}\neq s^{(b)}_{j',B}$ unless $a=b$, $j'=j$, and $A=B$. 

Taking the product over the different interfaces, we obtain the following BPS particle contribution to the partition function:
\be
Z^{\Gamma,\text{BPS particles}}_{ \mathcal{T}^{6d}_{r,W}}[\vec{\boldsymbol{w}}](\vec{\underline{\boldsymbol{s}}},\epsilon_+,\tau)
=
\prod_{a=1}^{r-1} Z_{M5/TN_W}^{\Gamma,(a)}
.
\label{eq:zzNS}
\ee
Note in particular that under the action of the outer automorphism group $\mathcal{O}(\widehat{\mathfrak{g}})$ discussed at the end of Section \ref{sec:sel}, Equation \eqref{eq:zzNS} transforms simply by acting on the parameters according to Equation \eqref{eq:OG}. \\

\noindent\underline{\textbf{Specializing to M-strings}.} Equation \eqref{eq:zzNS} takes a simple form when $W=1$. In this case $(s_{0,0}^{(0)},s_{0,0}^{(1)},\dots,s_{0,0}^{(r)})$ are the only Wilson line parameters, and $s_{0,0}^{(a)}= s_{0,0}^{(a-1)}+\mu$ as a consequence of the Stückelberg mechanism, where we take $\mu>0$. The BPS particle contribution to the partition then simplifies significantly, leading to: 
\be
Z^{\Gamma,\text{BPS particles}}_{ \mathcal{T}^{6d}_{r,1}}(\mu,\epsilon_+,\tau)
=
PE
\bigg[
r
\left(m-t+\frac{q}{1-q}\frac{(1-mt)(1-m/t)}{m}\right)
t \mathcal{H}^{\Gamma}_{\rho_0}(t)
\bigg],
\label{eq:zbpsmg}
\ee
which is a simple generalization of the result for $\mathbb{C}^2$ (Equation \eqref{eq:zbpsc2}).\\

\noindent\textbf{Current algebra contribution.}
Let us now turn to the contribution from the 2d sector $(\mathcal{H}\times \widehat{\mathfrak{g}}_1)^r$. If there are no BPS strings wrapping the $T^2$, the contribution to the partition function from the chiral algebra would simply be given by its character
\be
\chi_{\cH}^r\prod_{a=1}^r{\chi}^{\widehat{\mathfrak{g}}}_{\omega^{KK,(a)}}(\vec{\xi},\tau),
\label{eq:vw}
\ee
where
\be
\chi^{\widehat{\mathfrak{g}}_1}_\omega 
=
\Tr_{\vec{\lambda}_\omega}
\left(
q^{H_L} e^{2\pi i \vec{\xi}\cdot (C^{\Gamma})^{-1}\cdot \vec{J}_\Gamma}
\right)
 =
 q^{-\frac{\text{rk }\mathfrak{g}}{24}+h_\omega}\left(\boldsymbol{R}_\omega + \mathcal{O}(q)\right)
\ee
is the character of the level-1 integrable highest weight representation $\omega$ of $\widehat{\mathfrak{g}}$ of conformal dimension $h_\omega$, whose highest weight has  Dynkin labels $\vec{\lambda}_\omega$. We can equivalently denote this integrable representation in terms of the  corresponding irreducible representation $\boldsymbol{R}_{\omega}$ of $\mathfrak{g}$. This expression gets modified in the presence of BPS strings as we will see shortly.

The parameter $\omega^{KK,(a)}$ in Equation \eqref{eq:vw} labels a choice of integrable highest weight representation for the $\widehat{\mathfrak{g}}_1$ factor associated to the $a$-th M5 brane; this can ultimately be traced to a choice of monodromy on $\mathbb{C}^2/\Gamma$ for the anti-selfdual two-form $B^{(a)}_{\mu\nu}$, via the McKay correspondence. The expression \eqref{eq:vw} is simply the partition function for the $\mathcal{N}=(2,0)$ theory of $r$ M5 branes on $T^2\times\mathbb{C}^2/\Gamma$, with Vafa--Witten twist along $\mathbb{C}^2/\Gamma$; in \cite{Bruzzo:2013daa,Bruzzo:2014jza,DelZotto:2023rct,DelZotto:2023ryf} it was noticed that for $\Gamma = \mathcal{C}_N$ the same degrees of freedom also appear in the present equivariant setting. We assume that the contribution from the current algebra generalizes in the obvious way to arbitrary $\Gamma$, and in Appendix \ref{app:anomaly-canc} we will verify that its presence is required to cancel gauge anomalies of the worldsheet theory of the BPS strings to which it couples. On the other hand, the free chiral boson contribution corresponding to the Heisenberg algebra is completely decoupled and we will choose to treat it as a distinct component of the partition function.\\

\noindent\textbf{BPS strings contribution.} Finally, let us turn to the contribution of the BPS strings. Their worldsheet theory will be studied in detail in the next section, and here we content ourselves with some general remarks. Orbifold backgrounds allow for field configurations with fractional instanton charge, as is  manifest in Equation \eqref{eq:ch2V}, which we report here for convenience for a $\mathfrak{u}(W)^{(a)}$ gauge symmetry:
\be
\int_{\widetilde{\IC^2/\Gamma}}ch_2(\mathcal{V}^{(a)})=\sum_{j=1}^{\text{rk }\mathfrak{g}}u^{(a)}_j N_j + \frac{\sum_{j=0}^{\text{rk }\mathfrak{g}} a_j v^{(a)}_j}{|\Gamma|}.
\ee
Consistent with the Kronheimer--Nakajima \cite{KronheimerNakajima} picture of instantons on $\mathbb{C}^2/\Gamma$, we interpret the $\vec{v}^{(a)}=(v_0^{(a)},\dots,v_{\text{rk }\mathfrak{g}}^{(a)})$ as data specifying the topological charges associated to the instantons of the 6d gauge algebras, that is to say, of the BPS strings. Due to Equations \eqref{eq:uw}, \eqref{eq:c1diag}, and \eqref{eq:uw0}, within a given superselection sector (specifically for fixed $\vec{u}^{(a)}$), $\vec{v}^{(a)}$ is determined up to shifts by a vector with entries 
\be
\kappa^{(a)}\vec{a}
=
(a_0\kappa^{(a)},\dots,a_{\text{rk }\mathfrak{g}} \kappa^{(a)}),
\ee
which shifts the instanton number by $\kappa^{(a)}$ due to Equation~\eqref{eq:orderGamma}. Nonnegativity of the $v^{(a)}_j$ for BPS instantons implies that, for a given choice of $\mathfrak{u}(W)^{(a)}$ monodromy data $\vec{w}^{(a)}$ and first Chern class $\vec{u}^{(a)}=\vec{w}^{(0)}$, one can determine a vector $\vec{v}^{\star,(a)}$ corresponding to minimal instanton charge, and the allowed topological charges corresponding to BPS strings are given by 
\be
\vec{v}^{\,\star,(a)}+\kappa^{(a)}\vec{a}\qquad\text{for }\kappa^{(a)}\geq 0.
\ee
Following the terminology introduced in \cite{DelZotto:2023ryf}, we refer to the BPS strings corresponding to the minimal charge $\vec{v}^{\,\star,(a)}$ as \emph{frozen BPS strings}, since they do not possess any center of mass degree of freedom and are pinned at the orbifold singularity. We will denote the instanton charge of this frozen BPS string by $N^{\star,(a)}$, which is determined according to Equation \eqref{eq:ch2V}. It is straightforward to see that under the action of $\mathcal{O}(\widehat{\mathfrak{g}})$ one has simply $o({\boldsymbol{v}}^{\,\star}_j)= \boldsymbol{v}^\star_{o(j)} $, and $o(N^{\star,(a)}) = N^{\star,(a)}$. \\

Due to the interactions between the current algebra and the BPS strings, it is not possible to disentangle their contributions to the partition function. Rather, they are described jointly in terms a relative 2d $\mathcal{N}=(0,4)$ QFT 
which we denote by
\be
\cQ^{\Gamma,\vec{\boldsymbol{w}}}_{\boldsymbol{\vec{v}}}.
\ee
This theory contributes to the partition function via its elliptic genus
\be
\mathbb{E}_{\boldsymbol{\vec{v}}}^{\Gamma,\vec{\boldsymbol{w}}}[\boldsymbol{\omega}^{KK}]({\vec{\xi}},\vec{\underline{\boldsymbol{s}}},\epsilon_+,\tau),
\ee
which depends on a choice of superselection sector data $\boldsymbol{\omega}^{KK}$ and is discussed in detail in Section \ref{sec:EG}. Here we remark simply that in the case where $\vec{w}^{(a)}=\vec{w}^{(0)}$ for all $a$ (no frozen strings), for instanton charge zero the elliptic genus reduces to the contributions of the current algebra:
\be
\mathbb{E}_{\va{0}}^{\Gamma,\vec{\boldsymbol{w}}}[\boldsymbol{\omega}^{KK}]({\vec{\xi}},\vec{\underline{\boldsymbol{s}}},\epsilon_+,\tau)
=
\prod_{a=1}^r{\chi}^{\widehat{\mathfrak{g}}_1}_{\omega^{KK,(a)}}(\vec{\xi},\tau).
\ee
For more general choices of topological charges, the $\vec{\xi}$ are shifted due to the interactions with the BPS string as a consequence of anomaly inflow, as we will see in Section \ref{sec:EG}.\\

\noindent\textbf{The partition function.} Having discussed the various contributions, we are finally in the position to present the expression for the partition function of the theory $\mathcal{T}^{6d}_{r,W}$ on $T^2\ltimes \mathbb{C}^2/\Gamma$:
\be
\boxed{
\begin{aligned}
&
Z^\Gamma_{\mathcal{T}^{6d}_{r,W}}
[\vec{\boldsymbol{w}},{\boldsymbol{\omega}}^{KK}]
(\boldsymbol{\varphi},\vec{\xi},\vec{\underline{\boldsymbol{s}}},\epsilon_+,\tau)
=
\\
&
\hspace{.5in}
q^{\frac{r}{24}}
\chi_{\cH}^r
Z^{\Gamma,\text{BPS particles}}_{\mathcal{T}^{6d}_{r,W}}
[\vec{\boldsymbol{w}}]
(\vec{\underline{\boldsymbol{s}}},\epsilon_+,\tau)
\sum_{{\boldsymbol{\kappa}}\in\mathbb{Z}^{r-1}_{\geq 0}}
e^{-\boldsymbol{({\boldsymbol{N}}^\star+{\boldsymbol{\kappa}})}\cdot\boldsymbol{\varphi}}
\mathbb{E}^{\Gamma,\vec{\boldsymbol{w}}}_{\vec{\boldsymbol{v}}^\star+{\boldsymbol{\kappa}}\vec{a}}
[\boldsymbol{\omega}^{KK}]
(\vec{\xi},\vec{\underline{\boldsymbol{s}}},\epsilon_+,\tau).
\end{aligned}
}
\label{eq:PF}
\ee
We have seen above that the BPS particles' contribution to the partition function transforms covariantly under the action of $\mathcal{O}(\widehat{\mathfrak{g}})$, and furthermore ${\boldsymbol{N}}^\star$ is unchanged. The covariance of the full partition function then follows from covariance of the elliptic genus:
\be
\mathbb{E}^{\Gamma,o(\vec{\boldsymbol{w}})}_{o(\vec{\boldsymbol{v}}^\star)+{\boldsymbol{\kappa}}\vec{a}}
[o(\boldsymbol{\omega}^{KK})]
(o(\vec{\xi}),o(\vec{\underline{\boldsymbol{s}}}),\epsilon_+,\tau)
=
\mathbb{E}^{\Gamma,\vec{\boldsymbol{w}}}_{\vec{\boldsymbol{v}}^\star+{\boldsymbol{\kappa}}\vec{a}}
(\vec{\xi},\vec{\underline{\boldsymbol{s}}},\epsilon_+,\tau),
\ee
a fact which which we will verify in Section \ref{sec:EG}. Moreover the partition function is invariant upon reversing the ordering of M5 branes, which corresponds to inverting the labels $a\to r-a$ on all parameters with an upper index ${Q}^{(a)}$. \\

\noindent In the case of M-strings, the partition function simplifies to:
\bea
\nn
&&
Z^\Gamma_{\mathcal{T}^{6d}_{r,1}}
[{\boldsymbol{\omega}}^{KK}]
(\boldsymbol{\varphi},\vec{\xi},\mu,\epsilon_+,\tau)
=
\\
&&
\qquad\qquad
q^{\frac{r}{24}}
\chi_{\cH}^r
Z^{\Gamma,\text{BPS particles}}_{\mathcal{T}^{6d}_{r,1}}(\mu,\epsilon_+,\tau)
\sum_{{\boldsymbol{\kappa}}\in\mathbb{Z}^{r-1}_{\geq 0}}
e^{-\boldsymbol{\vec{\kappa}}\cdot\boldsymbol{\varphi}}
\mathbb{E}^{\Gamma}_{{\boldsymbol{\kappa}}\vec{a}}[\boldsymbol{\omega}^{KK}](\mu,{\vec{\xi}},\epsilon_+,\tau).
\label{eq:PFM}
\eea
Recall from Section \ref{sec:6dSCFT} that the 6d M-string SCFT of rank $r$, which arises from a stack of $r$ M5 branes localized at the center of a Taub--NUT space, can be deformed to the 6d $\mathcal{N}=(2,0) $ SCFT of the same rank. This is achieved simply by moving the M5 branes far away from the Taub--NUT center and decompactifying the TN circle. This relation can be seen at the level of partition functions: the one of the M-string SCFT is expected to coincide with the one of the $\cN= (2,0)$ theory in the limit $m \to \epsilon_+$ \footnote{Upon compactifying on a circle the M-string SCFT can be interpreted as the 5d $ U(r)$  $\mathcal{N}=1^*$ gauge theory. The formulation of the theory in the Omega-background, which is a deformation of the Donaldson--Witten twist, requires coupling the theory to a $Spin^c$ structure, which results in a shift of the the mass parameter for the adjoint hypermultiplet \cite{Manschot:2021qqe,Closset:2022vjj}.}. In this limit, the BPS particles' contribution (Equation \eqref{eq:zbpsmg}) becomes identically 1; moreover, only the sector with zero BPS string charge contributes due to the presence of fermionic zero modes that cause the elliptic genera of BPS strings to vanish. At the end of the day, the only nonvanishing contribution to the partition function comes from the chiral degrees of freedom supported on the M5 branes in the zero BPS string sector, \emph{i.e.}\ it coincides with the partition function of the $\mathfrak{u}(1)^r$ $\mathcal{N}=(2,0)$ theory on $\mathbb{C}^2/\Gamma$:
\be
 Z^{\Gamma}_{\mathcal{T}^{6d}_{r,1}}
 [\boldsymbol{\omega}^{KK}]
 (\boldsymbol{\varphi},\vec{\xi},\mu,\epsilon_+,\tau)
 \to
 q^{\frac{r}{24}}\prod_{a=1}^r \frac{{\chi}^{\widehat{\mathfrak{g}}}_{\omega^{KK,(a)}}(\vec\xi,\tau)}{\eta(\tau)}
 =
 \left(Z^{\Gamma}_{\cT^{6d}_{1,0}}(\vec{\xi},\tau)\right)^r.
\ee

\section{The BPS string worldsheet theories: UV quivers}
\label{sec:strings}
We now turn to an analysis of the worldsheet theories of the BPS strings of theories $\cT^{6d}_{r,W}$ on the target space $\mathbb{C}^2/\Gamma$. We begin in Section \ref{sec:quiv} by determining the worldsheet degrees of freedom of an arbitrary bound state of BPS strings for the 6d M-string SCFT $\cT^{6d}_{r,1}$;
in Section \ref{sec:TSUdecomp} we describe a gluing procedure for constructing the worldsheet theories of the BPS strings out of 2d analogues of the linear quiver theories of Gaiotto and Witten~\cite{Gaiotto:2008ak}; in Section \ref{sec:M-string_orb} we generalize these results to the case of arbitrary gauge rank $W$; finally, in Section \ref{sec:exs} we provide a number of examples of BPS string configurations that highlight interesting features that occur for different choices of orbifold singularity $\mathbb{C}^2/\Gamma$.

\subsection{M-string quivers for generic $\Gamma$ of ADE type}
\label{sec:quiv}
As reviewed in the previous sections, the 6d M-string SCFT $\mathcal{T}^{6d}_{r,1}$ possesses a spectrum of two-dimensional BPS strings which arise from bound states of M2 branes stretched between parallel M5 branes. From the point of view of the 6d gauge algebra they can be thought as bound states of point-like instantons. These are labeled by a vector $\vb*{\kappa}=(\kappa^{(1)},\dots,\kappa^{(r-1)})\in \mathbb{Z}^{r-1}_{\geq 0}$, where $\kappa^{(a)}$ corresponds to the instanton charge with respect to the $\mathfrak{u}(1)^{(a)}$ gauge field. The 2d worldsheet theory describing a bound state can be inferred from the geometry of the M-theory setup, which for us is $T^2\times \IC^2/\Gamma\times\IR\times TN_1$. For the case $\Gamma=\cC_N$, this singular geometry can be related by dualities to a system of intersecting branes in Type IIB on a non-singular background, from which is possible to read off the 2d $(0,4)$ quiver gauge theory describing arbitrary bound states of M-strings. This is the approach that was followed in \cite{DelZotto:2023rct}, where a crucial step in the duality chain consisted of replacing the $A$-type ALE singularity with a $N$-centered Taub--NUT space. While a similar approach can be performed in principle for $D$-type singularities, this does not generalize to the exceptional case. Nonetheless, for any finite subgroup $\Gamma$ of $ SU(2)$ there exist families of 2d $(0,4)$ quiver gauge theories which are natural candidates to describe the BPS strings. To see this, let us start by considering a number $\kappa$ of M2 branes in the following setup (leaving out the M5 branes for the moment):
\be
\nonumber
    \begin{tabular}{c|cc|cccc|c|cccc}
    & \multicolumn{2}{c|}{$T^2$} 
         & \multicolumn{4}{c|}{$\mathbb{C}^2/\Gamma$} 
         & \multicolumn{1}{c|}{$\IR$} 
         & \multicolumn{4}{c}{$TN_1$} \\

         &  0 & 1 & 2 & 3 & 4 & 5 & 6 & 7 & 8 & 9 &10 \\ \hline
        $\kappa$ $M2$& $\cross$ &$\cross$& & & & & $\cross$& & &&     
    \end{tabular}
\ee
If we then go to Type IIA by compactifying along the circle fiber of $TN_1$ we arrive at the configuration shown in Table \ref{tab:tt}.
\begin{table}[t]
    \centering
    \begin{tabular}{c|cc|cccc|c|ccc}
    & \multicolumn{2}{c|}{$T^2$} 
         & \multicolumn{4}{c|}{$\mathbb{C}^2/\Gamma$} 
         & $\mathbb{R}$ & \multicolumn{3}{c}{$\mathbb{R}^3$} \\

         &  0 & 1 & 2 & 3 & 4 & 5 & 6 & 7 & 8 & 9 \\ \hline
        $\kappa$ $D2$& $\cross$ &$\cross$ & & & & & $\cross$& & &    \\
         $D6$& $\cross$ &$\cross$ &$\cross$ &$\cross$ &$\cross$ &$\cross$ & $\cross$& & &    
    \end{tabular}
    \caption{ Type IIA brane configuration realizing $\kappa$ U(1)  instantons on $\mathbb{C}^2/\Gamma$.}
    \label{tab:tt}
\end{table}
This Type IIA setup corresponds to the brane system studied by Douglas and Moore~\cite{Douglas:1996sw}, and the worldvolume theory of the D2 branes is a 3d $\cN=4$ theory described by a Kronheimer--Nakajima quiver $\mathcal{KN}^{\Gamma}_{\kappa}$~\cite{KronheimerNakajima}, whose Higgs branch corresponds to the moduli space of $\kappa$ $U(1)$ instantons on $\IC^2/\Gamma$. The quiver corresponding to $\Gamma=\cI$ is depicted in Figure~\ref{fig:U(1)KNquiver} for the sake of illustration.\\
\begin{figure}[h]
	\centering
	\scalebox{0.8}{\begin{tikzpicture}[line width=0.8]
		\node[gauge]   (g0) at (0,0) {$\kappa$};
		\node[gauge]   (g1) at (2,0) {$2\kappa$};
        \node[gauge]   (g2) at (4,0) {$3\kappa$};
        \node[gauge]   (g3) at (6,0) {$4\kappa$};
        \node[gauge]   (g4) at (8,0) {$5\kappa$};
        \node[gauge]   (g5) at (10,0) {$6\kappa$};
        \node[gauge]   (g6) at (12,0) {$4\kappa$};
        \node[gauge]   (g7) at (14,0) {$2\kappa$};
        \node[gauge]   (g8) at (10,2) {$3\kappa$};
        \node[flavor] (f0) at (0,2) {$1$};
        \draw (g0)--(g1);
        \draw (g1)--(g2);
        \draw (g2)--(g3);
        \draw (g3)--(g4);
        \draw (g4)--(g5);
        \draw (g5)--(g6);
        \draw (g6)--(g7);
        \draw (g5)--(g8);
        \draw (f0)--(g0);
	\end{tikzpicture}}
    \caption{$3d$ $\cN=4$ Kronheimer--Nakajima quiver $\mathcal{KN}^{\cI}_{\kappa}$.}
    \label{fig:U(1)KNquiver}
\end{figure}

Let us now consider $r$ parallel M5 branes extended along directions $x_0,\dots,x_5$ and spaced along $x_6$. Between each pair of adjacent M5's, we can suspend $\kappa^{(a)}$ M2 branes, with $a=1,\dots,r-1$. In the Type IIA picture the M5 branes become NS5 branes, and the D2 branes suspended between them give rise to a collection of $r-1$ Kronheimer--Nakajima theories $\mathcal{KN}^{\Gamma}_{\kappa^{(a)}}$ interacting with each other through the NS5-brane interfaces. The boundary conditions preserve half of the supersymmetry, and upon reducing along $x_6$ we obtain a two-dimensional theory with $\mathcal{N}=(0,4)$ supersymmetry. In particular the interactions between neighboring quivers are described in terms of $\mathcal{N}=(0,4)$ bifundamental multiplets connecting various gauge and flavor nodes in a manner which is determined by performing an orbifold projection by $\Gamma$ on the 2d $\mathcal{N}=(0,4)$ quiver theory for M-strings on $\mathbb{C}^2$ as explained in \cite{DelZotto:2023ryf}; the resulting spectrum of Fermi and hypermultiplets is described below.
Notably, there are additional chiral degrees of freedom living on the NS5 branes and localized at the ALE singularity which give rise to a $\widehat{\mathfrak{g}}_1$ current algebra on each NS5 ~\cite{Dijkgraaf:2007sw,DelZotto:2023rct} and are fundamental for the cancelation of gauge anomalies of the gauge degrees of freedom on the BPS strings as we will see in Appendix \ref{app:anomaly-canc}.\\

\noindent The 2d $(0,4)$ theory can be schematically written as
\begin{equation}
\label{eq:schematic_Mstring_quiver}
\mathcal{KN}^{\Gamma}_{0}\bigg] \boldsymbol{NS}^\Gamma\bigg[\mathcal{KN}^{\Gamma}_{\kappa^{(1)}}\bigg]  \boldsymbol{NS}^\Gamma\bigg[ \mathcal{KN}^{\Gamma}_{\kappa^{(2)}}\bigg]  \boldsymbol{NS}^\Gamma \bigg[\quad \dots\quad \bigg]\boldsymbol{NS}^\Gamma \bigg[ \mathcal{KN}^{\Gamma}_{\kappa^{(r-1)}}\bigg] \boldsymbol{NS}^\Gamma\bigg[ \mathcal{KN}^{\Gamma}_{0} ,
\end{equation}
where the $\boldsymbol{NS}^\Gamma$ represent the 2d $(0,4)$ interfaces between the Kronheimer--Nakajima theories, each supporting a $\widehat{\mathfrak{g}}_1$ current algebra and a decoupled free boson. The theory so obtained can be described in terms of a 2d $(0,4)$ theory, depicted for $\Gamma=\cI$ in Figure \ref{fig:MstringQuiverE8}, with the $\boldsymbol{NS}^\Gamma$ interfaces represented as vertical blue lines between  Kronheimer--Nakajima quivers. We will refer to this type of quiver supplemented by $(\widehat{\mathfrak{g}}_1)^r$ current algebras at the interfaces as a \emph{$\Gamma$-dressed quiver}.
\begin{figure}[t]
	\centering
    \scalebox{0.7}{\begin{tikzpicture}[line width=0.6]
    \draw[blue] (-1,-2.5)--(-1,15);
    \node[blue] at (-1,15.5) {$\widehat{\mathfrak{e}}_{8,1}^{(0)}$};
    \node[blue] at (3,15.5) {$\widehat{\mathfrak{e}}_{8,1}^{(1)}$};
    \node[blue] at (7,15.5) {$\widehat{\mathfrak{e}}_{8,1}^{(2)}$};
    \node[blue] at (11,15.5) {$\widehat{\mathfrak{e}}_{8,1}^{(3)}$};
    \node[blue] at (13,15.5) {$\widehat{\mathfrak{e}}_{8,1}^{(r-1)}$};
    \node[blue] at (17,15.5) {$\widehat{\mathfrak{e}}_{8,1}^{(r)}$};
    %%%%%%%%%%%%%%%%%%%%%
		\node[gauge]   (g10) at (0,0) {$\kappa^{(1)}$};
		\node[gauge]   (g11) at (0,2) {$2\kappa^{(1)}$};
        \node[gauge]   (g12) at (0,4) {$3\kappa^{(1)}$};
        \node[gauge]   (g13) at (0,6) {$4\kappa^{(1)}$};
        \node[gauge]   (g14) at (0,8) {$5\kappa^{(1)}$};
        \node[gauge]   (g15) at (0,10) {$6\kappa^{(1)}$};
        \node[gauge]   (g16) at (0,12) {$4\kappa^{(1)}$};
        \node[gauge]   (g17) at (0,14) {$2\kappa^{(1)}$};
        \node[gauge]   (g18) at (2,10) {$3\kappa^{(1)}$};
        \node[flavor] (f10) at (0,-2) {$1$};
        \draw (g10)--(g11);
        \draw (g11)--(g12);
        \draw (g12)--(g13);
        \draw (g13)--(g14);
        \draw (g14)--(g15);
        \draw (g15)--(g16);
        \draw (g16)--(g17);
        \draw (g15)--(g18);
        \draw (f10)--(g10);
        %%%%%%%%%%%%%%%%%%%
        \draw[blue] (3,-2.5)--(3,15);
        %%%%%%%%%%%%%%%%%%%
        \node[gauge]   (g20) at (4,0) {$\kappa^{(2)}$};
		\node[gauge]   (g21) at (4,2) {$2\kappa^{(2)}$};
        \node[gauge]   (g22) at (4,4) {$3\kappa^{(2)}$};
        \node[gauge]   (g23) at (4,6) {$4\kappa^{(2)}$};
        \node[gauge]   (g24) at (4,8) {$5\kappa^{(2)}$};
        \node[gauge]   (g25) at (4,10) {$6\kappa^{(2)}$};
        \node[gauge]   (g26) at (4,12) {$4\kappa^{(2)}$};
        \node[gauge]   (g27) at (4,14) {$2\kappa^{(2)}$};
        \node[gauge]   (g28) at (6,10) {$3\kappa^{(2)}$};
        \node[flavor] (f20) at (4,-2) {$1$};
        \draw (g20)--(g21);
        \draw (g21)--(g22);
        \draw (g22)--(g23);
        \draw (g23)--(g24);
        \draw (g24)--(g25);
        \draw (g25)--(g26);
        \draw (g26)--(g27);
        \draw (g25)--(g28);
        \draw (f20)--(g20);
        %%%%%%%%%%%%%%%%%%%
        \draw[blue] (7,-2.5)--(7,15);
        %%%%%%%%%%%%%%%%%%%
        \node[gauge]   (g30) at (8,0) {$\kappa^{(3)}$};
		\node[gauge]   (g31) at (8,2) {$2\kappa^{(3)}$};
        \node[gauge]   (g32) at (8,4) {$3\kappa^{(3)}$};
        \node[gauge]   (g33) at (8,6) {$4\kappa^{(3)}$};
        \node[gauge]   (g34) at (8,8) {$5\kappa^{(3)}$};
        \node[gauge]   (g35) at (8,10) {$6\kappa^{(3)}$};
        \node[gauge]   (g36) at (8,12) {$4\kappa^{(3)}$};
        \node[gauge]   (g37) at (8,14) {$2\kappa^{(3)}$};
        \node[gauge]   (g38) at (10,10) {$3\kappa^{(3)}$};
        \node[flavor] (f30) at (8,-2) {$1$};
        \draw (g30)--(g31);
        \draw (g31)--(g32);
        \draw (g32)--(g33);
        \draw (g33)--(g34);
        \draw (g34)--(g35);
        \draw (g35)--(g36);
        \draw (g36)--(g37);
        \draw (g35)--(g38);
        \draw (f30)--(g30);
         %%%%%%%%%%%%%%%%%%%
        \draw[blue] (11,-2.5)--(11,15);
        \node (dots1) at (12,-1) {$\cdots$};
        \node (dots1) at (12,14.5) {$\cdots$};
        \draw[blue] (13,-2.5)--(13,15);
        %%%%%%%%%%%%%%%%%%%
        \node[gauge]   (gl0) at (14,0) {$\kappa^{(r-1)}$};
		\node[gauge]   (gl1) at (14,2) {$2\kappa^{(r-1)}$};
        \node[gauge]   (gl2) at (14,4) {$3\kappa^{(r-1)}$};
        \node[gauge]   (gl3) at (14,6) {$4\kappa^{(r-1)}$};
        \node[gauge]   (gl4) at (14,8) {$5\kappa^{(r-1)}$};
        \node[gauge]   (gl5) at (14,10) {$6\kappa^{(r-1)}$};
        \node[gauge]   (gl6) at (14,12) {$4\kappa^{(r-1)}$};
        \node[gauge]   (gl7) at (14,14) {$2\kappa^{(r-1)}$};
        \node[gauge]   (gl8) at (16,10) {$3\kappa^{(r-1)}$};
        \node[flavor] (fl0) at (14,-2) {$1$};
        \draw (gl0)--(gl1);
        \draw (gl1)--(gl2);
        \draw (gl2)--(gl3);
        \draw (gl3)--(gl4);
        \draw (gl4)--(gl5);
        \draw (gl5)--(gl6);
        \draw (gl6)--(gl7);
        \draw (gl5)--(gl8);
        \draw (fl0)--(gl0);
        %%%%%%%%%%%%%%%%%%%
        \draw[blue] (17,-2.5)--(17,15);
        %%%%%%%%%%%%%%%%%%%
        %(1)-(2)
        \draw[purple] (g10)--(g20);
        \draw[purple] (g11)--(g21);
        \draw[purple] (g12)--(g22);
        \draw[purple] (g13)--(g23);
        \draw[purple] (g14)--(g24);
        \draw[purple] (g15)to[out=30,in=150](g25);
        \draw[purple] (g16)--(g26);
        \draw[purple] (g17)--(g27);
        \draw[purple] (g18)to[out=-30,in=-150](g28);
         %(2)-(3)
         \draw[purple] (g20)--(g30);
        \draw[purple] (g21)--(g31);
        \draw[purple] (g22)--(g32);
        \draw[purple] (g23)--(g33);
        \draw[purple] (g24)--(g34);
        \draw[purple] (g25)to[out=30,in=150](g35);
        \draw[purple] (g26)--(g36);
        \draw[purple] (g27)--(g37);
        \draw[purple] (g28)to[out=-30,in=-150](g38);
        %(3)-...
         \draw[purple] (g30)--(11.5,0);
        \draw[purple] (g31)--(11.5,2);
        \draw[purple] (g32)--(11.5,4);
        \draw[purple] (g33)--(11.5,6);
        \draw[purple] (g34)--(11.5,8);
        \draw[purple] (g35)to[out=30,in=180](11.5,11);
        \draw[purple] (g36)--(11.5,12);
        \draw[purple] (g37)--(11.5,14);
        \draw[purple] (g38)to[out=-30,in=180](11.5,9.1);
        \draw[purple,dotted] (11.5,0)--(11.8,0);
        \draw[purple,dotted] (11.5,2)--(11.8,2);
        \draw[purple,dotted] (11.5,4)--(11.8,4);
        \draw[purple,dotted] (11.5,6)--(11.8,6);
        \draw[purple,dotted] (11.5,8)--(11.8,8);
        \draw[purple,dotted] (11.5,11)--(11.8,11);
        \draw[purple,dotted] (11.5,9.1)--(11.8,9.1);
        \draw[purple,dotted] (11.5,12)--(11.8,12);
        \draw[purple,dotted] (11.5,14)--(11.8,14);
        \draw[purple,dotted] (12.2,0)--(12.5,0);
        \draw[purple,dotted] (12.2,2)--(12.5,2);
        \draw[purple,dotted] (12.2,4)--(12.5,4);
        \draw[purple,dotted] (12.2,6)--(12.5,6);
        \draw[purple,dotted] (12.2,8)--(12.5,8);
        \draw[purple,dotted] (12.2,9.1)--(12.5,9.1);
        \draw[purple,dotted] (12.2,11)--(12.5,11);
        \draw[purple,dotted] (12.2,12)--(12.5,12);
        \draw[purple,dotted] (12.2,14)--(12.5,14);
        %...-(r-1)
        \draw[purple] (12.5,0)--(gl0);
        \draw[purple] (12.5,2)--(gl1);
        \draw[purple] (12.5,4)--(gl2);
        \draw[purple] (12.5,6)--(gl3);
        \draw[purple] (12.5,8)--(gl4);
        \draw[purple] (12.5,11)to[out=0,in=120](gl5);
        \draw[purple] (12.5,9.1)to[out=0,in=-150](gl8);
        \draw[purple] (12.5,12)--(gl6);
        \draw[purple] (12.5,14)--(gl7);
         %(1)-(2)
        \draw[green!70!black,dashed] (g10)--(g21);
        \draw[green!70!black,dashed] (g11)--(g22);
        \draw[green!70!black,dashed] (g12)--(g23);
        \draw[green!70!black,dashed] (g13)--(g24);
        \draw[green!70!black,dashed] (g14)--(g25);
        \draw[green!70!black,dashed] (g15)to[out=30,in=-155](g26);
        \draw[green!70!black,dashed] (g16)--(g27);
        \draw[green!70!black,dashed] (g11)--(g20);
        \draw[green!70!black,dashed] (g12)--(g21);
        \draw[green!70!black,dashed] (g13)--(g22);
        \draw[green!70!black,dashed] (g14)--(g23);
        \draw[green!70!black,dashed] (g15)--(g24);
        \draw[green!70!black,dashed] (g16)to[out=-25,in=150](g25);
        \draw[green!70!black,dashed] (g17)--(g26);
         %(2)-(3)
        \draw[green!70!black,dashed] (g20)--(g31);
        \draw[green!70!black,dashed] (g21)--(g32);
        \draw[green!70!black,dashed] (g22)--(g33);
        \draw[green!70!black,dashed] (g23)--(g34);
        \draw[green!70!black,dashed] (g24)--(g35);
        \draw[green!70!black,dashed] (g25)to[out=30,in=-155](g36);
        \draw[green!70!black,dashed] (g26)--(g37);
        \draw[green!70!black,dashed] (g21)--(g30);
        \draw[green!70!black,dashed] (g22)--(g31);
        \draw[green!70!black,dashed] (g23)--(g32);
        \draw[green!70!black,dashed] (g24)--(g33);
        \draw[green!70!black,dashed] (g25)--(g34);
        \draw[green!70!black,dashed] (g26)to[out=-25,in=150](g35);
        \draw[green!70!black,dashed] (g27)--(g36);
        %(3)-...
        \draw[green!70!black,dashed] (g30)--(11.5,1);
        \draw[green!70!black,dashed] (g31)--(11.5,3);
        \draw[green!70!black,dashed] (g32)--(11.5,5);
        \draw[green!70!black,dashed] (g33)--(11.5,7);
        \draw[green!70!black,dashed] (g34)--(11.5,9);
        \draw[green!70!black,dashed] (g35)to[out=30,in=180](11.5,11.5);
        \draw[green!70!black,dashed] (g36)--(11.5,13);
        \draw[green!70!black,dashed] (g31)--(11.5,1.1);
        \draw[green!70!black,dashed] (g32)--(11.5,3.1);
        \draw[green!70!black,dashed] (g33)--(11.5,5.1);
        \draw[green!70!black,dashed] (g34)--(11.5,7.1);
        \draw[green!70!black,dashed] (g35)to[out=-30,in=180](11.5,9.1);
        \draw[green!70!black,dashed] (g36)--(11.5,11.1);
        \draw[green!70!black,dashed] (g37)--(11.5,13.1);
        %...-(r-1)
        \draw[green!70!black,dashed] (12.2,1.2)--(gl1);
        \draw[green!70!black,dashed] (12.2,3.2)--(gl2);
        \draw[green!70!black,dashed] (12.2,5.2)--(gl3);
        \draw[green!70!black,dashed] (12.2,7.2)--(gl4);
        \draw[green!70!black,dashed] (12.2,9.2)--(gl5);
        \draw[green!70!black,dashed] (12.2,8.9)to[out=0,in=-135](gl8);
        \draw[green!70!black,dashed] (12.2,11.2)--(gl6);
        \draw[green!70!black,dashed] (12.2,13.2)--(gl7);
        \draw[green!70!black,dashed] (12.2,0.8)--(gl0);
        \draw[green!70!black,dashed] (12.2,2.8)--(gl1);
        \draw[green!70!black,dashed] (12.2,4.8)--(gl2);
        \draw[green!70!black,dashed] (12.2,6.8)--(gl3);
        \draw[green!70!black,dashed] (12.2,8.8)--(gl4);
        \draw[green!70!black,dashed] (12.2,10.8)--(gl5);
        \draw[green!70!black,dashed] (12.2,12.8)--(gl6);
        \draw[green!70!black,dashed] (12.2,11.05)to[out=0,in=140](gl8);
        %%%%%%%%%%%%%%%
        \draw[dashed] (g10)--(f20);
        \draw[dashed] (f10)--(g20);
        \draw[dashed] (g20)--(f30);
        \draw[dashed] (f20)--(g30);
        \draw[dashed] (12.5,-1.1)--(fl0);
        \draw[dashed] (12.5,-0.8)--(gl0);
        \draw[dashed] (f30)--(11.5,-1.1);
        \draw[dashed] (g30)--(11.5,-0.8);
        \node[flavor] (fleft) at (-2,0) {$1$};
        \node[flavor] (fright) at (18,0) {$1$};
        \draw[dashed] (g10)--(fleft);
        \draw[dashed] (gl0)--(fright);
	\end{tikzpicture}}
    \caption{The 2d $(0,4)$ M-string quiver for $\Gamma=\cI$. Black solid lines are the twisted hypermultiplets $X_{ij}^{(a)}$ or $W^{(a)}$; purple solid lines are the hypermultiplets $Y^{(a)}_j$; green dashed lines are the Fermi multiplets $\Psi^{(a)}_{ij}$ and $\widetilde{\Psi}^{(a)}_{ij}$; black dashed lines are the Fermi multiplets $\Sigma^{(a)}$ and $\Theta^{(a)}$. Vertical blue lines represent the interfaces $\boldsymbol{NS}^\Gamma$ that support the $\widehat{\mathfrak{g}}_1$ current algebra.}
    \label{fig:MstringQuiverE8}
\end{figure}

The quiver for \eqref{eq:schematic_Mstring_quiver} contains $(r-1)\cdot (\text{rk }\mathfrak{g}+1)$ unitary gauge nodes $G_j^{(a)}=U(v_j^{(a)})$ with $j=0,\dots,\text{rk }\mathfrak{g}$ and $a=1,\dots,r-1$, and $(r+1)$ flavor nodes $U(1)^{(a)}$, $a=0,\dots,r$. The ranks of the gauge groups are given by:
\begin{equation}\label{eq:node_ranks}
	v_j^{(a)}=a_j \kappa^{(a)} ,
\end{equation}
 where the $a_j$ are the comarks of $\Gamma$. The field content of the theory is given by the following 2d $\cN=(0,4)$ multiplets:
 \begin{itemize}
     \item Vector multiplets $V_j^{(a)}$ for gauge groups $G_j^{(a)}$;
     \item Twisted hypermultiplets $X_{ij}^{(a)}$, for $i<j$ such that $C^{\widehat{\mathfrak{g}}}_{ij}\neq 0$, in the bifundamental representation of $G_{i}^{(a)}\times G_{j}^{(a)}$; 
     \item Hypermultiplets $Y^{(a)}_j$ in the bifundamental representation of $G_j^{(a)}\times G_{j}^{(a+1)}$;
     \item Fermi multiplets $\Psi^{(a)}_{ij}$, $\widetilde{\Psi}^{(a)}_{ij}$, for $i<j$ such that $(C^{\widehat{\mathfrak{g}}})_{ij}\neq 0$, respectively in the bifundamental representation of $G_i^{(a)}\times G_{j}^{(a+1)}$ and $G_j^{(a)}\times G_{i}^{(a+1)}$; 
     \item Twisted hypermultiplets $W^{(a)}$ in the bifundamental representation of $G_0^{(a)}\times U(1)^{(a)}$;
     \item Fermi multiplets $\Sigma^{(a)}$ and $\Theta^{(a)}$, respectively in the bifundamental representation of $G_0^{(a)}\times U(1)^{(a-1)}$ and $G_0^{(a)}\times U(1)^{(a+1)}$. 
 \end{itemize}
For $W^{(a)}$, $\Sigma^{(a)}$ and $\Theta^{(a)}$, recall that the global symmetries $U(1)^{(a)}$ are all identified with $U(1)^{diag}$ by the St\"{u}ckelberg mechanism.
\color{black}
\\

Note that cancelation of non-abelian gauge anomalies requires the ranks of the gauge groups to be proportional to the comarks of $\widehat{\mathfrak{g}}$. Indeed, if we forget about the constraint \eqref{eq:node_ranks} for the moment and treat the ranks $v_j^{(a)}$ as generic, the non-abelian anomaly polynomial associated to the $(a,j)-$th node is \cite{Hanany:2018hlz}
\begin{equation}
\label{eq:non_abelian_anomaly_polynomial}
\overbrace{4 v^{(a)}_j}^{\textbf{vector}}
+
\overbrace{2 \sum_{k\neq j} C^{\widehat{\mathfrak{g}}}_{jk} v^{(a)}_k}^{\textbf{twisted hypers}}
+
\overbrace{2 v_j^{(a-1)} + 2 v_j^{(a+1)}}^{\textbf{hypers}}
-
\overbrace{\sum_{k\neq j}C^{\widehat{\mathfrak{g}}}_{jk}\qty(v^{(a-1)}_k+v^{(a+1)}_k)}^{\textbf{Fermi}}
=
0,
\end{equation}
which can be rewritten as:
\begin{equation}
\sum_{b=1}^{r-1}\sum_{k=0}^{\mathrm{rk}\,\mathfrak{g}-1}
C^{\mathfrak{su}(r)}_{ab}
C^{\widehat{\mathfrak{g}}}_{jk}
v^{(b)}_k
=
0,
\qquad a=1,\dots,r-1,\qquad j=0 , \dots,\mathrm{rk}\,\mathfrak{g}-1.
\end{equation}
The positive definiteness of the Cartan matrix $C^{\mathfrak{su}(r)}_{ab}$ implies
\begin{equation}\label{eq:Mstrings_anomalies_canc}
\sum_{k=0}^{\mathrm{rk}\,\mathfrak{g}-1}C^{\widehat{\mathfrak{g}}}_{jk} v_k^{(a)}=0,
\end{equation}
which holds provided that the $v_j^{(a)}$ satisfy Equation \eqref{eq:node_ranks}. More general values of the ranks do occur, however, for the M-string orbifold SCFTs discussed in Section \ref{sec:M-string_orb}.\\

If we only consider the fields specified above, the resulting quiver $\cQ^{\Gamma,anom}_{\vb*{\kappa}}$ is in fact inconsistent due to a gauge anomaly affecting the abelian gauge factors. However, as anticipated above, the anomalies can be canceled by coupling the quiver to the current algebras $\oplus_{a=1}^r\widehat{\mathfrak{g}}_1^{(a)}$. Let us denote the corresponding currents by $\cJ^{\widehat{\mathfrak{g}}_1,(a)}_j$, $a=1,\dots,r$, $j=1,\dots,\mathrm{rk}\,\mathfrak{g}$. We can couple these degrees of freedom  to the gauge connections of the 2d QFT via the following interaction terms:
\begin{equation}\label{eq:currents-gauge-coupling}
\int_{T^2} \sum_{a=1}^r\sum_{j=1}^{\mathrm{rk}\,\mathfrak{g}}\cJ_j^{\widehat{\mathfrak{g}}_1,(a)}
\qty[C^{\widehat{\mathfrak{g}}}\cdot \qty(\Tr \vec{A}^{(a)}-\Tr \vec{A}^{(a-1)})]_j ,
\end{equation}
which are the natural generalization to arbitrary $\Gamma$ of the couplings that were found in \cite{DelZotto:2023rct,DelZotto:2023ryf} for the case $\Gamma=\mathcal{C}_N$. As shown in Appendix~\ref{app:anomaly-canc}, these additional interaction terms guarantee the cancelation of all abelian gauge anomalies.\\

As we discussed in Section \ref{sec:sel}, for each of the two-form fields $B^{(a)}$ on the M5 branes, we need to specify a monodromy at infinity,
which corresponds to a choice of a level-1 i.h.w.r.\ $\omega^{KK,(a)}$ for $\widehat{\mathfrak{g}}_1$. Therefore, one of the effects of the interfaces is to promote the anomalous quiver theory $\cQ^{\Gamma,anom}_{\vb*{\kappa}}$ to a relative theory
\be
\cQ^{\Gamma}_{\vb*{\kappa}}
=
\qty(\widehat{\mathfrak{g}}_1^r\ltimes \cQ^{\Gamma,anom}_{\vb*{\kappa}}).
\ee
The relative nature of the theory implies that, rather than having a single well defined partition function, these theories possess collections of conformal blocks labeled by  tuples $\boldsymbol{\omega}^{KK}$ of level-1 i.h.w.r.'s of $\widehat{\mathfrak{g}}$. The conformal blocks transform into each other as the components of a vector valued meromorphic Jacobi form under $SL(2,\IZ)$ transformations on the complex modulus $\tau$ of the torus.

\subsection{$T_{\vb*{\rho}}(SU(N))$ decomposition for M-strings on $\mathbb{C}^2/\Gamma$}
\label{sec:TSUdecomp}

In the cases where the singularity $\mathbb{C}^2/\Gamma$ corresponds to a \emph{star-shaped} affine ADE quiver, that is for $\Gamma=\cQ_4,\cT,\cO$ and $\cI$, one can break down the quiver in terms of simpler constituents, \emph{i.e.}\ a set of external linear quivers coupled to a central gauge node. These external quivers are constructed out of a class of $3d$ $\cN=4$ SCFTs denoted as $T_{\vb*{\rho}}(SU(N))$~\cite{Gaiotto:2008ak}, which are given by:
\begin{equation}\label{eq:qqqv}
		\begin{tikzpicture}[line width=0.4]
			\node[gauge] (g1) at (0,0) {$N_1$};
			\node[gauge] (g2) at (2,0) {$N_2$};
			\node[gauge] (gl-1) at (6,0) {$N_{l-1}$};
			\node[flavor] (f) at (8,0) {$N$};
			\draw (g1)--(g2);
			\draw (g2)--(3,0);
			\draw[dashed] (3,0)--(5,0);
			\draw (5,0)--(gl-1);
			\draw (gl-1)--(f);
		\end{tikzpicture}
\end{equation}
for some $N\geq 2$ and a non-increasing partition $\vb*{\rho}=(N-N_{l-1},N_{l-1}-N_{l-2},\dots,N_{2}-N_1,N_1\geq 1)$ of $N$, of length $l\leq N$.
In fact, it is possible to view these theories as 3d $\mathcal{N}=4$ Kronheimer--Nakajima theories for $\Gamma=\mathbb{C}^2/\mathcal{C}_{l}$ with a rank $0$ gauge group associated to the affine node, as in Figure \ref{fig:embed}. The quiver depicted in \eqref{eq:qqqv} also includes a decoupled  $U(1)\in U(N)$ factor of the flavor symmetry.\\
\begin{figure}[t]
	\centering
	\scalebox{0.8}{\begin{tikzpicture}[line width=0.8]
      \coordinate (center) at (0,0);
         \draw (center) ++(-60:3) arc[start angle=-60, end angle=180, radius=3];
         \draw[dashed] (center) ++(180:3) arc[start angle=180, end angle=300, radius=3];
        \node[fill=white,circle, draw, minimum size=40pt,inner sep=1] (g0) at (0,3) {$0$};
        \node[fill=white,circle, draw, minimum size=40pt,inner sep=1]  (gW-1) at (-2.6,1.5) {$N_{l-1}$};
        \node[fill=white,circle, draw, minimum size=40pt,inner sep=1](g2) at (2.6,1.5) {$N_1$};
        \node[fill=white,circle, draw, minimum size=40pt,inner sep=1]  (g3) at (2.6,-1.5) {$N_2$};
	\node[flavor]   (fl) at (-4.8,3.1) {$\phantom{tt}N\phantom{tt}$};
	\draw (gW-1)--(fl);
	\end{tikzpicture}}\hspace{.5in}
	\caption{The $\cC_l$-type Kronheimer--Nakajima quiver corresponding to theory $T_{\vb*{\rho}}(SU(N))$.}
	\label{fig:embed}
\end{figure}

Given a collection of $T_{\vb*{\rho}}(SU(N))$ theories with the same $N$, we can glue them together by gauging their common flavor node $U(N)$ to obtain a new quiver, as represented schematically in Figure~\ref{fig:E8gluing} in the case $\Gamma= \cI$. The quiver inherits the decoupled $U(1)$ flavor symmetry from the tails.
\begin{figure}[t]
	\centering
	\scalebox{0.8}{\begin{tikzpicture}[line width=0.8]
			\def\yTOP{0}
			\def\yBOT{-4.5}
			\node[gauge]   (gl0) at (-0.55,\yTOP) {$\kappa$};
			\node[gauge]   (gl1) at (1.45,\yTOP) {$2\kappa$};
			\node[gauge]   (gl2) at (3.45,\yTOP) {$3\kappa$};
			\node[gauge]   (gl3) at (5.45,\yTOP) {$4\kappa$};
			\node[gauge]   (gl4) at (7.45,\yTOP) {$5\kappa$};
			\node[flavor]   (fl) at (9.45,\yTOP) {$6\kappa$};
			\draw (gl0)--(gl1);
			\draw (gl1)--(gl2);
			\draw (gl2)--(gl3);
			\draw (gl3)--(gl4);
			\draw (gl4)--(fl);
			\node at (1.5,\yTOP+1) {\footnotesize$T_{(\kappa,\kappa,\kappa,\kappa,\kappa,\kappa)}(SU(6\kappa))$};
			%%%%%%%%%%%%%%%%%%%%%%%%%%%%%%%%%%
			\node[gauge]   (gu) at (10,\yTOP+3.1) {$3\kappa$};
			\node[flavor]   (fu) at (10,\yTOP+1.1) {$6\kappa$};
			\draw (gu)--(fu);
			\node at (12,\yTOP+3.1) {\footnotesize$T_{(3\kappa,3\kappa)}(SU(6\kappa))$};
			%%%%%%%%%%%%%%%%%%%%%%%%%%%%%%%%%%
			\node[gauge] (gr1) at (12.55,\yTOP) {$4\kappa$};
			\node[gauge] (gr2) at (14.55,\yTOP) {$2\kappa$};
			\node[flavor]   (fr) at (10.55,\yTOP) {$6\kappa$};
			\draw (fr)--(gr1);
			\draw (gr1)--(gr2);
			\node at (14,\yTOP+1) {\footnotesize$T_{(2\kappa,2\kappa,2\kappa)}(SU(6\kappa))$};	
			%%%%%%%%%%%%%%%%%%%%%%%%%%%%%%%%%%%%
			\node at (10,\yTOP-1.6) {\Large $\mathrel{\rotatebox[origin=c]{-90}{$\rightarrow$}}$};
			%%%%%%%%%%%%%%%%%%%%%%%%%%%%%%%%%%%%
			\node[gauge]   (g0) at (0,\yBOT) {$\kappa$};
			\node[gauge]   (g1) at (2,\yBOT) {$2\kappa$};
			\node[gauge]   (g2) at (4,\yBOT) {$3\kappa$};
			\node[gauge]   (g3) at (6,\yBOT) {$4\kappa$};
			\node[gauge]   (g4) at (8,\yBOT) {$5\kappa$};
			\node[gauge]   (g5) at (10,\yBOT) {$6\kappa$};
			\node[gauge]   (g6) at (12,\yBOT) {$4\kappa$};
			\node[gauge]   (g7) at (14,\yBOT) {$2\kappa$};
			\node[gauge]   (g8) at (10,\yBOT+2) {$3\kappa$};
			\draw (g0)--(g1);
			\draw (g1)--(g2);
			\draw (g2)--(g3);
			\draw (g3)--(g4);
			\draw (g4)--(g5);
			\draw (g5)--(g6);
			\draw (g6)--(g7);
			\draw (g5)--(g8);
			\draw[red,dashed] (10,\yTOP+0.4) circle (1.5);
	\end{tikzpicture}}
	\caption{The $\cI$-type Kronheimer--Nakajima theory, $\mathcal{KN}^{\cI}_{\kappa}$, can be obtained by gluing together the theories $T_{(\kappa,\kappa,\kappa,\kappa,\kappa,\kappa)}(SU(6\kappa))$, $T_{(2\kappa,2\kappa,2\kappa)}(SU(6\kappa))$, and $T_{(3\kappa,3\kappa)}(SU(6\kappa))$ by gauging the common flavor node.}
	\label{fig:E8gluing}
\end{figure}
Gluing makes it possible to break down the computation of observables of the star-shaped quivers in terms of simpler constituents
(see \emph{e.g.} \cite{cremonesi2014coulomb} for the case of 3d $\mathcal{N}=4$ theories).
We will now explain how this approach can be employed in the case of our star-shaped 2d $(0,4)$ relative theories, which in Section~\ref{sec:gluingEG} will enable us to write down simpler expressions for their elliptic genera.
To proceed, we first of all need to identify the appropriate set of building blocks for our class of 2d $\mathcal{N}=(0,4)$ theories. 
Let us start by considering a collection of $3d$ $\cN=4$ theories $T_{\vb*{\rho}^{(a)}}(SU(N^{(a)}))$, $a=1,\dots,r-1$, where each partition $\vb*{\rho}^{(a)}$ has the same length $l$. We can see each $T_{\vb*{\rho}^{(a)}}(SU(N^{(a)}))$ as an external tail of the $a-$th copy of the Kronheimer--Nakajima quiver belonging to the M-string quiver. Now, following the procedure used in the previous sections to obtain the 2d theories $\cQ^{\Gamma}_{\vb*{\kappa}}$, we place $\frac{1}{2}$-BPS interfaces between the $T_{\vb*{\rho}^{(a)}}(SU(N^{(a)}))$ quivers supporting $\mathcal{N}=(0,4)$ degrees of freedom. These in particular will include an $\widehat{\su}(l)_1$ current algebra, as in ~\cite{Costello:2018fnz,Gaiotto:2016wcv}, as well as multiplets charged under the gauge and flavor nodes of the neighboring quivers. This is depicted in Figure \ref{fig:2dTSUquiver}. On each interface, the currents generating the $\widehat{\su}(l)_1$ algebra couple to the gauge fields of the neighboring quivers in an analogous way as Equation
\eqref{eq:currents-gauge-coupling}.
The resulting 2d $(0,4)$ theory is once again a relative theory, whose elliptic genus depends on a choice of $r-$dimensional vector of i.h.w.r.'s of $\widehat{\su}(l)_1$. Requiring the cancelation of non-abelian gauge anomalies, by an analogous computation to the one that led to \eqref{eq:Mstrings_anomalies_canc} in Section~\ref{sec:quiv}, we find that 
\begin{equation}
\vb*{\rho}^{(a)}= v^{(a)} \underbrace{(1,\dots,1)}_{l-\text{dimensional}}
\end{equation}
for some integer $v^{(a)}$ for $a=1,\dots,{r-1}$. The resulting class of 2d theories will be denoted by $T^l_{\vb*{v}}$, where $\vb*{v}=(v^{(1)},\dots,v^{(r-1)})$ and will serve as our building blocks.\\
%
%%%%%%%%%%%%%%%%%%%%%%
% T_v^l quiver
%%%%%%%%%%%%%%%%%%%%%%
\begin{figure}[h]
	\centering
	\scalebox{0.65}{\begin{tikzpicture}[line width=0.6]
			%%%%%%%%%%%%%%%%%%%%%%%
			% Variables
			%%%%%%%%%%%%%%%%%%%%%%%
			\def\x0{0}
			\def\xsh{2}
			\def\ysh{2}
			\def\ytop{9}
			\def\ybot{0}
			\def\ynode{\ytop-1}
			%%%%%%%%%%%%%%%%%%%%%%%
			% blue lines
			%%%%%%%%%%%%%%%%%%%%%%%%
			\draw[blue] (\x0,\ytop)--(\x0,\ybot);
			\node[blue] at (\x0,\ytop+0.5) {$\widehat{\su}(l)_{1}^{(1)}$};
			\draw[blue] (\x0+\xsh,\ytop)--(\x0+\xsh,\ybot);
			\node[blue] at (\x0+\xsh,\ytop+0.5) {$\widehat{\su}(l)_{1}^{(2)}$};
			\draw[blue] (\x0+2*\xsh,\ytop)--(\x0+2*\xsh,\ybot);
			\node[blue] at (\x0+2*\xsh,\ytop+0.5) {$\widehat{\su}(l)_{1}^{(3)}$};
			\draw[blue] (\x0+3.2*\xsh,\ytop)--(\x0+3.2*\xsh,\ybot);
			\node[blue] at (\x0+3.2*\xsh,\ytop+0.5) {$\widehat{\su}(l)_{1}^{(r-1)}$};
			\draw[blue] (\x0+4.2*\xsh,\ytop)--(\x0+4.2*\xsh,\ybot);
			\node[blue] at (\x0+4.2*\xsh,\ytop+0.5) {$\widehat{\su}(l)_{1}^{(r)}$};
			%%%%%%%%%%%%%%%%%%%%%
			% (1) 
			%%%%%%%%%%%%%%%%%%%%%
			\node[gauge]   (g10) at (\x0+\xsh/2,\ynode) {$v^{(1)}$};
			\node[gauge]   (g11) at (\x0+\xsh/2,\ynode-\ysh) {$2v^{(1)}$};
			\node[gauge]   (g1l-1) at (\x0+\xsh/2,\ynode-2.5*\ysh) {\tiny $(l-1) v^{(1)}$};
			\node[flavor] (f1) at  (\x0+\xsh/2,\ynode-3.5*\ysh) {$l v^{(1)}$};
			\draw (g10)--(g11);
			\draw (g1l-1)--(f1);
			\draw  (g11)-- (\x0+\xsh/2,\ynode-1.5*\ysh) ;
			\draw[dashed] (\x0+\xsh/2,\ynode-1.5*\ysh)--(\x0+\xsh/2,\ynode-2*\ysh) ;
			\draw (\x0+\xsh/2,\ynode-2*\ysh)--(g1l-1)  ;
			%%%%%%%%%%%%%%%%%%%
			% (2)
			%%%%%%%%%%%%%%%%%%%
			\node[gauge]   (g20) at (\x0+3*\xsh/2,\ynode) {$v^{(2)}$};
			\node[gauge]   (g21) at (\x0+3*\xsh/2,\ynode-\ysh) {$2v^{(2)}$};
			\node[gauge]   (g2l-1) at (\x0+3*\xsh/2,\ynode-2.5*\ysh) {\tiny $(l-1)v^{(2)}$};
			\node[flavor] (f2) at  (\x0+3*\xsh/2,\ynode-3.5*\ysh) {$l v^{(2)}$};
			\draw (g20)--(g21);
			\draw (g2l-1)--(f2);
			\draw  (g21)-- (\x0+3*\xsh/2,\ynode-1.5*\ysh) ;
			\draw[dashed] (\x0+3*\xsh/2,\ynode-1.5*\ysh)--(\x0+3*\xsh/2,\ynode-2*\ysh) ;
			\draw (\x0+3*\xsh/2,\ynode-2*\ysh)--(g2l-1)  ;
			%%%%%%%%%%%%%%%%%%%
			% DOTS
			%%%%%%%%%%%%%%%%%%%%
			\draw (g20)--(g21);
			\draw (g2l-1)--(f2);
			\draw  (g21)-- (\x0+3*\xsh/2,\ynode-1.5*\ysh) ;
			\draw[dashed] (\x0+3*\xsh/2,\ynode-1.5*\ysh)--(\x0+3*\xsh/2,\ynode-2*\ysh) ;
			\draw (\x0+3*\xsh/2,\ynode-2*\ysh)--(g2l-1)  ;
			%%%%%%%%%%%%%%%%%%%
			% (r-1)
			%%%%%%%%%%%%%%%%%%%
			\node[gauge]   (gr-10) at (\x0+3.2*\xsh+\xsh/2,\ynode) {\tiny $v^{(r-1)}$};
			\node[gauge]   (gr-11) at (\x0+3.2*\xsh+\xsh/2,\ynode-\ysh) {\tiny $2v^{(r-1)}$};
			\node[gauge]   (gr-1l-1) at (\x0+3.2*\xsh+\xsh/2,\ynode-2.5*\ysh) {\tiny$(l-1)v^{(r-1)}$};
			\node[flavor] (fr-1) at  (\x0+3.2*\xsh+\xsh/2,\ynode-3.5*\ysh) {\tiny$l v^{(r-1)}$};
			\draw (gr-10)--(gr-11);
			\draw (gr-1l-1)--(fr-1);
			\draw  (gr-11)-- (\x0+3.2*\xsh+\xsh/2,\ynode-1.5*\ysh) ;
			\draw[dashed] (\x0+3.2*\xsh+\xsh/2,\ynode-1.5*\ysh)--(\x0+3.2*\xsh+\xsh/2,\ynode-2*\ysh) ;
			\draw (\x0+3.2*\xsh+\xsh/2,\ynode-2*\ysh)--(gr-1l-1)  ;
			%%%%%%%%%%%%%%%%%%%%%%%%%%%
			% Horizontal Hypermultiplets
			%%%%%%%%%%%%%%%%%%%%%%%%%%%
			\draw[purple] (g10)--(g20);
			\draw[purple] (g11)--(g21);
			\draw[purple] (g1l-1)--(g2l-1);
			\draw[purple] (g20)--(\x0+2*\xsh+\xsh/4,\ynode);
			\draw[purple] (g21)--(\x0+2*\xsh+\xsh/4,\ynode-\ysh);
			\draw[purple] (g2l-1)--(\x0+2*\xsh+\xsh/4,\ynode-2.5*\ysh);
			\draw[purple][dashed] (\x0+2*\xsh+\xsh/4,\ynode)--(\x0+3.2*\xsh-\xsh/4,\ynode);
			\draw[purple][dashed]  (\x0+2*\xsh+\xsh/4,\ynode-\ysh)--(\x0+3.2*\xsh-\xsh/4,\ynode-\ysh);
			\draw[purple][dashed]  (\x0+2*\xsh+\xsh/4,\ynode-2.5*\ysh)--(\x0+3.2*\xsh-\xsh/4,\ynode-2.5*\ysh);
			\draw[purple] (\x0+3.2*\xsh-\xsh/4,\ynode)--(gr-10);
			\draw[purple] (\x0+3.2*\xsh-\xsh/4,\ynode-\ysh)--(gr-11);
			\draw[purple] (\x0+3.2*\xsh-\xsh/4,\ynode-2.5*\ysh)--(gr-1l-1);
			%(1)-(2)
			\draw[green!70!black,dashed] (g10)--(g21);
			\draw[green!70!black,dashed] (g11)--(g20);
			\draw[green!70!black,dashed] (g11)--(\x0+3*\xsh/2-0.7,\ynode-2*\ysh+0.7) ;
			\draw[green!70!black,dashed] (g21)--(\x0+1*\xsh/2+0.7,\ynode-2*\ysh+0.7) ;
			\draw[green!70!black,dashed] (g2l-1)--(\x0+0.5*\xsh+0.5,\ynode-1.5*\ysh-0.5);
			\draw[green!70!black,dashed] (g1l-1)--(\x0+3*\xsh/2-0.5,\ynode-1.5*\ysh-0.5);
			%(2)-(3)...dots...
			\draw[green!70!black,dashed] (g20)--(\x0+5*\xsh/2-0.6,\ynode-\ysh+0.6);
			\draw[green!70!black,dashed] (g21)--(\x0+5*\xsh/2-0.6,\ynode-2*\ysh+0.6);
			\draw[green!70!black,dashed] (g21)--(\x0+5*\xsh/2-0.6,\ynode-0.6);
			\draw[green!70!black,dashed] (g2l-1)--(\x0+5*\xsh/2-0.6,\ynode-1.5*\ysh-0.6);
			%
			%....dots...(r-2)-(r-1)
			\draw[green!70!black,dashed] (gr-11)--(\x0+3.2*\xsh+\xsh/2-1.4,\ynode-\ysh+1.4);
			\draw[green!70!black,dashed] (gr-1l-1)--(\x0+3.2*\xsh+\xsh/2-1.4,\ynode-2.5*\ysh+1.4);
			\draw[green!70!black,dashed] (gr-10)--(\x0+3.2*\xsh+\xsh/2-1.4,\ynode-1.4);
			\draw[green!70!black,dashed] (gr-11)--(\x0+3.2*\xsh+\xsh/2-1.4,\ynode-\ysh-1.4);
			%%%%%%
			% Black Fermi
			%%%%%%
			\draw[dashed] (g1l-1)--(f2) ;
			\draw[dashed] (g2l-1)--(f1) ;
				\draw[dashed] (g2l-1)--(\x0+5*\xsh/2-0.6,\ynode-3.5*\ysh+0.6) ;
					\draw[dashed] (f2)--(\x0+5*\xsh/2-0.6,\ynode-2.5*\ysh-0.6) ;
					\draw[dashed] (gr-1l-1)--(\x0+3.2*\xsh+\xsh/2-1.4,\ynode-3.5*\ysh+0.6) ;
						\draw[dashed] (fr-1)--(\x0+3.2*\xsh+\xsh/2-1.4,\ynode-2.5*\ysh-0.6) ;
\def\xTILDE{12};
\def\ytopTILDE{\ytop+0.5};
\def\ynodeTILDE{\ytopTILDE-2};
%%%%%%%%%%%%%%%%%%%%%%%
% blue lines
%%%%%%%%%%%%%%%%%%%%%%%%
\draw[blue] (\xTILDE,\ytopTILDE)--(\xTILDE,\ybot);
\node[blue] at (\xTILDE,\ytopTILDE+0.5) {$\widehat{\su}(l)_{1}^{(1)}$};
\draw[blue] (\xTILDE+\xsh,\ytopTILDE)--(\xTILDE+\xsh,\ybot);
\node[blue] at (\xTILDE+\xsh,\ytopTILDE+0.5) {$\widehat{\su}(l)_{1}^{(2)}$};
\draw[blue] (\xTILDE+2*\xsh,\ytopTILDE)--(\xTILDE+2*\xsh,\ybot);
\node[blue] at (\xTILDE+2*\xsh,\ytopTILDE+0.5) {$\widehat{\su}(l)_{1}^{(3)}$};
\draw[blue] (\xTILDE+3.2*\xsh,\ytopTILDE)--(\xTILDE+3.2*\xsh,\ybot);
\node[blue] at (\xTILDE+3.2*\xsh,\ytopTILDE+0.5) {$\widehat{\su}(l)_{1}^{(r-1)}$};
\draw[blue] (\xTILDE+4.2*\xsh,\ytopTILDE)--(\xTILDE+4.2*\xsh,\ybot);
\node[blue] at (\xTILDE+4.2*\xsh,\ytopTILDE+0.5) {$\widehat{\su}(l)_{1}^{(r)}$};
%%%%%%%%%%%%%%%%%%%%%
% (1) 
%%%%%%%%%%%%%%%%%%%%%
\node[gauge]   (gT10) at (\xTILDE+\xsh/2,\ynodeTILDE) {$v^{(1)}$};
\node[gauge]   (gT11) at (\xTILDE+\xsh/2,\ynodeTILDE-\ysh) {$2v^{(1)}$};
\node[gauge]   (gT1l-1) at (\xTILDE+\xsh/2,\ynodeTILDE-2.5*\ysh) {\tiny $(l-1) v^{(1)}$};
\node[flavor] (fT1) at  (\xTILDE+\xsh/2,\ynodeTILDE-3.5*\ysh) {$l v^{(1)}$};
\draw (gT10)--(gT11);
\draw (gT1l-1)--(fT1);
\draw  (gT11)-- (\xTILDE+\xsh/2,\ynodeTILDE-1.5*\ysh) ;
\draw[dashed] (\xTILDE+\xsh/2,\ynodeTILDE-1.5*\ysh)--(\xTILDE+\xsh/2,\ynodeTILDE-2*\ysh) ;
\draw (\xTILDE+\xsh/2,\ynodeTILDE-2*\ysh)--(gT1l-1)  ;
%%%%%%%%%%%%%%%%%%%
% (2)
%%%%%%%%%%%%%%%%%%%
\node[gauge]   (gT20) at (\xTILDE+3*\xsh/2,\ynodeTILDE) {$v^{(2)}$};
\node[gauge]   (gT21) at (\xTILDE+3*\xsh/2,\ynodeTILDE-\ysh) {$2v^{(2)}$};
\node[gauge]   (gT2l-1) at (\xTILDE+3*\xsh/2,\ynodeTILDE-2.5*\ysh) {\tiny $(l-1)v^{(2)}$};
\node[flavor] (fT2) at  (\xTILDE+3*\xsh/2,\ynodeTILDE-3.5*\ysh) {$l v^{(2)}$};
\draw (gT20)--(gT21);
\draw (gT2l-1)--(fT2);
\draw  (gT21)-- (\xTILDE+3*\xsh/2,\ynodeTILDE-1.5*\ysh) ;
\draw[dashed] (\xTILDE+3*\xsh/2,\ynodeTILDE-1.5*\ysh)--(\xTILDE+3*\xsh/2,\ynodeTILDE-2*\ysh) ;
\draw (\xTILDE+3*\xsh/2,\ynodeTILDE-2*\ysh)--(gT2l-1)  ;
%%%%%%%%%%%%%%%%%%%
% DOTS
%%%%%%%%%%%%%%%%%%%%
\draw (gT20)--(gT21);
\draw (gT2l-1)--(fT2);
\draw  (gT21)-- (\xTILDE+3*\xsh/2,\ynodeTILDE-1.5*\ysh) ;
\draw[dashed] (\xTILDE+3*\xsh/2,\ynodeTILDE-1.5*\ysh)--(\xTILDE+3*\xsh/2,\ynodeTILDE-2*\ysh) ;
\draw (\xTILDE+3*\xsh/2,\ynodeTILDE-2*\ysh)--(gT2l-1)  ;
%%%%%%%%%%%%%%%%%%%
% (r-1)
%%%%%%%%%%%%%%%%%%%
\node[gauge]   (gTr-10) at (\xTILDE+3.2*\xsh+\xsh/2,\ynodeTILDE) {\tiny $v^{(r-1)}$};
\node[gauge]   (gTr-11) at (\xTILDE+3.2*\xsh+\xsh/2,\ynodeTILDE-\ysh) {\tiny $2v^{(r-1)}$};
\node[gauge]   (gTr-1l-1) at (\xTILDE+3.2*\xsh+\xsh/2,\ynodeTILDE-2.5*\ysh) {\tiny$(l-1)v^{(r-1)}$};
\node[flavor] (fTr-1) at  (\xTILDE+3.2*\xsh+\xsh/2,\ynodeTILDE-3.5*\ysh) {\tiny$l v^{(r-1)}$};
\draw (gTr-10)--(gTr-11);
\draw (gTr-1l-1)--(fTr-1);
\draw  (gTr-11)-- (\xTILDE+3.2*\xsh+\xsh/2,\ynodeTILDE-1.5*\ysh) ;
\draw[dashed] (\xTILDE+3.2*\xsh+\xsh/2,\ynodeTILDE-1.5*\ysh)--(\xTILDE+3.2*\xsh+\xsh/2,\ynodeTILDE-2*\ysh) ;
\draw (\xTILDE+3.2*\xsh+\xsh/2,\ynodeTILDE-2*\ysh)--(gTr-1l-1)  ;
%%%%%%%%%%%%%%%%%%%%%%%%%%%
% Horizontal Hypermultiplets
%%%%%%%%%%%%%%%%%%%%%%%%%%%
\draw[purple] (gT10)--(gT20);
\draw[purple] (gT11)--(gT21);
\draw[purple] (gT1l-1)--(gT2l-1);
\draw[purple] (gT20)--(\xTILDE+2*\xsh+\xsh/4,\ynodeTILDE);
\draw[purple] (gT21)--(\xTILDE+2*\xsh+\xsh/4,\ynodeTILDE-\ysh);
\draw[purple] (gT2l-1)--(\xTILDE+2*\xsh+\xsh/4,\ynodeTILDE-2.5*\ysh);
\draw[purple][dashed] (\xTILDE+2*\xsh+\xsh/4,\ynodeTILDE)--(\xTILDE+3.2*\xsh-\xsh/4,\ynodeTILDE);
\draw[purple][dashed]  (\xTILDE+2*\xsh+\xsh/4,\ynodeTILDE-\ysh)--(\xTILDE+3.2*\xsh-\xsh/4,\ynodeTILDE-\ysh);
\draw[purple][dashed]  (\xTILDE+2*\xsh+\xsh/4,\ynodeTILDE-2.5*\ysh)--(\xTILDE+3.2*\xsh-\xsh/4,\ynodeTILDE-2.5*\ysh);
\draw[purple] (\xTILDE+3.2*\xsh-\xsh/4,\ynodeTILDE)--(gTr-10);
\draw[purple] (\xTILDE+3.2*\xsh-\xsh/4,\ynodeTILDE-\ysh)--(gTr-11);
\draw[purple] (\xTILDE+3.2*\xsh-\xsh/4,\ynodeTILDE-2.5*\ysh)--(gTr-1l-1);
%(1)-(2)
\draw[green!70!black,dashed] (gT10)--(gT21);
\draw[green!70!black,dashed] (gT11)--(gT20);
\draw[green!70!black,dashed] (gT11)--(\xTILDE+3*\xsh/2-0.7,\ynodeTILDE-2*\ysh+0.7) ;
\draw[green!70!black,dashed] (gT21)--(\xTILDE+1*\xsh/2+0.7,\ynodeTILDE-2*\ysh+0.7) ;
\draw[green!70!black,dashed] (gT2l-1)--(\xTILDE+0.5*\xsh+0.5,\ynodeTILDE-1.5*\ysh-0.5);
\draw[green!70!black,dashed] (gT1l-1)--(\xTILDE+3*\xsh/2-0.5,\ynodeTILDE-1.5*\ysh-0.5);
%(2)-(3)...dots...
\draw[green!70!black,dashed] (gT20)--(\xTILDE+5*\xsh/2-0.6,\ynodeTILDE-\ysh+0.6);
\draw[green!70!black,dashed] (gT21)--(\xTILDE+5*\xsh/2-0.6,\ynodeTILDE-2*\ysh+0.6);
\draw[green!70!black,dashed] (gT21)--(\xTILDE+5*\xsh/2-0.6,\ynodeTILDE-0.6);
\draw[green!70!black,dashed] (gT2l-1)--(\xTILDE+5*\xsh/2-0.6,\ynodeTILDE-1.5*\ysh-0.6);
%
%....dots...(r-2)-(r-1)
\draw[green!70!black,dashed] (gTr-11)--(\xTILDE+3.2*\xsh+\xsh/2-1.4,\ynodeTILDE-\ysh+1.4);
\draw[green!70!black,dashed] (gTr-1l-1)--(\xTILDE+3.2*\xsh+\xsh/2-1.4,\ynodeTILDE-2.5*\ysh+1.4);
\draw[green!70!black,dashed] (gTr-10)--(\xTILDE+3.2*\xsh+\xsh/2-1.4,\ynodeTILDE-1.4);
\draw[green!70!black,dashed] (gTr-11)--(\xTILDE+3.2*\xsh+\xsh/2-1.4,\ynodeTILDE-\ysh-1.4);
%%%%%%
% Black Fermi
%%%%%%
\draw[dashed] (gT1l-1)--(fT2) ;
\draw[dashed] (gT2l-1)--(fT1) ;
\draw[dashed] (gT2l-1)--(\xTILDE+5*\xsh/2-0.6,\ynodeTILDE-3.5*\ysh+0.6) ;
\draw[dashed] (fT2)--(\xTILDE+5*\xsh/2-0.6,\ynodeTILDE-2.5*\ysh-0.6) ;
\draw[dashed] (gTr-1l-1)--(\xTILDE+3.2*\xsh+\xsh/2-1.4,\ynodeTILDE-3.5*\ysh+0.6) ;
\draw[dashed] (fTr-1)--(\xTILDE+3.2*\xsh+\xsh/2-1.4,\ynodeTILDE-2.5*\ysh-0.6) ;
%%%%%%
% TILDE multiplets
%%%%%%
\node[flavor] (fT00) at (\xTILDE-0.5*\xsh,\ytopTILDE-1) {$1$}; 
\node[flavor] (fT10) at (\xTILDE+0.5*\xsh,\ytopTILDE-0.5) {$1$}; 
\node[flavor] (fT20) at (\xTILDE+1.5*\xsh,\ytopTILDE-0.5) {$1$}; 
\node[flavor] (fTr-10) at (\xTILDE+3.2*\xsh+\xsh/2 ,\ytopTILDE-0.5) {$1$}; 
\node[flavor] (fTr0) at (\xTILDE+3.2*\xsh+\xsh/2+2,\ytopTILDE-1) {$1$}; 
\draw (gT10)--(fT10);
\draw (gT20)--(fT20);
\draw (gTr-10)--(fTr-10);
\draw[dashed] (gT10)--(fT00);
\draw[dashed] (gT10)--(fT20);
\draw[dashed] (gT20)--(fT10);
\draw[dashed] (gTr-10)--(fTr0);
\draw[dashed] (gT20)--(\xTILDE+5/2*\xsh-0.4,\ytopTILDE-0.5-0.4);
\draw[dashed] (fT20)--(\xTILDE+5/2*\xsh-0.4,\ynodeTILDE+0.4);
\draw[dashed] (gTr-10)--(\xTILDE+3.2*\xsh+\xsh/2-1.4,\ytopTILDE-0.5-0.4);
\draw[dashed] (fTr-10)--(\xTILDE+3.2*\xsh+\xsh/2-1.4,\ynodeTILDE+0.4);
	\end{tikzpicture}}
	\caption{ $\cC_l$-dressed quivers for the $T^l_{\vb*{v}}$ theory on the left and for the $\widetilde{T}^l_{\vb*{v}}$ theory on the right.}
	\label{fig:2dTSUquiver}
\end{figure}
In fact, analogously to the 3d case we can always realize the theories $T^l_{\vb*{v}}$ as worldsheet theories of specific configurations of BPS strings on $\mathbb{C}^2/\mathcal{C}_l$ for the 6d theory $\mathcal{T}^{6d}_{r,l+1}$, which in particular guarantees the cancelation of gauge and mixed anomalies. We will return to this point in more generality at the end of Section \ref{sec:M-string_orb}.

Let us now discuss how to glue the building blocks to obtain the M-string QFTs $\cQ^{\Gamma}_{\vb*{\kappa}}$ for $\Gamma=\cQ_4,\cT,\cO,\cI$. For a given choice of $\Gamma$, we pick a collection of theories $T_{\vb*{v}_1}^{l_1},\dots,T_{\vb*{v}_K}^{l_K}$, where 
\be\label{eq:v_I-def}
\vb*{v}_I
=
\frac{a_c\vb*\kappa}{l_I}, \qquad  I=1,\dots,K,
\ee
and the data $K, a_c, \{l_1,\dots,l_K\}$ correspond respectively to the number of external legs of $\Gamma$, to the comark of the central node, and to the lengths of the external legs. These data are listed in Table~\ref{tab:quivers_data}.

\begin{table}[t]
    \centering
    \begin{tabular}{c|ccc}
$\Gamma$   & $K$ & $a_c$  & $\{l_1,\dots,l_K\}$\\\hline    $\cQ_4$   &  4  &  $2$   & $\{2,2,2,2\}$ \\
   $\cT$   &  3  &  $3$   &  $\{3,3,3\}$\\
   $\cO$   &  3  &  $4$   &  $\{2,4,4\}$\\
   $\cI$   &  3  &  $6$   &  $\{2,3,6\}$\\
    \end{tabular}
    \caption{Data associated to the external linear quivers for star-shaped $\Gamma$.}
    \label{tab:quivers_data}
\end{table}
Recall that to specify a superselection sector for $\cQ^{\Gamma}_{\vb*{\kappa}}$ we need to supply a vector $\vb*{\omega}^{KK}=(\omega^{KK,(1)},\dots,\omega^{KK,(r)})$ of i.h.w.r.'s of $\widehat{\mathfrak{g}}_1$ as discussed at the end of Section~\ref{sec:quiv}. In order to describe this in terms of the superselection sector data for the $T^{l_I}_{\vb*{v}_I}$ theories, we have to decompose each $\omega^{KK,(a)}$ in terms of integrable representations of $\widehat{\su}(l_1)_1,\dots,\widehat{\su}(l_K)_1$ according to the branching rules determined by the conformal embedding
\be\label{eq:conformal_embedding}
\widehat{\su}(l_1)_1\oplus\dots\oplus\widehat{\su}(l_K)_1\subset\widehat{\mathfrak{g}}_1,
\ee
where we adopt the convention that $\mathfrak{su}(l_K)$ is associated to the tail that contains the affine node. 
Let us therefore consider the subset
\begin{equation*}
\mathcal{S}[\omega^{KK,(a)}]\subset\{(\varpi_1^{(a)},\dots,\varpi_K^{(a)})\;|\; \varpi_I^{(a)}\text{ is an i.h.w.r.\ of } \su(l_I)_1\text{ for }I=1,\dots,K\} ,
\end{equation*}
such that
\begin{equation}\label{eq:omega_decompos}
\omega^{KK,(a)}=\bigoplus_{\mathcal{S}[\omega^{KK,(a)}]}(\varpi_1^{(a)},\dots,\varpi_K^{(a)}).
\end{equation}
Then, for a given choice of $\vb*{\omega}^{KK}$ for $\cQ^{\Gamma}_{\vb*{\kappa}}$, we consider a collection of sets $\mathcal{S}[\omega^{KK,(1)}],\dots$, $\mathcal{S}[\omega^{KK,(r-1)}]$, out of which we form a set of all possible combinations of $K$ $r-$tuples $\vb*{\varpi}_1=(\varpi_{1}^{(1)},\dots,\varpi_{1}^{(r)})$, $\dots$, $\vb*{\varpi}_K=(\varpi_{K}^{(1)},\dots,\varpi_{K}^{(r)})$. Each element of this set corresponds to an allowed choice of i.h.w.r.'s for $T^{l_1}_{\vb*{v}_1}, \dots,T^{l_K}_{\vb*{v}_K}$ that will contribute to the elliptic genus of $\cQ^{\Gamma}_{\vb*{\kappa}}$. The embedding \eqref{eq:conformal_embedding} also determines a map $\pi_\Gamma = (\pi_1,\dots,\pi_K)$ of the chemical potentials $\vec{\xi}$, which take values in the complexification of the dual of the Cartan of $\mathfrak{g}$, to the chemical potentials $\vec{\xi}_I = \pi_I(\vec{\xi})$ for the theories $T^{l_I}_{\vb*{v}_I}$ which are valued in the complexification of the dual of the Cartan of $\mathfrak{su}(l_I)$. The M-string quiver has $r$ additional $U(1)$ flavor nodes attached to the affine nodes via twisted hypermultiplets $W^{(a)}$, and Fermi multiplets $\Sigma^{(a)}$ and $\Theta^{(a)}$, in the way explained earlier in Section~\ref{sec:quiv}. We can add these extra matter multiplets to $T^{l_K}_{\vb*{v}_K}$, by coupling them to its nodes of rank $\kappa^{(a)}$, which upon gluing will be identified with the affine nodes of $\cQ^{\Gamma}_{\vb*{\kappa}}$. It is straightforward to check that gauge anomalies still vanish after adding these new degrees of freedom by a similar computation to the one described in Section~\ref{sec:quiv}. Let us  denote the resulting quiver, which is shown on the right side of Figure \ref{fig:2dTSUquiver}, by~$\widetilde{T}^{l_K}_{\vb*{v}_K}$.\\

The gluing procedure requires gauging the flavor nodes $U(a_c \kappa^{(1)})$, $\dots$, $U(a_c \kappa^{(r-1)})$, where the abelian factors are decoupled for the theories $T^{l_1}_{\vb*{v}_1}, \dots,T^{l_{K-1}}_{\vb*{v}_{K-1}}$, but not for $\widetilde T^{l_{K}}_{\vb*{v}_{K}}$. More precisely, we have to couple the tails to the following 2d $(0,4)$ quiver:
\begin{equation}\label{eq:glue_quiver}
		\begin{tikzpicture}[line width=0.4]
		\node at (-2,0) {$\cQ^{glue}_{U(a_c \vb*{\kappa})}:$};
		\node[flavor] (g1) at (0,0) {$a_c\kappa^{(1)}$};
		\node[flavor] (g2) at (2,0) {$a_c\kappa^{(2)}$};
		\node[flavor] (gl-1) at (6,0) {\tiny $a_c\kappa^{(r-1)}$};
		\draw (g1)--(g2);
		\draw (g2)--(3,0);
		\draw[dashed] (3,0)--(5,0);
		\draw (5,0)--(gl-1);
	\end{tikzpicture}
\end{equation}
where the $U(a_c \kappa^{(a)})$ nodes are connected by bifundamental hypermultiplets. The coupling ensures that the nodes of the gluing quiver are not anomalous and can be gauged consistently. We are now able to schematically write the gluing formula as: 
\begin{equation}\label{eq:TSUdecomp}
\cQ^{\Gamma}_{\vb*{\kappa}}[\vb*{\omega}^{KK}]
=
\sum_{\mathcal{S}[\omega^{KK,(1)}]}\cdots\sum_{\mathcal{S}[\omega^{KK,(r)}]} \qty[\cQ^{glue}_{U(a_c \vb*{\kappa})} \, T^{l_1}_{\vb*{v}_1}[\vb*{\varpi}_1]\cdots \widetilde{T}^{l_{K}}_{\vb*{v}_{K}}[\vb*{\varpi}_K]]\Big{/} U(a_c \vb*{\kappa}),
\end{equation}
where we keep track of the superselection sector data $\boldsymbol{\omega}^{KK}$. \\

For the sake of illustration, let us consider the rank $r=2$ $\cQ^{\cI}_{(\kappa)}$ theory. This decomposes into $T_{2}^{3\kappa}$, $T_{3}^{2\kappa}$ and $\widetilde{T}_{6}^{\kappa}$. The only level 1 i.h.w.r.\ of $\widehat{\mathfrak{e}}_8$ is the vacuum representation $\boldsymbol{1}_{\widehat{\mathfrak{e}}_8}=[1,0,0,0,0,0,0,0,0]$. Therefore, the only possible choice $\vb*{\omega}^{KK}=(\omega^{KK,(1)},\omega^{KK,(2)})$ of i.h.w.r.\ for $\cQ^{\Gamma}_{\vb*{\kappa}}$ is $\omega^{KK,(1)}=\omega^{KK,(2)}=\boldsymbol{1}_{\widehat{\mathfrak{e}}_8}$, and by employing the branching rule for the conformal embedding $\widehat{\su}(2)\oplus\widehat{\su}(3) \oplus\widehat{\su}(6)\subset \widehat{\mathfrak{e}}_8$ we find
\begin{align}
\nonumber
\mathcal{S}[\omega^{KK,(1)}]=\mathcal{S}[\omega^{KK,(2)}]
=
\Big\{
&\qty(\vb{1}_{\widehat{\su}(2)},\vb{1}_{\widehat{\su}(3)},\vb{1}_{\widehat{\su}(6)}),
\qty(\vb{2}_{\widehat{\su}(2)},\overline{\vb{3}}_{\widehat{\su}(3)},\vb{6}_{\widehat{\su}(6)}),\\
\nonumber
&\qty(\vb{2}_{\widehat{\su}(2)},\vb{3}_{\widehat{\su}(3)},\overline{\vb{6}}_{\widehat{\su}(6)}),
\qty(\vb{1}_{\widehat{\su}(2)},\vb{3}_{\widehat{\su}(3)},\vb{15}_{\widehat{\su}(6)}),\\
&\qty(\vb{1}_{\widehat{\su}(2)},\overline{\vb{3}}_{\widehat{\su}(3)},\overline{\vb{15}}_{\widehat{\su}(6)}),
\qty(\vb{2}_{\widehat{\su}(2)},\vb{1}_{\widehat{\su}(3)},\vb{20}_{\widehat{\su}(6)})\Big\},
\end{align}
so that
\begin{equation}
\cQ^{\cI}_{(\kappa)}
[(\boldsymbol{1}_{\widehat{\mathfrak{e}}_8},\boldsymbol{1}_{\widehat{\mathfrak{e}}_8})]
=
\sum_{\mathcal{S}[\omega^{KK,(1)}]}\sum_{\mathcal{S}[\omega^{KK,(2)}]}  \qty[\cQ^{glue}_{U(6 \kappa)}\,
T_{2}^{3\kappa}[\vb*{\varpi}_1]
T_{3}^{2\kappa}[\vb*{\varpi}_2]
\widetilde{T}_{6}^{\kappa}[\vb*{\varpi}_3]]\Big{/} U(6 \kappa)  .
\end{equation}

\subsection{Generalization to M-string orbifold SCFTs}\label{sec:M-string_orb}

Up to this point in this section, we have focused on the case of the M-string SCFT $\mathcal{T}^{6d}_{r,1}$. We now turn to the more general class of M-string orbifold SCFTs $\mathcal{T}^{6d}_{r,W}$ corresponding to M5 branes probing a $W$-centered Taub--NUT space, and obtain the worldsheet theory of the BPS strings of this SCFT probing a background $T^2\times\IC^2/\Gamma$ for any $\Gamma$.\\

\noindent Recall from Section \ref{sec:6dSCFT} that the theory $\mathcal{T}^{6d}_{r,W}$
possesses a gauge algebra 
\be
\mathfrak{g}^{6d}=\prod_{a=1}^{r-1} \mathfrak{u}(W)^{(a)}
\ee
and flavor symmetry 
\be
\mathfrak{f}^{6d}=\mathfrak{u}(W)^{(0)}\times\mathfrak{u}(W)^{(r)}.
\ee
To each factor $\mathfrak{u}(W)^{(a)}$, $a=0,\dots,r$ is associated a vector bundle $\cV_W^{(a)}$ with connection $A^{(a)}$. The St\"{u}ckelberg mechanism~\cite{DelZotto:2023ryf} gives mass to $r$ of the $r+1$ abelian gauge and flavor factors, leading to a gauge algebra 
\be
\widetilde{\mathfrak{g}}^{6d}=\prod_{a=1}^{r-1} \su(W)^{(a)}
\ee
and a flavor symmetry
\be
\widetilde{\mathfrak{f}}^{6d}=\su(W)^{(0)}\times\su(W)^{(r)}\times \mathfrak{u}(1)^{diag}.
\ee

Recall from Section \ref{sec:sel} that in specifying a superselection sector for the theory $\mathcal{T}^{6d}_{r,W}$ we make a choice of a tuple of representations $\boldsymbol{\rho}=(\rho^{(0)},\dots,\rho^{(r)})$ of $\Gamma$, where
\be
\rho^{(a)} = \sum_{j=0}^{\text{rk }\mathfrak{g}} w^{(a)}_j\rho_j.
\ee
This corresponds to a decomposition
\begin{equation}\label{eq:bundle_decomposition}
    \cV_W^{(a)}=\bigoplus_{j=1}^{\text{rk }\mathfrak{g}} u^{(a)}_j\cR_j ,
\end{equation}
where
\begin{equation}\label{eq:fluxes_no_M2}
    u_j^{(a)}=\int_{\Sigma_j} c_1(\cV^{(a)}_W)=w_j^{(a)}
\end{equation}
are the fluxes of $A^{(a)}$ along the exceptional divisors $\Sigma_j$ of $\IC^2/\Gamma$ and we can write the first Chern class of the bundle as
\begin{equation}
\label{eq:first_Chern_decompos}
    c_1(\cV^{(a)}_W)=\sum_{j=1}^{\text{rk }\mathfrak{g}} u_j^{(a)} c_1(\cR_j) ,
\end{equation}
where $c_1(\cR_j)$, $j=1,\dots,\text{rk }\mathfrak{g}$, form a basis of $H^2(\widetilde{\IC^2/\Gamma}, \IZ )$ which is dual to the basis $\Sigma_1,\dots,\Sigma_{\text{rk }\mathfrak{g}}$ of $H_2(\widetilde{\IC^2/\Gamma}, \IZ )$~\cite{KronheimerNakajima,Douglas:1996sw}.
From Equation \eqref{eq:bundle_decomposition} it follows that the instanton number of $\cV_W^{(a)}$ is given by
\begin{equation}
     N^{(a)}
     =
     \int_{\widetilde{\mathbb{C}^2/\Gamma}}
     ch_2(\mathcal{V}^{(a)})
     =
     \sum_{j=1}^{\text{rk }\mathfrak{g}} u_j^{(a)} N_j =\frac{1}{|\Gamma|} \sum_{i,j=1}^{\text{rk }\mathfrak{g}} \qty(C^{\mathfrak{g}})^{-1}_{jk} w_j^{(a)} a_k.
\label{eq:NA}
\end{equation}

The BPS strings, which arise in the Type IIA frame as bound states of D2 branes stretched between adjacent NS5 branes and extending along a real codimension four locus in the D6 brane worldvolume, carry instanton charge with respect to the gauge algebras $\mathfrak u(W)^{(a)}$.  A configuration of D2 branes transverse to the $\mathbb{C}^2/\Gamma$ orbifold is specified in terms of a vector of nonnegative integers $\vec{v}^{(a)}=(v_0^{(a)},\dots,v_{\text{rk }\mathfrak{g}}^{(a)})\in \IZ_{\geq 0}^{\text{rk }\mathfrak{g}}$ \cite{Douglas:1996sw,KronheimerNakajima}. In the presence of the D2 branes the topological data of the bundle $\cV^{(a)}_W$ associated to the 6d gauge algebra $\mathfrak{u}(W)^{(a)}$ gets modified: in particular, the coefficients $u_j^{(a)}$ that appear in the decomposition \eqref{eq:bundle_decomposition} are now given by:
\begin{equation}\label{eq:fluxes_w_M2branes}
    u_j^{(a)}=w_j^{(a)}-\sum_{k=0}^{\text{rk }\mathfrak{g}} C^{\widehat{\mathfrak{g}}}_{jk} v_k^{(a)},
\end{equation}
which satisfy the constraint
\be
u_j^{(a)} = w_j^{(0)} \qquad a=1,\dots,r-1
\label{eq:c2}
\ee
as discussed in Section \ref{sec:sel}. The instanton number \eqref{eq:ch2V} is now given by:
\begin{equation}
    N^{(a)}=\sum_{j=1}^{\text{rk }{\mathfrak{g}}} u_j^{(a)} N_j + \frac{\sum_{j=0}^{\text{rk }{\mathfrak{g}}} a_j v_j^{(a)}}{|\Gamma|},
\end{equation}
which can be recast as
\begin{equation}
    N^{(a)}=v^{(a)}_0+\frac{1}{|\Gamma|} \sum_{j,k=1}^{\text{rk }{\mathfrak{g}}} \qty(C^{\mathfrak{g}})^{-1}_{jk} w_j^{(a)} a_k.
\end{equation}

The data described above determines the field content of the 2d $(0,4)$ $\Gamma$-dressed quiver QFTs corresponding to the BPS strings. This proceeds very much in the same way as for the M-string SCFT case we analyzed in Section~\ref{sec:quiv}. Namely, we consider a collection of 3d $\mathcal{N}=4$ Kronheimer--Nakajima quiver gauge theories $\mathcal{KN}^{\Gamma,\vec{w}^{(a)}}_{\vec{v}^{(a)}}$ with generic flavor symmetry nodes, as depicted in Figure~\ref{fig:genericKNquiver} for $\Gamma=\cI$~\cite{KronheimerNakajima}, and we stack them along an interval with $\frac{1}{2}$-BPS interfaces, which arise in the Type IIA picture from NS5 branes. Schematically:
\begin{equation*}
	\mathcal{KN}^{\Gamma}_{0}
	\bigg] \boldsymbol{NS}^\Gamma\bigg[
	\mathcal{KN}^{\Gamma,\vec{w}^{(1)}}_{\vec{v}^{(1)}}
	\bigg]  \boldsymbol{NS}^\Gamma\bigg[
	\mathcal{KN}^{\Gamma,\vec{w}^{(2)}}_{\vec{v}^{(2)}}
	 \bigg]  \boldsymbol{NS}^\Gamma \bigg[
	 \dots
	  \bigg]\boldsymbol{NS}^\Gamma \bigg[ 
	\mathcal{KN}^{\Gamma,\vec{w}^{(r-1)}}_{\vec{v}^{(r-1)}}
	  \bigg] \boldsymbol{NS}^\Gamma\bigg[ 
	  \mathcal{KN}^{\Gamma}_{0} ,
\end{equation*}
\begin{figure}[t]
	\centering
	\scalebox{0.8}{\begin{tikzpicture}[line width=0.8]
		\node[gauge]   (g0) at (0,0) {$v_0^{(a)}$};
		\node[gauge]   (g1) at (2,0) {$v_1^{(a)}$};
        \node[gauge]   (g2) at (4,0) {$v_2^{(a)}$};
        \node[gauge]   (g3) at (6,0) {$v_3^{(a)}$};
        \node[gauge]   (g4) at (8,0) {$v_4^{(a)}$};
        \node[gauge]   (g5) at (10,0) {$v_5^{(a)}$};
        \node[gauge]   (g6) at (12,0) {$v_6^{(a)}$};
        \node[gauge]   (g7) at (14,0) {$v_7^{(a)}$};
        \node[gauge]   (g8) at (10,2) {$v_8^{(a)}$};
        \node[flavor] (f0) at (0,-2) {$w_0^{(a)}$};
        \node[flavor] (f1) at (2,-2) {$w_1^{(a)}$};
        \node[flavor] (f2) at (4,-2) {$w_2^{(a)}$};
        \node[flavor] (f3) at (6,-2) {$w_3^{(a)}$};
        \node[flavor] (f4) at (8,-2) {$w_4^{(a)}$};
        \node[flavor] (f5) at (10,-2) {$w_5^{(a)}$};
        \node[flavor] (f6) at (12,-2) {$w_6^{(a)}$};
        \node[flavor] (f7) at (14,-2) {$w_7^{(a)}$};
        \node[flavor] (f8) at (8,2) {$w_8^{(a)}$};
        \draw (g0)--(g1);
        \draw (g1)--(g2);
        \draw (g2)--(g3);
        \draw (g3)--(g4);
        \draw (g4)--(g5);
        \draw (g5)--(g6);
        \draw (g6)--(g7);
        \draw (g5)--(g8);
        \draw (f0)--(g0);
        \draw (f1)--(g1);
        \draw (f2)--(g2);
        \draw (f3)--(g3);
        \draw (f4)--(g4);
        \draw (f5)--(g5);
        \draw (f6)--(g6);
        \draw (f7)--(g7);
        \draw (f8)--(g8);
	\end{tikzpicture}}
    \caption{The $3d$ $\cN=4$ Kronheimer--Nakajima quiver $\mathcal{KN}^{\cI,\vec{w}^{(a)}}_{\vec{v}^{(a)}}$.}
    \label{fig:genericKNquiver}
\end{figure}
We then reduce along the interval to obtain a 2d $\mathcal{N}=(0,4)$ theory, which can be obtained from the one described in Section \ref{sec:quiv} by replacing the $(a,0)$-th $U(1)^{(a)}$ flavor nodes by $F^{(a)}_0=U(w^{(a)}_0)$ and adding the following at the $(a,j)$-th site for $j\neq 0$: flavor nodes  $F_j^{(a)}=U(w_j^{(a)})$, hypermultiplets $W^{(a)}_j$ in the bifundamental representation of $F_j^{(a)}\times G_j^{(a)}$, and Fermi multiplets $\Sigma^{(a)}_{j}$ and $\Theta^{(a)}_{j}$ in the bifundamental of $G_j^{(a)}\times F^{(a+1)}_j$ and $G_j^{(a)}\times F^{(a-1)}_j$ respectively. These additional multiplets are depicted in Figure~\ref{fig:generic_Mstring_orb_quiver_node} for the $(a,j)-$th node of the quiver.
\begin{figure}[h]
	\centering
	\scalebox{0.8}{\begin{tikzpicture}[line width=0.8]
		\node[gauge]   (gja) at (0,0) {$v_j^{(a)}$};
        \node[flavor] (fja) at (0,-2) {$w_j^{(a)}$};
        \draw (gja)--(fja);
        \draw[line width=3pt] (gja)--(0,2);
        \draw[dashed,line width=3pt] (0,2)--(0,2.5);
        \draw[blue] (-1.5,-3)--(-1.5,3);
        \draw[blue] (1.5,-3)--(1.5,3);
        \node[fill=white] at (0,1.2) {$n_j$};
        \draw[dashed,green!70!black, line width=3pt] (gja)--(2.5,2.5);
        \node[fill=white] at (0.85,0.85) {$n_j$};
         \draw[dashed,green!70!black, line width=3pt] (gja)--(-2.5,2.5);
         \node[fill=white] at (-0.85,0.85) {$n_j$};
        %%%%%%%%%%%%%%%%%%%%
        \node[gauge]   (gja-1) at (-3,0) {$v_j^{(a-1)}$};
        \node[flavor] (fja-1) at (-3,-2) {$w_j^{(a-1)}$};
        \draw (gja-1)--(fja-1);
        \node[gauge]   (gja+1) at (3,0) {$v_j^{(a+1)}$};
        \node[flavor] (fja+1) at (3,-2) {$w_j^{(a+1)}$};
        \draw (gja+1)--(fja+1);
        %%%%%%%%%%%%%%%%%%%%
        \draw[purple] (gja-1)--(gja);
        \draw[purple] (gja)--(gja+1);
        \draw[dashed] (gja)--(fja+1);
        \draw[dashed] (gja)--(fja-1);
        \draw[dashed] (gja-1)--(fja);
        \draw[dashed] (gja+1)--(fja);
	\end{tikzpicture}}
    \caption{Generic node of the $\mathcal{Q}_{\va*{v}}^{\Gamma, \va*{w}}$ $\Gamma$-dressed quiver. The conventions for nodes and edges are the same of the M-string quiver in~Figure \ref{fig:MstringQuiverE8}, but here we also use thicker lines to represent additional lines exiting the nodes. The label $n_j=-\sum_{k>j} C^{\widehat{\mathfrak{g}}}_{jk}\in \{1,2,3,4\} $, is the number of lines connecting the $j$-th node to other nodes.}
    \label{fig:generic_Mstring_orb_quiver_node}
\end{figure}
The gauge group ranks $v^{(a)}_j$ can now take more general values than for the M-string case, but are still constrained by Equation \eqref{eq:c2}. This constraint is again equivalent to the requirement that non-abelian gauge anomalies cancel, as can be verified by a similar computation as the one discussed earlier in Section~\ref{sec:quiv}. We also recall from the discussion in Section \ref{sec:sel} that Equations \eqref{eq:uw} and \eqref{eq:uw0} impose constraints on the possible values of $w^{(a)}_j$ which are allowed; see also Section 3.1 of \cite{DelZotto:2023ryf} for a more detailed discussion of these constraints in the case $\Gamma = \cC_N$.
\\

Analogously to the M-string SCFT case, abelian anomalies can be canceled by including the coupling~\eqref{eq:currents-gauge-coupling} between the $(\widehat{\mathfrak{g}}_1)^r$ currents and the gauge connections of the $U(v_j^{(a)})$ quiver nodes. Moreover, as in~\cite{DelZotto:2023ryf}, the appearance of mixed anomalies between the gauge and global symmetries of the quiver can be removed by adding the following couplings between the $(\widehat{\mathfrak{g}}_1)^r$ currents and the background gauge fields associated to the flavor nodes $U(w_j^{(a)})$:
\begin{equation}
\label{eq:mixed-anomaly-coupling}
    \int_{T^2} \sum_{a=1}^r\sum_{j=1}^{\mathrm{rk}\,\mathfrak{g}}\cJ_j^{\widehat{\mathfrak{g}}_1,(a)} \qty[\Tr A^{U(w_j^{(a-1)})}-\Tr A^{U(w_j^{(a)})}+\Tr A^{U(1)_{\mathfrak{m}}}],
\end{equation}
where $A^{U(1)_{\mathfrak{m}}}$ is the background connection for an additional global symmetry $U(1)_{\mathfrak{m}}$, whose conjugate chemical potential $\mathfrak{m}$ enters Equation \eqref{eq:st}.
The cancelation of abelian and mixed anomalies is shown in Appendix~\ref{app:anomaly-canc}. Notice that the formula above is only valid for $\Gamma\in\{\cQ_N,\cT,\cO,\cI\}$ while for $\Gamma=\cC_N$ there is an additional term inside the square brackets \cite{DelZotto:2023ryf} due to a coupling to the $U(1)_L$ isometry of $\mathbb{C}^2/\mathcal{C}_N$.\\

In summary, the two sets of nonnegative integers $\va*{v}=(v_j^{(a)})^{a=1,\dots,r-1}_{j=0,\dots,\mathrm{rk}\,\mathfrak{g}}$ and $\va*{w}=(w_j^{(a)})^{a=0,\dots,r}_{j=0,\dots,\mathrm{rk}\,\mathfrak{g}}$ satisfying \eqref{eq:c2} specify a bound state $(\va*{v},\va*{w})$ of BPS strings, whose dynamic is described by the 2d $(0,4)$ relative theory $\cQ_{\va*{v}}^{\Gamma, \va*{w}}$ we have constructed. 
As for the case of the M-string SCFT, a choice of monodromy at infinity for the 6d two-form fields corresponds to a choice of irreducible highest weight representations $\vb*{\omega}^{KK}=(\omega^{KK,(1)},\dots,\omega^{KK,(r)})$  for $(\widehat{\mathfrak{g}}_1)^r$ that specify a superselection sector for the theory $\cQ_{\va*{v}}^{\Gamma, \va*{w}}$.

\paragraph{$T^{\vb*{\sigma}}_{\vb*{\rho}}(SU(v))$ decomposition.} The gluing procedure we have presented in Section \ref{sec:TSUdecomp} for constructing the worldsheet theories of the BPS strings of the M-string SCFT out of linear 2d $\mathcal{N}=(0,4)$ quivers can be extended naturally to the M-string orbifold SCFTs. Indeed we can consider the more general class of $3d$ $\cN=4$ linear quiver theories $T^{\vb*{\sigma}}_{\vb*{\rho}}(SU(v))$ \cite{Gaiotto:2008ak} shown in Figure~\ref{fig:linear_quiv}, which depend on a pair of partitions $\vb*{\sigma},\vb*{\rho}$  of $v$ or equivalently on a collection of nonnegative integers $v_j$ and $w_j$, $j=1,\dots,l-1$ satisfying the constraint
\begin{equation}\label{eq:TSU_node_constraint}
    v_{j-1}+v_{j+1}+w_{j}\geq 2 v_j,
\end{equation}
\begin{figure}[h]
	\centering
	\begin{tikzpicture}[line width=0.8]
		\node[gauge]   (g1) at (-4,0) {$v_1$};
		\node[flavor] (f1) at (-4,-1.5) {$w_1$};
		\node[gauge]   (g2) at (-2,0) {$v_2$};
		\node[flavor] (f2) at (-2,-1.5) {$w_2$};
		\node   (gi) at (0,0) {$\cdots$};
		\node[gauge]   (gM) at (2,0) {$v_{l-1}$};
		\node[flavor] (fM) at (2,-1.5) {$w_{l-1}$};
		\draw (g1) -- (g2);
		\draw (g1) -- (f1);
		\draw (g2) -- (f2);
		\draw (g2) -- (gi);
		\draw (gi) -- (gM);
		\draw (gM) -- (fM);
	\end{tikzpicture}
    \caption{Linear quiver for the $T^{\vb*{\sigma}}_{\vb*{\rho}}(SU(v))$ theory.}
    \label{fig:linear_quiv}
\end{figure}
implying that these theories are \emph{good} theories in the sense of~\cite{Gaiotto:2008ak}.

\begin{figure}[t]
		\centering
		
			\scalebox{1}{\begin{tikzpicture}[line width=0.8]
				\def\xsh{3.5};
				\def\gap{2};
				\draw[blue] (-2.5,-2.7)--(-2.5,5.2);
				\node[blue] at (-2.5,5.5) {$\widehat{\mathfrak{su}}(l)^{(1)}_1$};
				\draw[blue] (1,-2.7)--(1,5.2);		
				\node[blue] at (1,5.5) {$\widehat{\mathfrak{su}}(l)^{(2)}_1$};
				\draw[blue] (1+\xsh,-2.7)--(1+\xsh,5.2);		
				\node[blue] at (1+\xsh,5.5) {$\widehat{\mathfrak{su}}(l)^{(3)}_1$};
				\draw[blue] (1+\gap+\xsh,-2.7)--(1+\gap+\xsh,5.2);		
				\node[blue] at (1+\gap+\xsh,5.5) {$\widehat{\mathfrak{su}}(l)^{(r-1)}_1$};
				\draw[blue] (1+\gap+2*\xsh,-2.7)--(1+\gap+2*\xsh,5.2);		
				\node[blue] at (1+\gap+2*\xsh,5.5) {$\widehat{\mathfrak{su}}(l)^{(r)}_1$};
				%%%%%    (0)		
				%%%%%%%%%
				\node[flavor] (f01) at (-\xsh,4) {$w_1^{(0)}$};
				\node[flavor] (f02) at (-\xsh,2) {$w_2^{(0)}$};
				\node   (f0i) at (-\xsh,0) {$\vdots$};
				\node[flavor] (f0M) at (-\xsh,-2) {$w_{l-1}^{(0)}$};
				%%%%%%%%%
				%%%%%    (1)		
				%%%%%%%%%
				\node[gauge]   (g11) at (0,4) {$v_1^{(1)}$};
				\node[flavor] (f11) at (-1.5,4) {$w_1^{(1)}$};
				\node[gauge]   (g12) at (0,2) {$v_2^{(1)}$};
				\node[flavor] (f12) at (-1.5,2) {$w_2^{(1)}$};
				\node   (g1i) at (0,0) {$\vdots$};
				\node[gauge]   (g1M) at (0,-2) {$v_{l-1}^{(1)}$};
				\node[flavor] (f1M) at (-1.5,-2) {$w_{l-1}^{(1)}$};
				\draw (g11) -- (g12);
				\draw (g11) -- (f11);
				\draw (g12) -- (f12);
				\draw (g12) -- (g1i);
				\draw (g1i) -- (g1M);
				\draw (g1M) -- (f1M);
				%%%%%%%%%
				%%%%%    (2)		
				%%%%%%%%%
				\node[gauge]   (g21) at (0+\xsh,4) {$v_1^{(2)}$};
				\node[flavor] (f21) at (-1.5+\xsh,4) {$w_1^{(2)}$};
				\node[gauge]   (g22) at (0+\xsh,2) {$v_2^{(2)}$};
				\node[flavor] (f22) at (-1.5+\xsh,2) {$w_2^{(2)}$};
				\node   (g2i) at (0+\xsh,0) {$\vdots$};
				\node[gauge]   (g2M) at (0+\xsh,-2) {$v_{l-1}^{(2)}$};
				\node[flavor] (f2M) at (-1.5+\xsh,-2) {$w_{l-1}^{(2)}$};
				\draw (g21) -- (g22);
				\draw (g21) -- (f21);
				\draw (g22) -- (f22);
				\draw (g22) -- (g2i);
				\draw (g2i) -- (g2M);
				\draw (g2M) -- (f2M);
				%%%%%%%%%%(dots)%%%%%%%%%%%%
				\node  (d1) at (1+\xsh+\gap/2,4) {\Large$\cdots$};
				\node  (d2) at (1+\xsh+\gap/2,2) {\Large$\cdots$};
				\node  (di)  at (1+\xsh+\gap/2,0) {\Large$\cdots$};
				\node (dM) at (1+\xsh+\gap/2,-2) {\Large$\cdots$};
				%%%%%%%%%
				%%%%%    (r-1)		
				%%%%%%%%%
				\node[gauge]   (gr-11) at (0+\gap+2*\xsh,4) {\tiny$v_1^{(r-1)}$};
				\node[flavor] (fr-11) at (-1.5+\gap+2*\xsh,4) {\tiny$w_1^{(r-1)}$};
				\node[gauge]   (gr-12) at (0+\gap+2*\xsh,2) {\tiny$v_2^{(r-1)}$};
				\node[flavor] (fr-12) at (-1.5+\gap+2*\xsh,2) {\tiny$w_2^{(r-1)}$};
				\node   (gr-1i) at (0+\gap+2*\xsh,0) {$\vdots$};
				\node[gauge]   (gr-1M) at (0+\gap+2*\xsh,-2) {\tiny$v_{l-1}^{(r-1)}$};
				\node[flavor] (fr-1M) at (-1.5+\gap+2*\xsh,-2) {\tiny$w_{l-1}^{(r-1)}$};
				\draw (gr-11) -- (gr-12);
				\draw (gr-11) -- (fr-11);
				\draw (gr-12) -- (fr-12);
				\draw (gr-12) -- (gr-1i);
				\draw (gr-1i) -- (gr-1M);
				\draw (gr-1M) -- (fr-1M);
				%%%%%%%%%
				%%%%%    (r)		
				%%%%%%%%%
				\node[flavor] (fr1) at (-1.5+\gap+3*\xsh,4) {$w_1^{(r)}$};
				\node[flavor] (fr2) at (-1.5+\gap+3*\xsh,2) {$w_2^{(r)}$};
				\node   (fri) at (-1.5+\gap+3*\xsh,0) {$\vdots$};
				\node[flavor] (frM) at (-1.5+\gap+3*\xsh,-2) {$w_{l-1}^{(r)}$};
				%%%%%%%%%%%%%%%
				% (0) -- (1) 
				%%%%%%%%%%%%%%%
				\draw[dashed] (f01)to[out=-30,in=-150](g11);
				\draw[dashed] (f02)to[out=-30,in=-150](g12);
				\draw[dashed] (f0M)to[out=-30,in=-150](g1M);
				%%%%%%%%%%%%%%%
				% (1)--(2)
				%%%%%%%%%%%%%%%
				\draw[purple] (g11)to[out=-30,in=-150](g21);
				\draw[dashed] (g11)--(f21);
				\draw[dashed] (f11)to[out=30,in=150](g21);
				\draw[green!70!black,dashed] (g11)--(g22);
				\draw[green!70!black,dashed] (g12)--(g21);
				%%%%%%%%%%%%%%%%%
				\draw[purple] (g12)to[out=-30,in=-150](g22);
				\draw[dashed] (g12)--(f22);
				\draw[dashed] (f12)to[out=30,in=150](g22);
				\draw[green!70!black,dashed] (g12)--(g2i);
				\draw[green!70!black,dashed] (g1i)--(g22);
				%%%%%%%%%%%%%%%%%%%%
				\draw[purple] (g1M)to[out=-30,in=-150](g2M);
				\draw[dashed] (g1M)--(f2M);
				\draw[dashed] (f1M)to[out=30,in=150](g2M);
				\draw[green!70!black,dashed] (g1M)--(g2i);
				\draw[green!70!black,dashed] (g1i)--(g2M);
				%%%%%%%%%%%%%%%%%%%%
				% (2)--(dots)
				%%%%%%%%%%%%%%%
				\draw[purple] (g21)to[out=-30,in=-150](d1);
				\draw[dashed] (g21)--(d1);
				\draw[dashed] (f21)to[out=30,in=150](d1);
				\draw[green!70!black,dashed] (g21)--(d2);
				\draw[green!70!black,dashed] (g22)--(d1);
				%%%%%%%%%%%%%%%%%
				\draw[purple] (g22)to[out=-30,in=-150](d2);
				\draw[dashed] (g22)--(d2);
				\draw[dashed] (f22)to[out=30,in=150](d2);
				\draw[green!70!black,dashed] (g22)--(di);
				%%%%%%%%%%%%%%%%%%%%
				\draw[purple] (g2M)to[out=-30,in=-150](dM);
				\draw[dashed] (g2M)--(dM);
				\draw[dashed] (f2M)to[out=30,in=150](dM);
				\draw[green!70!black,dashed] (g2M)--(di);
				%%%%%%%%%%%%%%%%%%%%
				% (dots)--(r-1)
				%%%%%%%%%%%%%%%
				\draw[purple] (d1)to[out=-30,in=-150](gr-11);
				\draw[dashed] (d1)--(fr-11);
				\draw[dashed] (d1)to[out=30,in=150](gr-11);
				\draw[green!70!black,dashed] (d1)--(gr-12);
				\draw[green!70!black,dashed] (d2)--(gr-11);
				%%%%%%%%%%%%%%%%%
				\draw[purple] (d2)to[out=-30,in=-150](gr-12);
				\draw[dashed] (d2)--(fr-12);
				\draw[dashed] (d2)to[out=30,in=150](gr-12);
				\draw[green!70!black,dashed] (di)--(gr-12);
				%%%%%%%%%%%%%%%%%%%%
				\draw[purple] (dM)to[out=-30,in=-150](gr-1M);
				\draw[dashed] (dM)--(fr-1M);
				\draw[dashed] (dM)to[out=30,in=150](gr-1M);
				\draw[green!70!black,dashed] (di)--(gr-1M);
				%%%%%%%%%%%%%%%
				% (r-1) -- (r) 
				%%%%%%%%%%%%%%%
				\draw[dashed] (gr-11)--(fr1);
				\draw[dashed] (gr-12)--(fr2);
				\draw[dashed] (gr-1M)--(frM);
		\end{tikzpicture}}
    \caption{The 2d $\mathcal{N}=(0,4)$ $\cC_l$-dressed quiver theory $T^{l,\vec{\boldsymbol{w}}}_{\vec{\boldsymbol{v}}}$.}
    \label{fig:gammaquiv}
\end{figure}
Let us denote by $T^{l,\vec{\boldsymbol{w}}}_{\vec{\boldsymbol{v}}}$ the 2d $\mathcal{N}=(0,4)$ QFT that arises from stacking a collection of $r-1$ theories of $T^{\vb*{\sigma}}_{\vb*{\rho}}(SU(v))$ type and coupling them to a $\widehat{\mathfrak{su}}(l)^r_1$ current algebra. The $\cC_l$-dressed quiver for the theory $T^{l,\vec{\boldsymbol{w}}}_{\vec{\boldsymbol{v}}}$ is shown in Figure \ref{fig:gammaquiv}. We will show in Appendix \ref{app:TVW} that any theory $T^{l,\vec{\boldsymbol{w}}}_{\vec{\boldsymbol{v}}}$ can be realized as a frozen BPS string for a theory $\mathcal{T}^{6d}_{r,W}$ on a $\mathbb{C}^2/\cC_l$ singularity by adding an affine node with gauge group $U(0)$. In particular we will show that
\begin{equation}
	T^{l,\vec{\boldsymbol{w}}}_{\vec{\boldsymbol{v}}}=\cQ^{\cC_l,\vec{\vb{w}}}_{\vec{\vb{v}}},
\end{equation}
where
\begin{equation}
	\vec{\boldsymbol{\mathrm{ v}}} = \mqty(\boldsymbol{0}\\ \boldsymbol{v}_1\\\vdots\\\boldsymbol{v}_{l-1})\quad\text{and}\quad
	\vec{\boldsymbol{\mathrm{ w}}} = \mqty(W-\sum_{j=1}^{l-1}\boldsymbol{w}_j\\\boldsymbol{w}_1\\\vdots\\\boldsymbol{w}_{l-1}).
\end{equation}
This ensures that the theories $T^{l,\vec{\boldsymbol{w}}}_{\vec{\boldsymbol{v}}}$  are free from gauge anomalies.  As a consequence, the 3d constraint \eqref{eq:TSU_node_constraint} is automatically satisfied for every gauge node of the theory $T^{l,\vec{\boldsymbol{w}}}_{\vec{\boldsymbol{v}}}$ by virtue of Equation \eqref{eq:c2}. We remark that the theories $T^l_{\vb*{v}}$ and $\widetilde{T}^l_{\vb*{v}}$ discussed in Section~\ref{sec:TSUdecomp} are just special cases of theories $T^{l,\vec{\boldsymbol{w}}}_{\vec{\boldsymbol{v}}}$, with:
\begin{equation}
	\va*{v}=\mqty(\vb*{v}\\2\vb*{v}\\\vdots\\ (l-1) \vb*{v})
\end{equation}
and
\begin{equation}
	\va*{w}=
	\begin{cases}
		\mqty(\smqty(0\\ \vdots\\ \\0),\smqty(0\\ \vdots\\\\l v^{(1)}),\cdots,\smqty(0\\\vdots\\\\l v^{(r-1)}),\smqty(0\\ \vdots\\ \\0))\quad\text{for }T_{\vb*{v}}^l,\\[15pt]
		\mqty(\smqty(1\\0\\\vdots\\\\0),\smqty(1\\0\\ \vdots\\\\l v^{(1)}),\cdots,\smqty(1\\0\\ \vdots\\\\l v^{(r-1)}),\smqty(1\\0\\ \vdots\\ \\0))\quad\text{for }\widetilde{T}_{\vb*{v}}^l,
	\end{cases}.
\end{equation}
\\

In the context of the gluing technique of Section~\ref{sec:TSUdecomp}, the theories $T^{l,\vec{\boldsymbol{w}}}_{\vec{\boldsymbol{v}}}$ form a more general set of building blocks from which we can obtain general $\Gamma$-dressed quiver theories $\cQ^{\Gamma,\va*{w}}_{\va*{v}}$ with $\Gamma=\cQ_4$, $\cT$, $\cO$ or $\cI$. The gluing technique can be generalized as follows. For a given choice of $\Gamma$, we pick a collection of theories $T_{\va*{v}_1}^{l_1,\va*{w}_1},\dots,T_{\va*{v}_K}^{l_K,\va*{w}_K}$, where the data $K$ and $\{l_1,\dots,l_K\}$ are the same as in Section~\ref{sec:TSUdecomp}, and are listed in Table~\ref{tab:quivers_data}.  The gauge and flavor node ranks for each tail are determined from the BPS string data $\va*{v}$ and $\va*{w}$ as follows:
\begin{equation}
	(v_I^{(a)})_j=v^{(a)}_{P^{\Gamma}_I(j)} \quad\text{and}\quad (w_I^{(a)})_j=w^{(a)}_{P^{\Gamma}_I(j)}
\end{equation}
for $j=1,\dots, l_I-1$ with $I=1,\dots,K$ and $a=1,\dots, r$, where the $P^{\Gamma}_I(j)$ are specific assignments of indices which depends on the particular choice of $\Gamma$, \emph{e.g.}, for $\Gamma=\cI$ we have
\begin{equation}
	P^{\cI}_I(j)=\begin{cases}
		j-1 \quad\text{for } I=1\\
		8-j \quad\text{for } I=2\\
		9-j\quad\text{for } I=3\\
	\end{cases} \ .
\end{equation}
We then shift the $(l-1)$-th flavor fugacities as follows
\begin{equation}\label{eq:flavor_shift}
	(w^{(a)}_I)_{l-1}\to	(w_{I}^{(a)})_{l-1}+ v_c^{(a)} \quad\text{for } a=1,\dots,r-1 \ ,
\end{equation}
where $v_c^{(a)}$ is the rank of the $a$-th central node of the $\Gamma$-dressed quiver $\cQ^{\Gamma,\va*{w}}_{\va*{v}}$, while leaving the others invariant. The dressed quiver of Figure~\ref{fig:gammaquiv} gets then modified as shown in Figure~\ref{fig:central_nodes}.
\begin{figure}[t]
	\centering
		\scalebox{1}{\begin{tikzpicture}[line width=0.8]
			\def\xsh{3.5};
			\def\gap{2};
			\draw[blue] (-2.5,-4)--(-2.5,0.2);
			\draw[blue] (1,-4)--(1,0.2);		
			\draw[blue] (1+\xsh,-4)--(1+\xsh,0.2);		
			\draw[blue] (1+\gap+\xsh,-4)--(1+\gap+\xsh,0.2);		
			\draw[blue] (1+\gap+2*\xsh,-4)--(1+\gap+2*\xsh,0.2);		
			\node[flavor] (f0M) at (-\xsh,-2) {$w_{l-1}^{(0)}$};
			%%%%%%%%%
			%%%%%    (1)		
			%%%%%%%%%
			\node   (g1i) at (0,0) {$\vdots$};
			\node[gauge]   (g1M) at (0,-2) {$v_{l-1}^{(1)}$};
			\node[flavor] (f1M) at (-1.5,-2) {$w_{l-1}^{(1)}$};
			\node[flavor] (c1) at (0,-3.5) {$v_c^{(1)}$};
			\draw (g1i) -- (g1M);
			\draw (g1M) -- (f1M);
			\draw (g1M)--(c1);
			%%%%%%%%%
			%%%%%    (2)		
			%%%%%%%%%
			\node   (g2i) at (0+\xsh,0) {$\vdots$};
			\node[gauge]   (g2M) at (0+\xsh,-2) {$v_{l-1}^{(2)}$};
			\node[flavor] (f2M) at (-1.5+\xsh,-2) {$w_{l-1}^{(2)}$};
			\node[flavor] (c2) at (\xsh,-3.5) {$v_{c}^{(2)}$};
			\draw (g2i) -- (g2M);
			\draw (g2M) -- (f2M);
			\draw (g2M)--(c2);
			%%%%%%%%%%(dots)%%%%%%%%%%%%
			\node  (di)  at (1+\xsh+\gap/2,0) {};
			\node (dM) at (1+\xsh+\gap/2,-2) {\Large$\cdots$};
			\node (dc) at (1+\xsh+\gap/2,-3.5) {};
			%%%%%%%%%
			%%%%%    (r-1)		
			%%%%%%%%%
			\node   (gr-1i) at (0+\gap+2*\xsh,0) {$\vdots$};
			\node[gauge]   (gr-1M) at (0+\gap+2*\xsh,-2) {\tiny$v_{l-1}^{(r-1)}$};
			\node[flavor] (fr-1M) at (-1.5+\gap+2*\xsh,-2) {\tiny$w_{l-1}^{(r-1)}$};
			\node[flavor] (cr-1) at (\gap+2*\xsh,-3.5) {\tiny$v_{c}^{(r-1)}$};
			\draw (gr-1i) -- (gr-1M);
			\draw (gr-1M) -- (fr-1M);
			\draw (gr-1M)--(cr-1);
			%%%%%%%%%
			%%%%%    (r)		
			%%%%%%%%%
			\node[flavor] (frM) at (-1.5+\gap+3*\xsh,-2) {$w_{l-1}^{(r)}$};
			%%%%%%%%%%%%%%%
			% (0) -- (1) 
			%%%%%%%%%%%%%%%
			\draw[dashed] (f0M)to[out=-30,in=-150](g1M);
			%%%%%%%%%%%%%%%
			% (1)--(2)
			%%%%%%%%%%%%%%%
			\draw[purple] (g1M)to[out=-30,in=-150](g2M);
			\draw[dashed] (g1M)--(f2M);
			\draw[dashed] (f1M)to[out=30,in=150](g2M);
			\draw[green!70!black,dashed] (g1M)--(g2i);
			\draw[green!70!black,dashed] (g1i)--(g2M);
			\draw[dashed] (g1M)to[out=-45,in=180](c2);
			\draw[dashed] (g2M)to[out=-135,in=0](c1);
			%%%%%%%%%%%%%%%%%%%%
			% (2)--(dots)
			%%%%%%%%%%%%%%%%%%%%
			\draw[purple] (g2M)to[out=-30,in=-150](dM);
			\draw[dashed] (g2M)--(dM);
			\draw[dashed] (f2M)to[out=30,in=150](dM);
			\draw[green!70!black,dashed] (g2M)--(di);
			\draw[dashed] (g2M)--(dc);
			\draw[dashed] (c2)--(dM);
			%%%%%%%%%%%%%%%%%%%%
			% (dots)--(r-1)
			%%%%%%%%%%%%%%%
			\draw[purple] (dM)to[out=-30,in=-150](gr-1M);
			\draw[dashed] (dM)--(fr-1M);
			\draw[dashed] (dM)to[out=30,in=150](gr-1M);
			\draw[green!70!black,dashed] (di)--(gr-1M);
			\draw[dashed] (dM)to[out=-45,in=180](cr-1);
			\draw[dashed] (gr-1M)to[out=-135,in=0](dc);
			%%%%%%%%%%%%%%%
			% (r-1) -- (r) 
			%%%%%%%%%%%%%%%
			\draw[dashed] (gr-1M)--(frM);
	\end{tikzpicture}}
\caption{Detail of the shift \eqref{eq:flavor_shift} on the $(l-1)$--th flavor node of the $\cC_l$-dressed quiver $T^{l,\va*{w}}_{\va*{v}}$.}
\label{fig:central_nodes}
\end{figure}
We can then gauge the central flavor nodes $U(v^{(a)}_{c})$ for each of the tails $T^{l,\va*{w}}_{\va*{v}}$ involved in the gluing  by coupling the multiplets charged under them to the $a$-th gauge node of the gluing quiver shown in Figure \ref{eq:gluing_quiver2}.

\begin{figure}[t]
\centering
\scalebox{0.9}{
  \begin{tikzpicture}[line width=0.8]
        \node[gauge]   (g2) at (4,0) {$v_c^{(1)}$};
        \node[gauge]   (g3) at (6,0) {$v_c^{(2)}$};
         \node   (g4) at (8,0) {$\cdots$};
        \node[gauge]   (g5) at (10,0) {\footnotesize $v_{c}^{(r-2)}$};
        \node[gauge]   (g6) at (12,0) {\footnotesize $v_{c}^{(r-1)}$};
        \node[flavor] (f1) at (2,-1.8) {$w_c^{(0)}$};
        \node[flavor] (f2) at (4,-1.8) {$w_c^{(1)}$};
        \node[flavor] (f3) at (6,-1.8) {$w_c^{(2)}$};
        \node[flavor] (f5) at (10,-1.8) {\footnotesize $w_c^{(r-2)}$};
        \node[flavor] (f6) at (12,-1.8) {\footnotesize$w_{c}^{(r-1)}$};
        \node[flavor] (f7) at (14,-1.8) {$w_c^{(r)}$};
        \draw[purple] (g2)--(g3);
        \draw[purple] (g3)--(g4);
        \draw[purple] (g4)--(g5);
        \draw[purple] (g5)--(g6);
        \draw (f2)--(g2);
        \draw (f3)--(g3);
        \draw (f5)--(g5);
        \draw (f6)--(g6);
        \draw[dashed] (g2) -- (f3);
        \draw[dashed] (g3) -- (f4);
        \draw[dashed] (g4) -- (f5);
 \draw[dashed] (g5) -- (f6);
 \draw[dashed] (g6) -- (f7);
 \draw[dashed] (f1) -- (g2);
        \draw[dashed] (f2) -- (g3);
        \draw[dashed] (f3) -- (g4);
        \draw[dashed] (f4) -- (g5);
 \draw[dashed] (f5) -- (g6);
 \end{tikzpicture}
}
\caption{The gluing quiver $\cQ^{glue,\vb*{w}_c}_{\vb*{v}_c}$.}
\label{eq:gluing_quiver2}
\end{figure}

\subsection{Examples of BPS string configurations} 
\label{sec:exs}

In the previous sections we have described how to construct the 2d $(0,4)$ worldsheet theories describing the BPS strings for any member of the class of 6d SCFTs $\cT^{6d}_{r,W}$, and for any choice of monodromies on $\mathbb{C}^2/\Gamma$ of the gauge connection allowed by Equation \eqref{eq:c2}. This gives rise to a large zoo of 2d QFTs $\cQ^{\Gamma,\vec{\boldsymbol{w}}}_{\vec{\boldsymbol{v}}}$,  and it is convenient at this point to give some concrete examples of the configurations which can occur. Here we will focus for definiteness on the case $r=2$ and introduce a simplified notation to describe the corresponding 2d quivers, which we explain in Figure \ref{fig:M-string_orbifold_quiver}.
\begin{figure}[h]
	\centering
    \begin{tikzpicture}[line width=0.4]
        \node (r0) at (10,6) {$a=0,2$};
        \node (r1) at (0,6) {$a=1$};
        \node[flavor] (f01) at (8,4) {$w^{(a)}_1$};
        \node[flavor] (f02) at (8,2) {$w^{(a)}_2$};
        \node[flavor] (f03) at (8,0) {$w^{(a)}_3$};
        \node[flavor] (f04) at (8,-2) {$w^{(a)}_4$};
        \node[flavor] (f05) at (8,-4) {$w^{(a)}_5$};
        \node[flavor] (f06) at (10,0) {$w^{(a)}_6$};
        \node[flavor](f00) at (12,0) {$w^{(a)}_0$};
        \draw[blue] (6,6) -- (6,-6);     
        \node[blue] at (6,6.5)   {$\widehat{\mathfrak{e}}_{6,1}$}      ;
        \node[flavor](f11) at (-2,4) {$w^{(1)}_1$};
        \node[flavor](f12) at (-2,2) {$w^{(1)}_2$};
        \node[flavor] (f13) at (-2,0) {$w^{(1)}_3$};
        \node[flavor] (f14) at (-2,-2) {$w^{(1)}_4$};
        \node[flavor] (f15) at (-2,-4) {$w^{(1)}_5$};
        \node[flavor] (f16) at (2,2) {$w^{(1)}_6$};
        \node[flavor] (f10) at (4,2) {$w^{(1)}_0$};
        \node[gauge]   (g1)  at (0,4) {$v^{(1)}_1$};
        \node[gauge]   (g2)  at (0,2) {$v^{(1)}_2$};
        \node[gauge]   (g3)  at (0,0) {$v^{(1)}_3$};
        \node[gauge]   (g4)  at (0,-2) {$v^{(1)}_4$};
        \node[gauge]   (g5)  at (0,-4) {$v^{(1)}_5$};
        \node[gauge]    (g6)  at (2,0) {$v^{(1)}_6$};
        \node[gauge]    (g0)  at (4,0) {$v^{(1)}_0$};
        \draw (g1) -- (g2);
        \draw (g2) -- (g3);
        \draw (g3) -- (g4);
        \draw (g4) -- (g5);
        \draw  (g3) -- (g6);
        \draw  (g6) -- (g0);
        \draw (g1) -- (f11);
        \draw (g2) -- (f12);
        \draw (g3) -- (f13);
        \draw  (g4) -- (f14);
        \draw (g5) -- (f15);
        \draw  (g6) -- (f16);
        \draw  (g0) -- (f10);
        \draw[dashed,double,double distance = 1.5pt] (g0) to[out=25,in=155] (f00);
        \draw[dashed,double,double distance = 1.5pt] (g1) -- (f01);
        \draw[dashed,double,double distance = 1.5pt] (g2) to[out=25,in=155] (f02);      \draw[dashed,double,double distance = 1.5pt] (g6) to[out=-25,in=-155] (f06);
        \draw[dashed,double,double distance = 1.5pt] (g3) to[out=-25,in=-155] (f03);
        \draw[dashed,double,double distance = 1.5pt] (g4) -- (f04);
        \draw[dashed,double,double distance = 1.5pt] (g5) -- (f05);
	\end{tikzpicture}
    \caption{ A simplified representation of BPS string quivers for rank $r=2$ M-string orbifold SCFTs on a $\Gamma=\CT$ singularity, where the ranks of the flavor and gauge nodes satisfy equation \eqref{eq:c2}. We employ an analogous notation for other choices of $\Gamma$ as well. The two sets of external flavor nodes at $a=0$ and $a=2$ are condensed for convenience into a single set on the right of the vertical blue line, since $\vec{w}^{(0)}=\vec{w}^{(2)}$. The blue line in the middle represents the two NS5 interfaces supporting current algebras, and the quiver on the left encodes the gauge degrees of freedom on the worldsheet of the BPS string. The solid lines between nodes are twisted bifundamental hypermultiplets.  A dashed double line that connects the gauge node $U(v_j^{(1)})$ on the left with the flavor node $U(w^{(a)}_j)$, $a=0,2$, on the right, represents a pair of Fermi multiplets, one in the bifundamental representations of $U(v^{(1)}_j)\times U(w^{(0)}_j)$ and one in the bifundamental of $U(v^{(1)}_j)\times U(w^{(2)}_j)$. Whenever $v_j^{(1)}=0$ or $w^{(a)}_j=0$, for some $a=0,1,2$ and $j=0,\dots,\mathrm{rk}\,\mathfrak{g}$, we represent the the corresponding node as an empty dotted node, see Figure \ref{fig:list_allowed_M-string_orb}.
    Moreover, decoupled flavor nodes, which are disconnected from the rest of the quiver but nevertheless couple to the current algebras, can also appear and will be highlighted in green.}\label{fig:M-string_orbifold_quiver}
    \end{figure}

\begin{figure}[!htbp]
        \centering
\scalebox{0.97}{
        \begin{tabular}{c|c}
       Configuration \textbf{1a}, $W=2$ & Configuration \textbf{1b}, $W=2$\\
       $\Gamma = \cQ_4,$ 
       $\mathfrak{g}^{6d} = \mathfrak{u}(1), $
       $\mathfrak{f}^{6d} = \mathfrak{u}(2)^{\oplus 2}$
       &
       $\Gamma = \cT,$ 
       $\mathfrak{g}^{6d} = \mathfrak{u}(1), $
       $\mathfrak{f}^{6d} = \mathfrak{u}(2)^{\oplus 2}$
       \\
         \scalebox{0.8}{\begin{tikzpicture}[line width=0.4]
        \node[smallflavor,green!70!black](f00) at (3,1) {$2$};
       \node[emptyflavor] (f01) at (5,1) {};
        \node[emptyflavor] (f02) at (4,0) {};
        \node[emptyflavor] (f03) at (3,-1) {};
        \node[emptyflavor] (f04) at (5,-1) {};
        \draw[blue] (2,2.5) -- (2,-2.5);     
        \node[emptyflavor] (f10) at (-1,2) {};
        \node[emptyflavor](f11) at (1,2) {};
        \node[smallflavor](f12) at (-1,0) {$1$};
        \node[emptyflavor] (f13) at (-1,-2) {};
        \node[emptyflavor] (f14) at (1,-2) {};
        \node[emptygauge]   (g0)  at (-1,1) {};
        \node[smallgauge]   (g1)  at (1,1) {$1$};
        \node[smallgauge]   (g2)  at (0,0) {$2$};
        \node[smallgauge]   (g3)  at (-1,-1) {$1$};
        \node[smallgauge]   (g4)  at (1,-1) {$1$};
        \draw[dotted] (g0) -- (g2);
        \draw (g1) -- (g2);
        \draw (g3) -- (g2);
        \draw (g4) -- (g2);
        \draw[dotted] (g0) -- (f10);
        \draw[dotted] (g1) -- (f11);
        \draw (g2) -- (f12);
        \draw[dotted] (g3) -- (f13);
        \draw[dotted] (g4) -- (f14);
	\end{tikzpicture}}
    &\scalebox{0.8}{\begin{tikzpicture}[line width=0.4]
       \node[emptyflavor] (f01) at (4,2) {};
        \node[emptyflavor] (f02) at (4,1) {};
        \node[emptyflavor] (f03) at (4,0) {};
        \node[emptyflavor] (f04) at (4,-1) {};
        \node[emptyflavor] (f05) at (4,-2) {};
        \node[emptyflavor] (f06) at (5,0) {};
        \node[smallflavor,green!70!black](f00) at (6,0) {$2$};
        \draw[blue] (3,2.5) -- (3,-2.5);             
        \node[emptyflavor](f11) at (-1,2) {};
        \node[emptyflavor](f12) at (-1,1) {};
        \node[emptyflavor] (f13) at (-1,0) {};
        \node[emptyflavor] (f14) at (-1,-1) {};
        \node[emptyflavor] (f15) at (-1,-2) {};
        \node[smallflavor] (f16) at (1,1) {$1$};
        \node[emptyflavor] (f10) at (2,1) {};
        \node[smallgauge]   (g1)  at (0,2) {$1$};
        \node[smallgauge]   (g2)  at (0,1) {$2$};
        \node[smallgauge]   (g3)  at (0,0) {$3$};
        \node[smallgauge]   (g4)  at (0,-1) {$2$};
        \node[smallgauge]   (g5)  at (0,-2) {$1$};
        \node[smallgauge]    (g6)  at (1,0) {$2$};
        \node[emptygauge]    (g0)  at (2,0) {};
        \draw (g1) -- (g2);
        \draw (g2) -- (g3);
        \draw (g3) -- (g4);
        \draw (g4) -- (g5);
        \draw (g3) -- (g6);
        \draw[dotted]  (g6) -- (g0);
        \draw[dotted] (g1) -- (f11);
        \draw[dotted]  (g2) -- (f12);
        \draw[dotted] (g3) -- (f13);
        \draw[dotted]  (g4) -- (f14);
        \draw[dotted]  (g5) -- (f15);
        \draw  (g6) -- (f16);
        \draw[dotted]  (g0) -- (f10);
	\end{tikzpicture}} \\ \hline 
  Configuration \textbf{2a}, $W=8$ &  Configuration \textbf{2b}, $W=2$\\
         $\Gamma = \cT,$ 
       $\mathfrak{g}^{6d} = \mathfrak{u}(2)\oplus \mathfrak{u}(1), $
       $\mathfrak{f}^{6d} = \mathfrak{u}(4)^{\oplus 2}$
       &
         $\Gamma = \cC_6,$ 
       $\mathfrak{g}^{6d} = \mathfrak{u}(2), $
       $\mathfrak{f}^{6d} = \mathfrak{u}(2)^{\oplus 2}$
       \\
\scalebox{0.8}{\begin{tikzpicture}[line width=0.4]
       \node[emptyflavor] (f01) at (4,2) {};
        \node[emptyflavor] (f02) at (4,1) {};
        \node[emptyflavor] (f03) at (4,0) {};
        \node[emptyflavor] (f04) at (4,-1) {};
        \node[emptyflavor] (f05) at (4,-2) {};
        \node[smallflavor,green!70!black] (f06) at (5,0) {$4$};
        \node[emptyflavor](f00) at (6,0) {};
        \draw[blue] (3,2.5) -- (3,-2.5);             
        \node[emptyflavor](f11) at (-1,2) {};
        \node[emptyflavor](f12) at (-1,1) {};
        \node[smallflavor] (f13) at (-1,0) {$2$};
        \node[emptyflavor] (f14) at (-1,-1) {};
        \node[emptyflavor] (f15) at (-1,-2) {};
        \node[smallflavor,green!70!black] (f16) at (1,1) {$1$};
        \node[emptyflavor] (f10) at (2,1) {};
        \node[smallgauge]   (g1)  at (0,2) {$1$};
        \node[smallgauge]   (g2)  at (0,1) {$2$};
        \node[smallgauge]   (g3)  at (0,0) {$3$};
        \node[smallgauge]   (g4)  at (0,-1) {$2$};
        \node[smallgauge]   (g5)  at (0,-2) {$1$};
        \node[emptygauge]    (g6)  at (1,0) {};
        \node[emptygauge]    (g0)  at (2,0) {};
        \draw (g1) -- (g2);
        \draw (g2) -- (g3);
        \draw (g3) -- (g4);
        \draw (g4) -- (g5);
        \draw[dotted]  (g3) -- (g6);
        \draw[dotted]  (g6) -- (g0);
        \draw[dotted] (g1) -- (f11);
        \draw[dotted]  (g2) -- (f12);
        \draw (g3) -- (f13);
        \draw[dotted]  (g4) -- (f14);
        \draw[dotted]  (g5) -- (f15);
        \draw[dotted]  (g6) -- (f16);
        \draw[dotted]  (g0) -- (f10);
	\end{tikzpicture}}
    & \scalebox{0.8}{\begin{tikzpicture}[line width=0.4]
        \node[emptyflavor] (f01) at (1,2) {};
        \node[emptyflavor] (f02) at (1,1) {};
        \node[emptyflavor] (f03) at (1,0) {};
        \node[emptyflavor] (f04) at (1,-1) {};
        \node[emptyflavor] (f05) at (1,-2) {};
        \node[smallflavor,green!70!black] (f00) at (2,0) {$2$};
        \draw[blue] (0,2.5) -- (0,-2.5);     
        \node[smallgauge]   (g1)  at (-3,2) {$1$};
        \node[smallgauge]   (g2)  at (-3,1) {$2$};
        \node[smallgauge]   (g3)  at (-3,0) {$3$};
        \node[smallgauge]   (g4)  at (-3,-1) {$2$};
        \node[smallgauge]   (g5)  at (-3,-2) {$1$};
        \node[emptygauge]   (g0)  at (-1,0) {};
        \node[emptyflavor](f11) at (-4,2) {};
        \node[emptyflavor](f12) at (-4,1) {};
        \node[smallflavor](f13) at (-4,0) {$2$};
        \node[emptyflavor](f14) at (-4,-1) {};
        \node[emptyflavor](f15) at (-4,-2) {};
        \node[emptyflavor](f10) at (-2,0) {};
        \draw[dotted] (g0) -- (g1);
        \draw  (g1) -- (g2);
        \draw  (g2) -- (g3);
        \draw  (g3) -- (g4);
        \draw  (g4) -- (g5);
        \draw[dotted] (g5) -- (g0);
        \draw[dotted] (g1) -- (f11);
        \draw[dotted]  (g2) -- (f12);
        \draw (g3) -- (f13);
        \draw[dotted] (g4) -- (f14);
        \draw[dotted] (g5) -- (f15);
        \draw[dotted] (g0) -- (f10);
	\end{tikzpicture}} \\\hline
     Configuration \textbf{3a}, $W=4$ & Configuration \textbf{3b}, $W=3$\\
              $\Gamma = \cQ_4,$ 
       $\mathfrak{g}^{6d} = \mathfrak{u}(1)^{\oplus 4}, $
       $\mathfrak{f}^{6d} = \mathfrak{u}(2)^{\oplus 2}$ &
              $\Gamma = \cT,$ 
       $\mathfrak{g}^{6d} = \mathfrak{u}(1)^{\oplus 3}, $
       $\mathfrak{f}^{6d} = \mathfrak{u}(1)^{\oplus 2}$\\
         \scalebox{0.8}{\begin{tikzpicture}[line width=0.4]
        \node[emptyflavor](f00) at (3,1) {};
       \node[emptyflavor] (f01) at (5,1) {};
        \node[smallflavor] (f02) at (4,0) {$2$};
        \node[emptyflavor] (f03) at (3,-1) {};
        \node[emptyflavor] (f04) at (5,-1) {};
        \draw[blue] (2,2.5) -- (2,-2.5);     
        \node[smallflavor] (f10) at (-1,2) {$1$};
        \node[smallflavor](f11) at (1,2) {$1$};
        \node[emptyflavor](f12) at (-1,0) {};
        \node[smallflavor] (f13) at (-1,-2) {$1$};
        \node[smallflavor] (f14) at (1,-2) {$1$};
        \node[smallgauge]   (g0)  at (-1,1) {$1$};
        \node[smallgauge]   (g1)  at (1,1) {$1$};
        \node[smallgauge]   (g2)  at (0,0) {$1$};
        \node[smallgauge]   (g3)  at (-1,-1) {$1$};
        \node[smallgauge]   (g4)  at (1,-1) {$1$};
        \draw (g0) -- (g2);
        \draw (g1) -- (g2);
        \draw (g3) -- (g2);
        \draw (g4) -- (g2);
        \draw (g0) -- (f10);
        \draw (g1) -- (f11);
        \draw[dotted]  (g2) -- (f12);
        \draw (g3) -- (f13);
        \draw (g4) -- (f14);
        \draw[dashed,double,double distance = 1.5pt] (g2) -- (f02);
	\end{tikzpicture}}
    & \scalebox{0.8}{\begin{tikzpicture}[line width=0.4]
       \node[emptyflavor] (f01) at (4,2) {};
        \node[emptyflavor] (f02) at (4,1) {};
        \node[smallflavor] (f03) at (4,0) {$1$};
        \node[emptyflavor] (f04) at (4,-1) {};
        \node[emptyflavor] (f05) at (4,-2) {};
        \node[emptyflavor] (f06) at (5,0) {};
        \node[emptyflavor](f00) at (6,0) {};
        \draw[blue] (3,2.5) -- (3,-2.5);             
        \node[smallflavor](f11) at (-1,2) {$1$};
        \node[emptyflavor](f12) at (-1,1) {};
        \node[emptyflavor] (f13) at (-1,0) {};
        \node[emptyflavor] (f14) at (-1,-1) {};
        \node[smallflavor] (f15) at (-1,-2) {$1$};
        \node[emptyflavor] (f16) at (1,1) {};
        \node[smallflavor] (f10) at (2,1) {$1$};
        \node[smallgauge]   (g1)  at (0,2) {$1$};
        \node[smallgauge]   (g2)  at (0,1) {$1$};
        \node[smallgauge]   (g3)  at (0,0) {$1$};
        \node[smallgauge]   (g4)  at (0,-1) {$1$};
        \node[smallgauge]   (g5)  at (0,-2) {$1$};
        \node[smallgauge]    (g6)  at (1,0) {$1$};
        \node[smallgauge]    (g0)  at (2,0) {$1$};
        \draw (g1) -- (g2);
        \draw (g2) -- (g3);
        \draw (g3) -- (g4);
        \draw (g4) -- (g5);
        \draw (g3) -- (g6);
        \draw  (g6) -- (g0);
        \draw (g1) -- (f11);
        \draw[dotted]  (g2) -- (f12);
        \draw[dotted] (g3) -- (f13);
        \draw[dotted]  (g4) -- (f14);
        \draw (g5) -- (f15);
        \draw[dotted]  (g6) -- (f16);
        \draw (g0) -- (f10);
       \draw[dashed,double,double distance = 1.5pt] (g3) to[out=-30,in=-150] (f03);
	\end{tikzpicture}}\\
    \hline
    Configuration \textbf{4}, $W=8$&\\
       $\Gamma = \cQ_4,$ 
       $\mathfrak{g}^{6d} = \mathfrak{u}(2)^{\oplus 4}, $
       $\mathfrak{f}^{6d} = \mathfrak{u}(4)^{\oplus 2}$ &
       \\
    \scalebox{0.8}{\begin{tikzpicture}[line width=0.4]
        \node[emptyflavor](f00) at (3,1) {};
       \node[emptyflavor] (f01) at (5,1) {};
        \node[smallflavor,green!70!black] (f02) at (4,0) {$4$};
        \node[emptyflavor] (f03) at (3,-1) {};
        \node[emptyflavor] (f04) at (5,-1) {};
        \draw[blue] (2,2.5) -- (2,-2.5);     
        \node[smallflavor] (f10) at (-1,2) {$2$};
        \node[smallflavor](f11) at (1,2) {$2$};
        \node[emptyflavor](f12) at (-1,0) {};
        \node[smallflavor] (f13) at (-1,-2) {$2$};
        \node[smallflavor] (f14) at (1,-2) {$2$};
        \node[smallgauge]   (g0)  at (-1,1) {$1$};
        \node[smallgauge]   (g1)  at (1,1) {$1$};
        \node[emptygauge]   (g2)  at (0,0) {};
        \node[smallgauge]   (g3)  at (-1,-1) {$1$};
        \node[smallgauge]   (g4)  at (1,-1) {$1$};
        \draw[dotted] (g0) -- (g2);
        \draw[dotted]  (g1) -- (g2);
        \draw[dotted]  (g3) -- (g2);
        \draw[dotted]  (g4) -- (g2);
        \draw (g0) -- (f10);
        \draw (g1) -- (f11);
        \draw[dotted]  (g2) -- (f12);
        \draw (g3) -- (f13);
        \draw (g4) -- (f14);
	\end{tikzpicture}} &
        \end{tabular}}
        \caption{Sample list of allowed BPS string configurations for the rank 6d theories $\mathcal{T}^{6d}_{2,W}$ on $\IC^2/\Gamma$.}
        \label{fig:list_allowed_M-string_orb}
    \end{figure}

The examples depicted in Figure~\ref{fig:list_allowed_M-string_orb} illustrate a number of interesting features of the theories $\cQ^{\Gamma,\va*{w}}_{\va*{v}}$:
\begin{itemize}
\item For any $\Gamma$, it is always possible to make a choice of monodromies $\vec{w}^{(a)}$ leading to a fractional BPS string of instanton charge
\begin{equation}
    N^{(a)}=\frac{|\Gamma|-1}{|\Gamma|} ,\text{ for }a=1,\dots,r-1,
\end{equation}
which has the properties that its moduli space is isomorphic to the one of a BPS string of instanton charge 1. This is the 2d $\mathcal{N}=(0,4)$ analogue of a feature that was observed in the mathematics literature already in \cite{KronheimerNakajima}. We illustrate this in two specific examples for $\Gamma = \cQ_4, $ and $ \cT$, which are portrayed respectively in Configurations \textbf{1a} and \textbf{1b}.
\item Configurations \textbf{2a} and \textbf{2b} illustrate the fact that the same 2d quiver can appear in two completely different 6d theories as the worldsheet theory of a fractional BPS string. In the case at hand, we have displayed a quiver that appears both for the $\mathfrak{g}=\mathfrak{u}(8)$ 6d SCFT on a $\Gamma = \cT$ singularity and for the $\mathfrak{g}=\mathfrak{u}(2)$ 6d SCFT on a $\Gamma = \cC_6$ singularity. While the quiver is the same, the details of how it couples to the two 6d theories are different:  in the two examples at hand the quiver couples respectively to  current algebras $\widehat{\mathfrak{g}}= \widehat{\mathfrak{e}}_6$ and $\widehat{\mathfrak{su}}(6)$. Moreover, the instantonic numbers are different, namely for Configuration \textbf{2a} and \textbf{2b} the instanton numbers are respectively $\frac{37}{8}$ and $\frac{3}{2}$.
 \item While the phenomenon illustrated above is quite frequent, there also exist 2d quivers that can only appear for a specific choice of $\Gamma$. Configuration \textbf{3a} and \textbf{3b} are two such examples, respectively for $\Gamma = \cQ_4$ and $\cT$, which have instanton charge $\frac{5}{2}$ and $\frac{7}{3}$ respectively.
 \item Finally, in Configuration \textbf{4} we provide an example of a charge $4$ BPS instanton for $\Gamma = \cQ_4$ which behaves as a collection of four decoupled frozen $\mathfrak{g}=\mathfrak{u}(2)$ instantons on a $\Gamma=\mathcal{C}_2$ singularity.
 \end{itemize}
 We also remark that the various examples of Figure \ref{fig:list_allowed_M-string_orb} can be constructed using the gluing technique discussed in the previous sections. For instance, Configuration \textbf{3b} can be constructed by gluing three copies of the theory $T_{\vec{\boldsymbol{v}}}^{3,\vec{\boldsymbol{w}}}$, where
\bea
\vec{\boldsymbol{w}}
=
\left\{
\begin{pmatrix}
	0\\
	0
\end{pmatrix}
,
\begin{pmatrix}
	1\\
	1
\end{pmatrix}
,
\begin{pmatrix}
	0\\
	0
\end{pmatrix}
\right\},
\qquad
\vec{\boldsymbol{v}}=
\left\{
\begin{pmatrix}
	1\\
	1
\end{pmatrix}
\right\}.
\eea

\section{The elliptic genus}\label{sec:ell}
In this section we turn to the computation of the elliptic genera of the BPS strings for arbitrary $\Gamma$: In Section \ref{sec:EG} we provide expressions for the elliptic genus, while in Section \ref{sec:gluingEG} we discuss a prescription for computing the elliptic genera for $\Gamma\in\{\cQ_{N},\cT,\cO,\cI\}$ by gluing of the tail theories $T^{l,\va*{w}}_{\va*{v}}$.

\subsection{Elliptic genus from localization}
\label{sec:EG}
The contribution of a bound state of $(\va*{v},\va*{w})$ BPS strings to the 6d partition function \eqref{eq:PF} is given in terms of the elliptic genus of their worldsheet theory $\cQ^{\Gamma,\va*{w}}_{\va*{v}}$, which is
defined as follows:
\begin{equation}\label{eq:ADE_elliptic_genus}
\mathbb{E}^{\Gamma,\va*{w}}_{\va*{v}}
[\vb*{\omega}^{KK}]
(\vec{\xi},\underline{\vec{\vb*{s}}},\epsilon_+,\tau)
=
\Tr_{\vb*{\omega}^{KK}} (-1)^F q^{H_L}\overline{q}^{H_R}e^{2\pi i \epsilon_+ (J_R+J_I)} e^{2\pi i \vec{\boldsymbol\xi}\cdot \vec{\boldsymbol J}^{\widehat{\mathfrak{g}}}}e^{2\pi i \underline{\vb*{s}}\cdot \underline{\vb*{J}}}.
\end{equation}
The trace is taken over the conformal family of $(\widehat{\mathfrak{g}}_1)^r$ labeled by the $r$-tuple of i.h.w.r.'s $\boldsymbol{\omega}^{KK}$. The operators $\vec{\boldsymbol J}^{\widehat{\mathfrak{g}}}$ denote the generators for the Cartan of the $(\widehat{\mathfrak{g}}_1)^r$ current algebra, which appears as a global symmetry of $\cQ^{\Gamma,\va*{w}}_{\va*{v}}$. Moreover, $J_R$ and $J_I$ denote respectively the Cartan generators for $SU(2)_R$ and $SU(2)_I$, while the $\underline{J}^{(a)}$ are currents for $\prod_{j=0}^{\text{rk }\mathfrak{g}} U(w_j)^{(a)}$. The shifted chemical potentials $\vec{\boldsymbol \xi}=(\vec\xi^{(1)},\dots,\vec\xi^{(r)})$ which appear in the trace will be defined shortly, in Equation \eqref{eq:xi-term}. Note that, differently from the elliptic genus \eqref{eq:elliptic_genus_on_C2} for $\mathbb{C}^2$, we cannot turn on a holonomy $\epsilon_-$ for the Cartan of the $SU(2)_L$ symmetry which for $\Gamma\neq\mathcal{C}_N$ is completely broken by the orbifold.\\

By localization \cite{Benini:2013xpa}, the elliptic genus can be expressed as a multidimensional contour integral over flat connections on $T^2$:
\begin{equation}
\label{eq:ADE_elliptic_genus_loc}
\begin{aligned}
	\mathbb{E}^{\Gamma,\va*{w}}_{\va*{v}}[\vb*{\omega}^{KK}]
	=
	&\pqty{\prod_{j=0}^{\text{rk }\mathfrak{g}}\prod_{a=1}^{r-1}\frac{1}{v_j^{(a)}!}}\int\qty[\prod_{j=0}^{\text{rk }\mathfrak{g}}\qty(\prod_{a=1}^{r-1}Z_{V_j^{(a)}}Z_{W_j^{(a)}}Z_{\Sigma_{j}^{(a)}}Z_{\Theta_j^{(a)}})\qty(\prod_{a=1}^{r-2} Z_{Y_j^{(a)}} )] \\
	&\qquad\times\prod_{j=1}^{\text{rk }\mathfrak{g}}\prod_{i=0}^{j-1}\qty[\qty(\prod_{a=1}^{r-1}Z_{X_{i,j}^{(a)}}) \qty(\prod_{a=1}^{r-2} Z_{\Psi_{i,j}^{(a)}} Z_{\widetilde{\Psi}_{i,j}^{(a)}})]^{-C^{\widehat{\mathfrak{g}}}_{ij}}\qty(\prod_{a=1}^r 
	Z^{\widehat{\mathfrak{g}}_1^{(a)}}_{\omega^{KK,(a)}}),
\end{aligned}
\end{equation}
where $z_{j,1}^{(a)}, \dots,z_{j,v_{j}^{(a)}}^{(a)}$ are the holonomies along $T^2$ for the gauge groups $G_j^{(a)}=U(v_j^{(a)})$.
The integral is evaluated by summing over Jeffrey-Kirwan residues~\cite{Benini:2013xpa}. The integrand in \eqref{eq:ADE_elliptic_genus_loc} is the one-loop determinant of the Gaussian path integral over the bosonic zero-modes and factorizes in terms of contributions from each of the multiplets of the theory. Their explicit expressions are given by:
\begin{align}
\label{eq:Z_V}	
Z_{V_j^{(a)}}&=\qty(\prod_{k=1}^{v_j^{(a)}} \frac{\dd{z_{j,k}^{(a)}}}{2\pi i}\frac{2\pi \eta^3\theta_1(2 \epsilon_+)}{\eta^2})\qty(\prod_{\substack{ k,l=1\\ k\neq l}}^{v_j^{(a)}} \frac{\theta_1(z_{j,k}^{(a)}-z_{j,l}^{(a)})\theta_1(2\epsilon_++z_{j,k}^{(a)}-z_{j,l}^{(a)})}{\eta^2});
\\
\label{eq:Z_Y}Z_{Y_{j}^{(a)}}&=\prod_{k=1}^{v_j^{(a)}}\prod_{l=1}^{v_{j}^{(a+1)}} \frac{\eta^2}{\theta_1(\epsilon_++z_{j,l}^{(a+1)}-z_{j,k}^{(a)})\theta_1(\epsilon_++z_{j,k}^{(a)}-z_{j,l}^{(a+1)})};
\\
\label{eq:Z_X}Z_{X_{i,j}^{(a)}}&=\prod_{k=1}^{v_i^{(a)}}\prod_{l=1}^{v_j^{(a)}} \frac{\eta^2}{\theta_1(\epsilon_++z_{j,l}^{(a)}-z_{i,k}^{(a)})\theta_1(\epsilon_++z_{i,k}^{(a)}-z_{j,l}^{(a)})};
\\
\label{eq:Z_Psi}	Z_{\Psi_{i,j}^{(a)}}&=\prod_{k=1}^{v_i^{(a)}}\prod_{l=1}^{v_{j}^{(a+1)}}\frac{\theta_1(z_{j,l}^{(a+1)}-z_{i,k}^{(a)})}{\eta};
\\
\label{eq:Z_tildePsi}Z_{\widetilde{\Psi}_{i,j}^{(a)}}&=\prod_{k=1}^{v_j^{(a)}}\prod_{l=1}^{v_{i}^{(a+1)}} \frac{\theta_1(z_{i,l}^{(a+1)}-z_{j,k}^{(a)})}{\eta};\\
\label{eq:Z1loop_W}	
Z_{W_j^{(a)}}&=\prod_{k=1}^{v_j^{(a)}} \prod_{K=1}^{w_j^{(a)}} \frac{\eta^2}{\theta_1(\epsilon_++s_{j,K}^{(a)}-z_{j,k}^{(a)})\theta_1(\epsilon_++z_{j,k}^{(a)}-s_{j,K}^{(a)})};\\
\label{eq:Z1loop_Sigma}	
Z_{\Sigma_j^{(a)}}&=\prod_{k=1}^{v_j^{(a)}} \prod_{K=1}^{w_j^{(a+1)}}\frac{\theta_1(s_{j,K}^{(a+1)}-z_{j,k}^{(a)})}{\eta};\\
\label{eq:Z1loop_Theta}	
Z_{\Theta_j^{(a)}}&=\prod_{k=1}^{v_j^{(a)}} \prod_{K=1}^{w_j^{(a-1)}}\frac{\theta_1(z_{j,k}^{(a)}-s_{j,K}^{(a-1)})}{\eta}.
\end{align}
The one-loop factors are expressed in terms of Dedekind eta and Jacobi theta functions
\begin{align}
\eta&=\eta(\tau)=q^{\frac{1}{24}}\prod_{k=1}^{\infty}(1-q^k),\\
\theta_1(z)&=\theta_1(z,\tau)=i q^{\frac{1}{8}}e^{-\pi i z}\prod_{k=1}^{\infty}(1-q^k)(1-q^{k-1}e^{2\pi i z})(1-q^ke^{-2\pi i z}).
\end{align}
Lastly, the contributions of the current algebras are given by
\begin{equation}\label{eq:Z_Gamma}
Z^{\widehat{\mathfrak{g}}_1^{(a)}}_{\omega^{KK,(a)}}={\chi}^{\widehat{\mathfrak{g}}_1}_{\omega^{KK,(a)}}\qty(\va{\xi}^{(a)},\tau),
\end{equation}
where
\begin{equation}\label{eq:xi-term}
	\xi^{(a)}_{j}=\xi_j+\sum_{k=0}^{\text{rk }\mathfrak{g}}C_{jk}^{\widehat{\mathfrak{g}}}\qty(Z_k^{(a)} -Z_k^{(a-1)})-S^{(a)}_j+S^{(a-1)}_j +\mathfrak{m},
\end{equation}
for $j=1,\dots,\text{rk }\mathfrak{g}$ and $a=0,\dots,r$, and
\begin{equation}\label{eq:Zj_and_Sj}
	Z_k^{(a)}=\sum_{l=1}^{v_k^{(a)}} z_{k,l}^{(a)},\quad S^{(a)}_j=\sum_{K=1}^{w_j^{(a)}} s_{j,K}^{(a)}
\end{equation}
for $k=0,\dots,\text{rk }\mathfrak{g}$. The chemical potentials $\xi_j$ are shifted in \eqref{eq:xi-term} as a consequence of the couplings \eqref{eq:currents-gauge-coupling} and \eqref{eq:mixed-anomaly-coupling} of the $(\widehat {\mathfrak{g}}_{1})^r$ currents to the gauge and background connections respectively. In Appendix \ref{app:anomaly-canc} we will use the explicit expression for the integrand of Equation \eqref{eq:ADE_elliptic_genus_loc} to verify that the worldsheet CFT is free from abelian gauge anomalies and mixed anomalies, where the cancelation of mixed anomalies occurs thanks to the chemical potential shifts \eqref{eq:xi-term}. \\

The expression \eqref{eq:ADE_elliptic_genus_loc} is manifestly covariant under the action \eqref{eq:OG} of the outer automorphism group $\cO(\widehat{\mathfrak{g}}))$ on the parameters ${\boldsymbol{\omega}}^{KK}$, $\vec{\boldsymbol{w}}$, $\vec{\boldsymbol{v}}$, $\vec{\xi}$, $\vec{\underline{\boldsymbol{s}}}$, where $o(\boldsymbol{v}_j) = \boldsymbol{v}_{o(j)}$ follows from Equation \eqref{eq:uw}. This completes the proof of the statement that the full partition function \eqref{eq:PF} of the 6d SCFT $\cT^{6d}_{r,W}$ transforms covariantly. A technical point that deserves mention is that for general $\Gamma$ the integrand of Equation \eqref{eq:ADE_elliptic_genus_loc} has a more intricate structure than in the $\Gamma=\cC_N$ case and requires the use of the full-fledged Jeffrey-Kirwan formalism as well as possibly integration over non-simple poles associated to the trivalent/tetravalent nodes in the quiver. As a consequence, the resulting expressions for the elliptic genus cannot in general be expressed in terms of combinatorial formulas as is the case for $\Gamma=\cC_N$ \cite{DelZotto:2023rct,DelZotto:2023ryf} and tend to be significantly more complicated.\\

In the $T^2\times\IC^2$ background, the St\"{u}ckelberg mechanism relates the flavor holonomies $s^{(a)}_{A}$, $A=1,\dots, W$ of each $\mathfrak{u}(W)^{(a)}$ as follows~\cite{Haghighat:2013tka}:
\begin{equation}\label{eq:Stuck_constraint_on_s_A}
    \sum_{A=1}^{W}s_A^{(a+1)}= \sum_{A=1}^{W}s_A^{(a)}+W\mathfrak{m}.
\end{equation}
After orbifolding by $\Gamma$, the gauge algebras $\mathfrak{u}(W)^{(a)}$ break into $\bigoplus_{j=0}^{\text{rk }\mathfrak{g}} \mathfrak{u}(w_j^{(a)})$. The embedding implies the following relation between the respective holonomies:
\begin{equation}
\frac{1}{W}\sum_{A=1}^{W}s_A^{(a)}
=
\frac{1}{h^{\vee}_{\Gamma}}\sum_{j=0}^{\text{rk }\mathfrak{g}}a_jS_{j}^{(a)}.
\end{equation}
This ensures that
\begin{equation}
    \sum_{j=0}^{\text{rk }\mathfrak{g}} a_j \xi^{(a)}_j =  0 ,
\end{equation}
where $\xi^{(a)}_0$ is defined by extending \eqref{eq:xi-term} to $j=0$. We remark that the constraint \eqref{eq:Stuck_constraint_on_s_A} can be written in terms of the $\mathfrak{u}(w_j^{(a)})$ holonomies as 
\begin{equation}\label{eq:Stuckelberg_constr}
    \sum_{j=0}^{\text{rk }\mathfrak{g}} a_j\qty(S^{(a)}_j-S^{(a-1)}_j)= h^{\vee}_{\Gamma}\mathfrak{m}.
\end{equation}

\subsection{Gluing formula for the elliptic genus}
\label{sec:gluingEG}
In Section~\ref{sec:TSUdecomp} we introduced a new class of 2d $\mathcal{N}=(0,4)$ relative QFTs called $T^l_{\vb*{v}}$, whose $\cC_l$-dressed quiver is depicted in Figure~\ref{fig:2dTSUquiver}. As explained there, these theories are obtained by placing $\tfrac{1}{2}$-BPS interfaces between multiple copies of the $3d$ $\cN=4$ Gaiotto--Witten theories $T_{\vb*{\rho}}(SU(v))$. In the same section we have also outlined a gluing prescription to obtain the theories $\cQ^{\Gamma}_{\vb*{\kappa}}$  with $\Gamma=\cQ_4$, $\cT,$ $\cO,$ $\cI$, from combinations of $T^l_{\vb*{v}}$ theories, see Equation \eqref{eq:TSUdecomp}.  In Section~\ref{sec:M-string_orb} we extended these theories in order to accommodate the more general Gaiotto--Witten quiver $T_{\vb*{\rho}}^{\vb*{\sigma}}(SU(v))$, and obtained the theories $T_{\va*{v}}^{l,\va*{w}}$, whose dressed quiver is shown in Figure~\ref{fig:gammaquiv}. We have also shown how the gluing technique applies to these theories as well and can be used to obtain general theories $\cQ^{\Gamma, \va*{w}}_{\va*{v}}$. The gluing procedure simplifies the computation of the elliptic genera of the BPS string quivers by decomposing them into a set of simpler building blocks.\\

\noindent The gluing procedure can be summarized as follows:
\begin{enumerate}
	\item Decompose $\cQ^{\Gamma, \va*{w}}_{\va*{v}}$ into a collection $T^{l_1,\va*{w}_1}_{\va*{v}_1},\dots,T^{l_K,\va*{w}_K}_{\va*{v}_K}$ as shown in Section~\ref{sec:M-string_orb}, with each $T^{l_I,\va*{w}_I}_{\va*{v}_I}$, $I=1,\dots,K$, representing an external tail of the Dynkin diagram of type $\Gamma$. $K$ is the number of edges attached to the central node of $\Gamma$, $l_1,\dots,l_K$  are the lengths of the external tails. Both quantities are listed in Table~\ref{tab:quivers_data} for $\Gamma=\cQ_4$, $\cT$, $\cO$, $\cI$. 
	\item The elliptic genera $\IE^{\Gamma,\va*{w}}_{\va*{v}}[\vb*{\omega}^{KK}]$ depends on a choice of i.h.w.r.'s $\vb*{\omega}^{KK}$ of $(\widehat{\mathfrak{g}}_1)^r$. Decompose each $\omega^{KK,(a)}$, $a=1,\dots,r$, as 
	\begin{equation*}
		\omega^{KK,(a)}=\bigoplus_{\mathcal{S}[\omega^{KK,(a)}]}(\varpi_1,\dots,\varpi_K),
	\end{equation*}
	where $(\varpi_1,\dots,\varpi_K)$ denotes a $K-$tuple of integrable representations of $\widehat{\su}(l_1)_1\oplus \dots\oplus \widehat{\su}(l_K)_1$, and $\mathcal{S}[\omega^{KK,(a)}]$ is determined according to the branching rules of the conformal embedding \begin{equation*}
		\widehat{\su}(l_1)_1\oplus\dots\oplus\widehat{\su}(l_K)_1\subset\widehat{\mathfrak{g}}_1,
	\end{equation*}
	which also determines the map $\pi_\Gamma$ between the chemical potentials $\vec\xi$ for $\mathfrak{g}$ and those for the $\widehat{\su}(l_I)$ as explained in Section~\ref{sec:TSUdecomp}.
	
	\item From the sets $\mathcal{S}[\omega^{KK,(a)}]$, $a=1,\dots,r$, form all possible $K$-tuples $(\boldsymbol{\varpi}_1,\dots,\boldsymbol{\varpi}_K)$ built as follows:
	\begin{equation}
	\qty{\begin{pmatrix}
			\vb*{\varpi}_1&=&(\varpi^{(1)}_1,\dots,\varpi^{(r)}_1)\\
			&\vdots&\\
			\vb*{\varpi}_K&=&(\varpi^{(1)}_K,\dots,\varpi^{(r)}_K)
			\end{pmatrix}
			\;\left.\vphantom{\mqty(\\ \\ \\ \\)}\right|\;
			\begin{matrix}
					(\varpi^{(1)}_1,\dots,\varpi^{(1)}_K),& \dots &,	(\varpi^{(r)}_1,\dots,\varpi^{(r)}_K)\\
					\mathbin{\rotatebox[origin=c]{-90}{$\in$}}& &	\mathbin{\rotatebox[origin=c]{-90}{$\in$}}\\
					\mathcal{S}[\omega^{KK,(1)}],&\dots&,\mathcal{S}[\omega^{KK,(r)}]
			\end{matrix}
	}
	\end{equation}
	For any such tuple, compute the elliptic genera $\IE^{l_I,\va*{w}_I}_{\va*{v}_I}[\vb*{\varpi}_I]$ of $T^{l_I,\va*{w}_I}_{\va*{v}_I}$ for $I=1,\dots,K$, which can be obtained as the elliptic genera of specific frozen BPS string configurations on $\mathbb{C}^2/\cC_l$ singularities, as shown in Appendix \ref{app:TVW}.
	\item Finally, glue the elliptic genera of the tails to obtain the elliptic genus $\IE^{\Gamma,\va*{w}}_{\va*{v}}[\vb*{\omega}^{KK}]$ of $\cQ^{\Gamma,\va*{w}}_{\va*{v}}$. The gluing formula \eqref{eq:TSUdecomp} translates to:
	\begin{multline}	\label{eq:gluingEG}
		\IE^{\Gamma,\va*{w}}_{\va*{v}}
		[\vb*{\omega}^{KK}]
		(\vec{\xi},\underline{\va*{s}},\epsilon_+,\tau)
		=
		\sum_{\mathcal{S}[\omega^{KK,(1)}]}\cdots\sum_{\mathcal{S}[\omega^{KK,(r)}]}\\
		\times	\int
		Z^{glue}
		\prod_{I=1}^{K}
		\IE^{l_I,\va*{w}_I}_{\va*{v}_I}[\vb*{\varpi}_I](\pi_I(\vec{\xi})+\mathfrak{m},\underline{\vb*{x}},\underline{\va*{s}},\epsilon_+,\tau)
	\end{multline}
	where
	\begin{equation}
		\label{eq:glue_factor}
		\begin{aligned}
			Z^{glue}&=
			\frac{
				{\displaystyle{\prod_{a=1}^{r-1}}}
				\Bigg[
				\frac{1}{v_c^{(a)}!}
				\left({\displaystyle{\prod_{k=1}^{v_c^{(a)}}}} \frac{\dd{x_{k}^{(a)}}}{2\pi i}\frac{2\pi \eta^3\theta_1(2 \epsilon_+)}{\eta^2}\right)
				{\displaystyle{\prod_{\substack{ k,p=1\\ k\neq p}}^{v_c^{(a)}}}} \frac{\theta_1(x_{k}^{(a)}-x_{p}^{(a)})\theta_1(2\epsilon_++x_{k}^{(a)}-x_{p}^{(a)})}{\eta^2}\Bigg]
			}{
				\times
				{\displaystyle{\prod_{a=1}^{r-2}}}
				\Bigg[
				{\displaystyle{\prod_{k=1}^{v_c^{(a)}}\prod_{p=1}^{v_c^{(a+1)}}}}
				\frac{\theta_1(\epsilon_++x_{p}^{(a+1)}-x_{k}^{(a)})\theta_1(\epsilon_++x_{k}^{(a)}-x_{p}^{(a+1)})}{\eta^2}
				\Bigg]
			}\\
			&\times \prod_{a=1}^{r-1}
			\qty[\frac{
				\qty(\displaystyle{\prod_{k=1}^{v_c^{(a)}}\prod_{p=1}^{w_c^{(a+1)}}}  \tfrac{\theta_1(s_{c,p}^{(a+1)}-x_k^{(a)})}{\eta} )
				\qty(\displaystyle{\prod_{k=1}^{v_c^{(a)}}\prod_{p=1}^{w_c^{(a-1)}} } \tfrac{\theta_1(x_k^{(a)}-s_{c,p}^{(a-1)})}{\eta} )
			}{
				\displaystyle{\prod_{k=1}^{v_c^{(a)}}\prod_{p=1}^{w_c^{(a)}} } \tfrac{\theta_1(\epsilon_++s_{c,p}^{(a)} -x_k^{(a)})\theta_1(\epsilon_++x_k^{(a)}-s_{c,p}^{(a)})}{\eta^2}
			}]
		\end{aligned}
	\end{equation}
is the contribution of the quiver $\cQ_{\vb*{v}_c}^{glue,\va*{w}_c}$ introduced in Figure~\ref{eq:gluing_quiver2}, and the parameters $x_k^{(a)}$ are holonomies for the $U(v_c^{(a)})$ gauge fields while the parameters $s_{c,p}^{(a)}$ are holonomies for the $U(w_c^{(a)})$ central flavour nodes. The integral in Equation \eqref{eq:gluingEG} is evaluated by summing over Jeffrey-Kirwan residues.
\end{enumerate}
 
\section{The BPS string worldsheet theories: IR NLSM}
\label{sec:stringsIR}

At low energies, we expect a bound state of BPS strings labeled by the vector of integers $\vec{\boldsymbol{v}}$ to be described in terms of a nonlinear sigma model on the moduli space of  $\mathfrak{u}(W)^{r-1}$ instantons on $\mathbb{C}^2/\Gamma$ 
\be
\prod_{a=1}^{r-1}\cM^\Gamma_{\vec{w}^{(a)},\vec{v}^{(a)}},
\ee
for a given choice of monodromy data $(\vec{w}^{(1)},\dots,\vec{w}^{(r-1)})$ for the gauge fields. In Section \ref{sec:abelian_anomaly} we determine the central charges and 't Hooft anomaly coefficients for the NLSM, while a more in-depth discussion of the infrared properties of this class of theories will be presented in \cite{wip}.

\subsection{Central charges and modular properties}\label{sec:abelian_anomaly}
From the explicit expressions for the elliptic genus obtained in Section \ref{sec:EG} it is straightforward to determine the modular anomaly and IR central charges of the theories $\cQ^{\Gamma,\va*{w}}_{\va*{v}}$. We start by determining its gravitational anomaly $c_L-c_R$, which can be read off from the leading order power in the $q$-series expansion of the elliptic genus evaluated in the vacuum sector of $(\widehat{\mathfrak{g}}_1)^{r}$. The leading order power is given by
\be\label{eq:leading_order}
\frac{2c_L-2c_R-3 r \,\text{rk}\,\mathfrak{g}}{24},
\ee
and from \eqref{eq:ADE_elliptic_genus_loc} we find:
\begin{equation}
	c_L-c_R= r\, \text{rk }\mathfrak{g} -\frac{1}{2} \sum_{a,b=1}^{r-1} C_{ab}^{\mathfrak{su}(r)}	\qty(\vec{v}^{(a)}\cdot C^{\widehat{\mathfrak{g}}}\cdot \vec{v}^{(b)}).
\end{equation}

Next let us determine the right-moving central charge $c_R$, which can be computed as in \cite{Kim:2014dza} from
\be
    c_R=3 \Tr \gamma^3 R^2 ,
\ee
where $\gamma^3$ is the 2d chirality matrix and $R=2 J_I$. The 2d $(0,4)$ multiplets are summarized in Table~\ref{tab:(0,4)multiplets} together with their field content and their $SU(2)_R\times SU(2)_I$ transformation.
\begin{table}[t]
    \centering
    \begin{tabular}{cc}
       Multiplet  & Field content and $SU(2)_R\times SU(2)_I$ transformation  \\ \hline\hline
       Vector  & Vector $A_\mu$: $(\vb{1},\vb{1})$ + left fermions $\lambda^{\dot{\alpha} A}$: $(\vb{2},\vb{2})$ \\
       Fermi & Left complex fermion $\psi$: $(\vb{1},\vb{1})$\\
       Hyper & Complex scalars $x_{\dot{\alpha}}$: $(\vb{2},\vb{1})$ + right complex fermions $\chi^{A}$: $(\vb{1},\vb{2})$ \\
       Twisted hyper & Complex scalars $y_A$: $(\vb{1},\vb{2})$ + right complex fermions $\xi^{\dot{\alpha}}$: $(\vb{2},\vb{1})$
    \end{tabular}
    \caption{2d $(0,4)$ multiplets with their field content and $R$-symmetry transformation. The indices $\dot{\alpha}, A$ label respectively the components of the doublet of $SU(2)_R$ and $SU(2)_I$.}
    \label{tab:(0,4)multiplets}
\end{table}
The only contributions to  $c_R$ arise from the fermions in the vector multiplets $V^{(a)}_j$ and in the twisted hypermultiplets $X^{(a)}_{ij}$ and $W^{(a)}_j$, namely the degrees of freedom coming from the 3d $\cN=4$ Kronheimer--Nakajima quivers. Putting all together we obtain:
\begin{equation}
\label{eq:Rcentral_charge}
c_R=6\sum_{a=1}^{r-1}\pqty{\sum_{j=0}^{\text{rk }\mathfrak{g}} w_j^{(a)}v_j^{(a)}-\frac{1}{2}\sum_{i,j=0}^{\text{rk }\mathfrak{g}}{C}^{\widehat{\mathfrak{g}}}_{ij} v^{(a)}_iv^{(a)}_j}=6 \sum_{a=1}^{r-1} \dim_\IH\cM^\Gamma_{\vec{w}^{(a)},\vec{v}^{(a)}} ,
\end{equation}
where
\begin{equation}
\label{eq:mod_space_dim}
\dim_\IH\cM_{\vec{w}^{(a)},\vec{v}^{(a)}}^\Gamma=\frac{1}{2}\vec{v}^{(a)}\cdot\qty(\vec{w}^{(a)}+\vec{u}^{(a)})
\end{equation}
is the quaternionic dimension of the moduli space $\cM_{\vec{w}^{(a)},\vec{v}^{(a)}}^\Gamma$ of $(\vec{v}^{(a)},\vec{w}^{(a)})$ $U(W)$ instantons on $\IC^2/\Gamma$~\cite{KronheimerNakajima} and $\vec{u}^{(a)}$ is given in Equation \eqref{eq:fluxes_w_M2branes}.\\

Finally let us turn to the levels of the global symmetries of the BPS strings. One convenient way to compute them is by determining the modular transformation properties of its elliptic genus under an $S$-transformation. This acts on the continuous parameters of the elliptic genus as follows:
\be
(\vec{\xi},\vec{\underline{\boldsymbol{s}}},\epsilon_+,\tau)
\mapsto
\left(\frac{\vec{\xi}}{\tau},\frac{\vec{\underline{\boldsymbol{s}}}}{\tau},\frac{\epsilon_+}{\tau},-\frac{1}{\tau}\right).
\ee
The integrand of the elliptic genus is written in terms of modular objects that transform as follows:
\bea
\eta\qty(-\frac{1}{\tau})
&=&
\sqrt{-i\tau}\eta(\tau),
\label{eq:dedm}
\\
\theta_1\qty(\frac{z}{\tau},-\frac{1}{\tau})
&=&
(-i)\sqrt{-i\tau} e^{\frac{2\pi i}{\tau} \frac{1}{2}z^2},
\label{eq:thm}
\\
{\chi}^{\widehat{\mathfrak{g}}_1}_{\omega}\qty(\frac{\va{\xi}}{\tau},-\frac{1}{\tau})
&=&
e^{\frac{1}{2}\frac{2\pi i}{\tau}\va{\xi}\cdot \qty(C^{\mathfrak{g}})^{-1}\cdot\va{\xi}} \sum_{\upsilon} \mathcal{S}_{\omega,\upsilon} {\chi}^{\widehat{\mathfrak{g}}_1}_{\upsilon}\qty(\va{\xi},\tau),
\eea
where $\mathcal{S}_{\omega,\upsilon}$ is the modular S-matrix of $\widehat{\mathfrak{g}}_1$. In particular the modular transformation of the $\widehat{\mathfrak{g}}_1$ characters imply that the elliptic genus transforms as a vector-valued Jacobi form, whose components correspond to the different possible choices of superselection sectors.
Specifically, we find that the elliptic genus transforms as
\begin{multline}
	\mathbb{E}^{\Gamma,\va*{w}}_{\va*{v}}
	[\vb*{\omega}^{KK}]
	\left(\frac{\vec{\xi}}{\tau},\frac{\vec{\underline{\boldsymbol{s}}}}{\tau},\frac{\epsilon_+}{\tau},-\frac{1}{\tau}\right)
	=
	e^{\frac{2\pi i}{\tau}
	f_{\va*{v}}^{\Gamma,\va*{w}}(\va{\xi},\vec{\underline{\boldsymbol{s}}},\epsilon_+)}
	\times
	\\
	\sum_{\vb*{\upsilon}^{KK}=(\upsilon^{KK,(1)},\dots,\upsilon^{KK,(r)})} \qty(\prod_{a=1}^{r} \mathcal{S}_{\omega^{KK,(a)}, \upsilon^{KK,(a)}} )
	\mathbb{E}^{\Gamma,\va*{w}}_{\va*{v}}
	[\vb*{\upsilon}^{KK}]
	(\vec{\xi},\vec{\underline{\boldsymbol{s}}},\epsilon_+,\tau),
	\label{eq:Str}
\end{multline}
where the quadratic polynomial $f_{\va*{v}}^{\Gamma,\va*{w}}(\va{\xi},\vec{\underline{\boldsymbol{s}}},\epsilon_+)$ encodes the 't Hooft anomalies for the global symmetries of $\cQ^{\Gamma,\va*{w}}_{\va*{v}}$ \cite{Benini:2013cda}. Its computation is detailed in Appendix~\ref{app:anomaly-canc}, where we find that
\begin{equation}
    f_{\va*{v}}^{\Gamma,\va*{w}}(\va{\xi},\vec{\underline{\boldsymbol{s}}},\epsilon_+)
    =
    \frac{1}{2}\sum_{a=1}^{r}\vec{\xi}_{\mathfrak{m},\vec{\underline{\boldsymbol{s}}}}^{(a)}\cdot (C^{\mathfrak{g}})^{-1}\cdot \vec{\xi}^{(a)}_{\mathfrak{m},\vec{\underline{\boldsymbol{s}}}}
    +
    k_R \epsilon_+^2
    +
    \sum_{a=0}^r \sum_{j=0}^{\text{rk }\mathfrak{g}} \frac{k_{\mathfrak{u}(w_j^{(a)})}}{2}\sum_{K=1}^{w_j^{(a)}} \qty(s_{j,K}^{(a)})^2,
    \label{eq:ff}
\end{equation}
with
\be
(\vec{\xi}^{(a)}_{\mathfrak{m},\vec{\underline{\boldsymbol{s}}}})_j
=
\xi_j+\mathfrak{m}-S^{(a)}_j+S^{(a-1)}_j.
\ee
From Equation \eqref{eq:ff} one can readily read off the anomaly coefficients  $k_R$ and $k_{\mathfrak{u}(w_j^{(a)})}$ for $SU(2)_R$ and $\mathfrak{u}(w_j^{(a)})$:
\begin{align}
\label{eq:Rlevel}    k_R&=-\frac{c_R}{6}+\frac{1}{2}\sum_{j=0}^{\text{rk }\mathfrak{g}} \vb*{v}_j\cdot C^{\su(r)}\cdot \vb*{v}_j,\\
    k_{\mathfrak{u}(w_j^{(a)})}&=\begin{cases}
    -\qty(C^{\su(r)}\cdot \vb*{v}_j)^{(a)} & \text{ for } a=1,\dots,r-1\\
    v_{j}^{(1)} & \text{ for } a=0\\
    v_{j}^{(r-1)} & \text{ for } a=r\\
    \end{cases}.
    \label{eq:levels}
\end{align}
One can in fact completely decouple the dependence on the chemical potentials $\vec{\xi}$ from the 6d gauge symmetry parameters $\vec{S}^{(1)},\dots,\vec{S}^{(r-1)}$: indeed, one can write
\bea
\nonumber
\frac{1}{2}\sum_{a=1}^{r}\vec{\xi}_{\mathfrak{m},\vec{\underline{\boldsymbol{s}}}}^{(a)}\cdot (C^{\mathfrak{g}})^{-1}\cdot \vec{\xi}^{(a)}_{\mathfrak{m},\vec{\underline{\boldsymbol{s}}}}
&=&
\frac{r}{2}
\left(\vec{\xi}+\mathfrak{m}+\frac{\vec{S}^{(0)}-\vec{S}^{(r)}}{r}\right)
\cdot
(C^{\mathfrak{g}})^{-1}
\cdot
\left(\vec{\xi}+\mathfrak{m}+\frac{\vec{S}^{(0)}-\vec{S}^{(r)}}{r}\right)
\\
&+&
\xi-\text{independent terms}
,
\label{eq:shifted}
\eea
where a redefinition of $\vec{\xi}$ can be used to absorb the dependence on the global symmetry parameters on the right-hand side of the first line. Equation \eqref{eq:shifted} suggests a possible decomposition of the elliptic genera in terms of the affine subalgebra of $\widehat{\mathfrak{g}}_r$ which is compatible with the choice of monodromies for the 6d flavor symmetry groups $\mathfrak{f}^{6d}$. This possibility will be explored further in \cite{wip} and in the examples in the next section.

\section{Examples}
\label{sec:examples}

In this section we study a variety of examples of BPS string configurations for theories $\cT^{6d}_{r,W}$ on different orbifold singularities and determine their elliptic genera. In Section~\ref{sec:frozen} we consider a number of frozen BPS string configurations for 6d theories with rank $r=2$, beginning with the case of a frozen BPS string on $\mathbb{C}^2/\cC_2$, which is closely related to the theory $T_{(1)}^2$; after this we move on to examples on $\cQ_4$ and $\cT$ singularities. In Section \ref{sec:full} we turn to examples with integer string charge, beginning with a general discussion of charge 1 configurations and a more detailed discussion of the $\cQ_4$ orbifold case. We conclude this section by looking at a higher rank configuration on a $\cQ_4$ singularity, namely a bound state of two BPS strings for the theory $\cT^{6d}_{3,1}$. Additional examples and a more detailed discussion of the infrared physics of the corresponding strings will be presented in \cite{wip}.

\subsection{Frozen BPS strings for $r = 2$}
\label{sec:frozen}

%%%%%%%%%%%%%%%%%%
\subsection{The $T_{(1)}^2$ theory and the frozen BPS string on $\mathbb{C}^2/\cC_2$}
\label{sec:T12}

\begin{figure}[h]
	\centering
	\scalebox{1}{\begin{tikzpicture}[line width=0.4]
			\node at (-4.75,3) {$T_{(1)}^2$};
			\draw[blue] (-5.6,2) -- (-5.6,-1);     
			\draw[blue] (-3.9,2) -- (-3.9,-1); 
			\node[blue] at (-3.9,2.4) {$\widehat{\su}(2)_1$};
			\node[blue] at (-5.6,2.4) {$\widehat{\su}(2)_1$};
			
			\node[smallgauge]   (g)  at (-4.75,1) {$1$};
			\node[smallflavor]   (f)  at (-4.75,0) {$2$};
			\node at (-4.75,-0.7) {$x_1^{(1)},x_2^{(1)}$};
			\draw (g)-- (f);
			%%%%%%%%%%%%%%%%%%%%%%%%%%%%
			\node at (-3,0.5) {$\hookrightarrow$};
			%%%%%%%%%%%%%%%%%%%%%%%%%%%%
			\node at (0,3.5) {$\cQ^{\cC_2,\va*{w}}_{\va*{v}}$};
			\draw[blue] (1.1,2.5) -- (1.1,-1.5);     
			\draw[blue] (-1.1,2.5) -- (-1.1,-1.5); 
			\node[blue] at (1.1,2.9) {$\widehat{\su}(2)_1$};
			\node[blue] at (-1.1,2.9) {$\widehat{\su}(2)_1$};
			
			\node[emptygauge]   (g10)  at (0,1.5) {};
			\node[smallgauge]   (g11)  at (0,0.0) {$1$};
			\node[smallflavor] (f11) at (0,-1) {$2$};
			\node at (0,-1.7) {$s^{(1)}_{1,1},s^{(1)}_{1,2}$};
			\node[smallflavor,green!70!black]   (f00)  at (-1.7,1.5) {$2$};
			\node[green!70!black] at (-1.9,2.1) {$s^{(0)}_{0,1},s^{(0)}_{0,2}$};
			\node[emptyflavor]   (f01)  at (-1.7,0) {};
			\node[smallflavor,green!70!black]   (f20)  at (1.7,1.5) {$2$};
			\node[green!70!black] at (1.9,2.1) {$s^{(2)}_{0,1},s^{(2)}_{0,2}$};
			\node[emptyflavor]   (f21)  at (1.7,0) {};
			\draw (g11)-- (f11);
			\draw[dotted] (g10) to[out=-110,in=110] (g11);
			\draw[dotted] (g10) to[out=-70,in=70] (g11);
	\end{tikzpicture}}
	\caption{The $\cC_2$-dressed quiver of $T^2_{(1)}$, shown on the left, and the $\cC_2$-dressed quiver of theory $\cQ^{\cC_2,\va*{w}}_{\va*{v}}$ into which it embeds.}
	\label{quiver:1-T_2^1}
\end{figure}
Here we consider the simplest non trivial example of a $T_{\vb*{v}}^l$ theory, namely $T_{(1)}^2$, whose $\cC_2$-dressed quiver is depicted on the left side of Figure~\ref{quiver:1-T_2^1}. We will later use it as a building block in the following examples to construct more complicated theories by gluing. The elliptic genus of $T_{(1)}^2$ is given by the integral
\bea
&&\IE^{2}_{(1)}[\boldsymbol{\varpi}]
(\widehat{\xi}, \underline{\vb*{x}}, \epsilon_+,\tau)
=
\nn
\\
&&
-i
\int
dz\eta^5\theta_1(2\epsilon_+)
\frac{
\chi^{\widehat{\su}(2)_1}_{\varpi^{(1)}}(\widehat\xi+2z-x^{(1)}_1-x^{(1)}_2)
\chi^{\widehat{\su}(2)_1}_{\varpi^{(2)}}(\widehat\xi-2z+x^{(1)}_1+x^{(1)}_2)}
{\prod_{k=1}^2\theta_1(\epsilon_++z-x^{(1)}_k) \theta_1(\epsilon_+-z+x^{(1)}_k)}
\eea
which picks up residues at $z=\epsilon_+ +x^{(1)}_k$ for $k=1,2$ and evaluates to
\bea
\nonumber
	\IE^2_{(1)}
	[\boldsymbol{\varpi}]
	(\widehat{\xi}, \underline{\vb*{x}}, \epsilon_+,\tau)
	&=&
	\frac{\eta^2}{\theta_1(x^{(1)})}
	\Bigg(
	\frac{\chi^{\widehat{\su}(2)_1}_{\varpi^{(1)}}(\widehat{\xi} +2 \epsilon_+-x^{(1)} ) \chi^{\widehat{\su}(2)_1}_{\varpi^{(2)}}(\widehat{\xi} -2 \epsilon_++x^{(1)} )}{\theta_1(2 \epsilon_+ -x^{(1)})}
	\\
	&&
	-\frac{\chi^{\widehat{\su}(2)_1}_{\varpi^{(1)}}(\widehat{\xi} +2 \epsilon_+ +x^{(1)} ) \chi^{\widehat{\su}(2)_1}_{\varpi^{(2)}}( \widehat{\xi}-2 \epsilon_+-x^{(1)} )}{\theta_1(2 \epsilon_+ +x^{(1)})}
	\Bigg),
\eea
where $\varpi^{(1)},\varpi^{(2)}$ can be taken to be either of the two i.h.w.r.'s of $\widehat{\su}(2)_1$, and the two holonomies $x_1^{(1)}, x_2^{(1)}$ for the flavor node $U(2)^{(1)}$ enter only through the combination $x^{(1)}=x_2^{(1)}-x_1^{(1)}$, as one expects from the decoupling of the abelian factor of the flavor symmetry $U(2)^{(1)}$. \\

The theory $T_{(1)}^2$ can be embedded into the worldsheet theory $\cQ^{\cC_2,\va*{w}}_{\va*{v}}$ of a frozen BPS string configuration of $\cT^{6d}_{2,2}$ on $T^2\times \IC^2/\cC_2$ with
\begin{equation}
	\va*{w}=\qty{\mqty(2\\0),\mqty(0\\2),\mqty(2\\0)}\quad, \quad \va*{v}=\qty{\mqty(0\\1)},
\end{equation}
as depicted in Figure~\ref{quiver:1-T_2^1}, providing a concrete example of the embedding discussed in Appendix~\ref{app:TVW}. 
This configuration was already considered in~\cite{DelZotto:2023ryf} where its elliptic genus was also computed. The moduli space has quaternionic dimension $1$ and the level for the $SU(2)_R$ symmetry is $k_R = 0$.\\

The embedding $T_{(1)}^2\hookrightarrow\cQ^{\cC_2,\va*{w}}_{\va*{v}}$ gives the following map between fugacities:
\begin{align*}
	\xi&=\widehat{\xi}+\mathfrak{m},\\
	s_{1,k}^{(1)} &= x_k^{(1)}\quad\text{with } k=1,2.
\end{align*}
The extra fugacities $s^{(a)}_{0,1}$ and $s^{(a)}_{0,2}$, with $a=0,2$, do not appear explicitly in the elliptic genus of $\cQ^{\cC_2,\va*{w}}_{\va*{v}}$; they enter implicitly through $\mathfrak{m}$ via the constraint \eqref{eq:Stuckelberg_constr}. We can therefore write the elliptic genus of $\cQ^{\cC_2,\va*{w}}_{\va*{v}}$ as
\begin{equation}
	\IE^{\cC_2,\va*{w}}_{\va*{v}}[\vb*{\omega}^{KK}](\xi,  \underline{\va*{s}}, \epsilon_+,\tau) =	\IE_{(1)}^2[\vb*{\omega}^{KK}](\widehat{\xi}+\mathfrak{m},\underline{\vb*{x}},\epsilon_+,\tau).
\end{equation}

%%%%%%%%%%%%%%%%%%
\subsubsection{Frozen BPS strings on $\mathbb{C}^2/\cQ_4$}
\paragraph{Instanton charge $4$.} We next consider the simple configuration of BPS strings on $\IC^2/\cQ_{4}$ of instanton charge $4$ displayed in Figure \ref{quiver:4TSU2}, which corresponds to Configuration 4 of  of Table~\ref{fig:list_allowed_M-string_orb} and arises in the 6d theory $\cT^{6d}_{2,8}$.
\begin{figure}[h]
	\centering
	\scalebox{0.8}{\begin{tikzpicture}[line width=0.4]
			\node[emptyflavor](f00) at (3,1) {};
			\node[emptyflavor] (f01) at (5,1) {};
			\node[smallflavor,green!70!black] (f02) at (4,0) {$4$};
			\node[green!70!black] at (4.7,0) {$\underline{s}_{2}^{(a)}$};
			\node[emptyflavor] (f03) at (3,-1) {};
			\node[emptyflavor] (f04) at (5,-1) {};
			\draw[blue] (2,2.5) -- (2,-2.5);     
			\node[blue] at (2,2.9) {$\widehat{\so}(8)_1$};
			\node at (4,2.9) {$a=0,2$};
			\node at (0,2.9) {$a=1$};
			\node[smallflavor] (f10) at (-1,2) {$2$};
			\node at (-1.6,2) {$\underline{s}^{(1)}_{0}$};
			\node[smallflavor](f11) at (1,2) {$2$};
			\node at (0.4,2) {$\underline{s}^{(1)}_{1}$};
			\node[emptyflavor](f12) at (-1,0) {};
			\node[smallflavor] (f13) at (-1,-2) {$2$};
			\node at (-1.6,-2) {$\underline{s}^{(1)}_{4}$};
			\node[smallflavor] (f14) at (1,-2) {$2$};
			\node at (0.4,-2) {$\underline{s}^{(1)}_{3}$};
			\node[smallgauge]   (g0)  at (-1,1) {$1$};
			\node[smallgauge]   (g1)  at (1,1) {$1$};
			\node[emptygauge]   (g2)  at (0,0) {};
			\node[smallgauge]   (g3)  at (-1,-1) {$1$};
			\node[smallgauge]   (g4)  at (1,-1) {$1$};
			\draw[dotted] (g0) -- (g2);
			\draw[dotted]  (g1) -- (g2);
			\draw[dotted]  (g3) -- (g2);
			\draw[dotted]  (g4) -- (g2);
			\draw (g0) -- (f10);
			\draw (g1) -- (f11);
			\draw[dotted]  (g2) -- (f12);
			\draw (g3) -- (f13);
			\draw (g4) -- (f14);
	\end{tikzpicture}}
	\caption{$\cQ_4$-dressed quiver $\cQ^{\cQ_4,\va*{w}}_{\va*{v}}$ for a frozen BPS string of charge $4$ on a $\cQ_4$ singularity.}
	\label{quiver:4TSU2}
\end{figure}
The configuration corresponds to the following data:
\begin{equation}\label{eq:4TSU2_data}
	\va*{w}=\qty{\mqty(0\\0\\4\\0\\0),\mqty(2\\2\\0\\2\\2),\mqty(0\\0\\4\\0\\0)}\quad, \quad \va*{v}=\qty{\mqty(1\\1\\0\\1\\1)},
\end{equation} 
from which we read off $c_R = 24$ corresponding to a moduli space of quaternionic dimension four. Note that the $\mathfrak{su}(4)^{\oplus 2}$ component of the 6d flavor symmetry $\mathfrak{f}^{6d}$ is decoupled from the strings.
This configuration can be obtained by a trivial gluing of four copies of $T_{(1)}^2$ with parameters
\begin{align}
\label{eq:mapping}
	\widehat{\xi}_I&=	\pi_I(\vec{\xi})+\mathfrak{m}\\
	(x^{(1)}_{k})_I &=
		s^{(1)}_{I,k}
\end{align}
where $I\in\{0,1,3,4\}$, $k=1,2$, and
\be
\label{eq:pi}
\pi_{I}(\vec\xi)
=
\begin{cases}
-\xi_1-2\xi_2-\xi_3-\xi_4 & \text{for }I={0}
\\\xi_I & \text{for }I={1,3,4}
\end{cases}
\ee
are the chemical potentials for the $T^{2}_{(1)}$ quivers which are determined by the embedding
\begin{equation}\label{eq:4su2INTOso8}
	\su(2)_0\oplus\su(2)_1\oplus\su(2)_3\oplus\su(2)_4\subset\so(8).
\end{equation}
Here again the extra fugacities $\underline{s}_2^{(a)}$, $a=0,2$, of $\cQ^{\cQ_4,\va*{w}}_{\va*{v}}$ only appear implicitly through $\mathfrak{m}$ due to Equation \eqref{eq:Stuckelberg_constr}. Using the notational conventions explained in Section \ref{sec:gluingEG}, the elliptic genus of $\cQ^{\cQ_4,\va*{w}}_{\va*{v}}$ is given simply by:
\begin{equation}\label{eq:EG_4TSU(2)}
	\IE^{\cQ_4,\va*{w}}_{\va*{v}}[\vb*{\omega}^{KK}] (\vec{\xi},\underline{\va*{s}},\epsilon_+,\tau)
	=
	\sum_{\mathcal{S}[\omega^{KK,(1)}]}
	\sum_{\mathcal{S}[\omega^{KK,(2)}]}
	\prod_{I=0,1,3,4} \IE_{(1)}^2[\vb*{\varpi}_I]\qty(\widehat{\xi}_I, \underline{{\vb*{x}}}_I, \epsilon_+,\tau ),
\end{equation}
where the sets $\mathcal{S}[\omega]$ are given by:
\bea
\label{eq:cS1}
\mathcal{S}[\boldsymbol{1}]
&=&
\{(\boldsymbol{1},\boldsymbol{1},\boldsymbol{1},\boldsymbol{1})
,
(\boldsymbol{2},\boldsymbol{2},\boldsymbol{2},\boldsymbol{2})\},
\\
\mathcal{S}[\boldsymbol{8}^v]
&=&
\{(\boldsymbol{1},\boldsymbol{1},\boldsymbol{2},\boldsymbol{2})
,
(\boldsymbol{2},\boldsymbol{2},\boldsymbol{1},\boldsymbol{1})\},
\\
\mathcal{S}[\boldsymbol{8}^s]
&=&
\{(\boldsymbol{1},\boldsymbol{2},\boldsymbol{1},\boldsymbol{2})
,
(\boldsymbol{2},\boldsymbol{1},\boldsymbol{2},\boldsymbol{1})\},
\\
\label{eq:cS4}
\mathcal{S}[\boldsymbol{8}^c]
&=&
\{(\boldsymbol{1},\boldsymbol{2},\boldsymbol{2},\boldsymbol{1})
,
(\boldsymbol{2},\boldsymbol{1},\boldsymbol{1},\boldsymbol{2})\}.
\eea
In other words, this configuration coincides with four frozen BPS strings on a $\mathbb{C}^2/\cC_2$ singularity corresponding to the external nodes of the $D_4$ affine Dynkin diagram, which do not interact with each other except for their common coupling to the $\widehat{\mathfrak{so}}(8)$ current algebras and $\mathfrak{u}(1)^{diag}$ flavor symmetry. In particular, a specific superselection sector for the BPS strings on $\cQ_4$ corresponds to a sum over sectors for the constituent $\mathbb{C}^2/\cC_2$ frozen strings. Equation \eqref{eq:EG_4TSU(2)} is a special instance of the gluing formula \eqref{eq:gluingEG} in the particular case where the rank of the central node is zero.

\paragraph{Instanton charge $\tfrac{5}{2}$.} 
Let us now turn to the example given in Configuration 3a of Table~\ref{fig:list_allowed_M-string_orb}, which corresponds to the following data: 
\begin{equation}\label{eq:4TSU2_data}
	\va*{w}=\qty{\mqty(0\\0\\2\\0\\0),\mqty(1\\1\\0\\1\\1),\mqty(0\\0\\2\\0\\0)}\quad, \quad \va*{v}=\qty{\mqty(1\\1\\1\\1\\1)}
\end{equation}
and arises for the 6d theory $\cT^{6d}_{2,4}$. The $\cQ_4$-dressed quiver $\cQ^{\cQ_4,\va*{w}}_{\va*{v}}$ is shown in Figure~\ref{quiver:D4unos}.
The quaternionic dimension of the instanton moduli space in this case is 3, corresponding to $c_R = 18$; the level $k_R$ is 2.
\begin{figure}[h]
	\centering
	\scalebox{0.8}{\begin{tikzpicture}[line width=0.4]
			\def\sh{0.7};
			\node[blue] at (2,2.9) {$\widehat{\so}(8)_1$};
			\node at (4,2.9) {$a=0,2$};
			\node at (0,2.9) {$a=1$};
			\node[emptyflavor](f00) at (3,1) {};
			\node[emptyflavor] (f01) at (5,1) {};
			\node[smallflavor] (f02) at (4,0) {$2$};
			\node at (4+\sh,0) {$\underline{s}^{(a)}_{2}$} ;
			\node[emptyflavor] (f03) at (3,-1) {};
			\node[emptyflavor] (f04) at (5,-1) {};
			\draw[blue] (2,2.5) -- (2,-2.5);     
			\node[smallflavor] (f10) at (-1,2) {$1$};
			\node at (-1-\sh,2) {$s^{(1)}_{0,1}$};
			\node[smallflavor](f11) at (1,2) {$1$};
			\node at (1-\sh,2) {$s^{(1)}_{1,1}$};
			\node[emptyflavor](f12) at (-1,0) {};
			\node[smallflavor] (f13) at (-1,-2) {$1$};
			\node at (-1-\sh,-2) {$s^{(1)}_{4,1}$};
			\node[smallflavor] (f14) at (1,-2) {$1$};
			\node at (1-\sh,-2) {$s^{(1)}_{3,1}$};
			\node[smallgauge]   (g0)  at (-1,1) {$1$};
			\node[smallgauge]   (g1)  at (1,1) {$1$};
			\node[smallgauge]   (g2)  at (0,0) {$1$};
			\node[smallgauge]   (g3)  at (-1,-1) {$1$};
			\node[smallgauge]   (g4)  at (1,-1) {$1$};
			\draw (g0) -- (g2);
			\draw (g1) -- (g2);
			\draw (g3) -- (g2);
			\draw (g4) -- (g2);
			\draw (g0) -- (f10);
			\draw (g1) -- (f11);
			\draw[dotted]  (g2) -- (f12);
			\draw (g3) -- (f13);
			\draw (g4) -- (f14);
			\draw[dashed,double,double distance = 1.5pt] (g2) -- (f02);
	\end{tikzpicture}}
 	\caption{$\cQ_4$-dressed quiver $\cQ^{\cQ_4,\va*{w}}_{\va*{v}}$ for a frozen BPS string of charge $\tfrac{5}{2}$ on a $\cQ_4$ singularity.}
 	\label{quiver:D4unos}
\end{figure}

This configuration can again be obtained by gluing four copies of  $T_{(1)}^2$ associated to the four exterior nodes of the $D_4$ affine Dynkin diagram by coupling a $U(1)$ factor of each $U(2)$ flavor symmetry to the following gluing quiver:
\begin{equation*}
	\begin{tikzpicture}[line width=0.4]
		\node (q) at (0,0) 	{$\cQ^{glue}:$};
		\node[smallflavor] (f0) at (2,0) {$2$};
		\node at (2,-0.7) {$\underline{s}_2^{(0)}$};
		\node[smallgauge] (g) at (3,0) {$1$};
		\node at (3,-0.5) {$z$};
		\node[smallflavor] (f2) at (4,0) {$2$};
		\node at (4,-0.7) {$\underline{s}_2^{(2)}$};
		\draw[dashed] (f0)--(g);
		\draw[dashed] (f2)--(g);
	\end{tikzpicture}
\end{equation*}
The four copies of $T_{(1)}^2 $  embed into the quiver $\cQ^{\cQ_4,\va*{w}}_{\va*{v}}$  of Figure~\ref{quiver:D4unos}, analogously to the previous example, with the following identification of parameters:
\begin{align}
	\widehat{\xi}_I&=	\pi_I(\vec{\xi})+\mathfrak{m}\\
	(x^{(1)}_{1})_I &=	s^{(1)}_{I,1}
\end{align}
with $\pi_I(\vec{\xi})$ given by Equation \ref{eq:pi}. The remaining fugacities $(x^{(1)}_{2})_I$, $I=0,1,3,4$, are identified with the gauge fugacity $z$ of the $U(1)$ node of the gluing quiver $\cQ^{glue}$:
\begin{equation}
	(x^{(1)}_{2})_I = z.
\end{equation}

\noindent The elliptic genus is given by:
\begin{equation}\label{eq:EG_D4unos}
	\IE^{\cQ_4,\va*{w}}_{\va*{v}}[\vb*{\omega}^{KK}] (\vec{\xi},\underline{\va*{s}},\epsilon_+,\tau)
	=
	\sum_{\mathcal{S}[\omega^{KK,(1)}]}
	\sum_{\mathcal{S}[\omega^{KK,(2)}]}
	\int Z^{glue} 
	\prod_{I\in\{0,1,3,4\}}
	\IE_{(1)}^2[\vb*{\varpi}_I]
	\qty(\widehat{\xi}_I, \underline{\vb*{x}}_I, \epsilon_+,\tau ),
\end{equation}
where the $\mathcal{S}[\omega]$ are given in Equations \eqref{eq:cS1}--\eqref{eq:cS4} and
\begin{equation}
	Z^{glue}
	= 
	\frac{dz}{i \eta^3} 
	\theta_1(2\epsilon_+) 
	\prod_{k=1}^2       \theta_1(s^{(2)}_{2,k}-z)       \theta_1(z-s^{(0)}_{2,k}).
\end{equation}
The integral picks up residues from simple poles at $z = 2\epsilon_++s^{(1)}_{j,1}$ with $j=0,1,3,4$. The following equations show the leading behavior of the elliptic genus in the various sectors, where for simplicity we set to zero all flavor fugacities $\underline{\vec{\boldsymbol{s}}}$ (and also $\mathfrak{m}$ as a consequence):
\begin{align*}
	\IE_{\va*{v}}^{\cQ_4,\va*{w}}[(\vb{1},\vb{1})]&=
q^{-1/2}\varphi_{\boldsymbol{1}}
	\qty(   
	1  +  2\cdot\boldsymbol{28}q)+	 q^{\frac{1}{2}}\qty(\varphi_{\boldsymbol{2}^{\oplus 4}} \boldsymbol{2}^{\oplus 4}
	+ \varphi_{\boldsymbol{28}}\boldsymbol{28})
	+
	\order{q^{3/2}},
	\\
	%%%%%%%%%%%%
	% VEC-VEC
	%%%%%%%%%%%%%
	\IE_{\va*{v}}^{\cQ_4,\va*{w}} 	[(\vb{8}^v,\vb{8}^v)] & = 
	q^{\frac{1}{2}}\qty(
	\varphi_{\boldsymbol{35}^v}\boldsymbol{1}
	+
	\varphi_{\boldsymbol{2}^{\oplus 4}}  \boldsymbol{2}^{\oplus 4}
	+
	\varphi_{\boldsymbol{28}}  \boldsymbol{28}
	+\varphi_{\boldsymbol{1}} 	\boldsymbol{8}^v\otimes\boldsymbol{8}^v)
	+
	\order{q^{3/2}},
	\\
	%%%%%%%
	% VAC-VEC
	%%%%%%%
	\IE_{\va*{v}}^{\cQ_4,\va*{w}} 	[(\vb{1},\vb{8}^v)] & = 
\varphi_{\boldsymbol{8}^v} \boldsymbol{8}^v
	+
	\order{q},
	\\
	%%%%%%%%%%%%
	% SP-CONJ
	%%%%%%%%%%%%%
	\IE_{\va*{v}}^{\cQ_4,\va*{w}} 	[(\vb{8}^s,\vb{8}^c)] & = 
	q^{\frac{1}{2}}[
	\varphi_{\boldsymbol{8}^v} \boldsymbol{56}^v+\varphi_{\boldsymbol{56}^v}\boldsymbol{8}^v]
	+
	\order{q^{3/2}}.
\end{align*}
The expressions for the elliptic genus in the remaining sectors can be obtained from the above by exploiting the triality permuting the labels $(v,s,c)$, which is unbroken by the choice of monodromy data in Equation \eqref{eq:4TSU2_data}.
Notice that we can almost completely express the elliptic genera in terms of $\mathfrak{so}(8)$ representations $\boldsymbol{R}$, except for the occurrence of a representation $\boldsymbol{2}^{\oplus 4}$ of $\su(2)^{\oplus 4}$ which reflects the breaking of the $\mathfrak{so}(8)$ symmetry due to the choice of monodromy $\vec{w}^{(0)}$ for the flavor symmetry  $\mathfrak{f}^{6d}$. We observe that the same decomposition holds when the flavor symmetry parameters $\underline{s}^{(a)}_2$ are kept generic, whereas upon turning on generic gauge chemical potentials the $\mathfrak{so}(8)$ representations break up into their $\mathfrak{su}(2)^{\oplus 4} $ constituents.

The $t$-dependent coefficients $\varphi_{\boldsymbol{R}}(t,\tau)$ encode the spacetime dependence of the elliptic genus and are given by:
\bea
	\varphi_{\boldsymbol{1}}
	&=&
	-2\frac{t^3(3-4t^2+3t^4)}{(1-t^2)^2(1+t^2)^3}
	+
	2\frac{t(3-22t^2+30t^4-22t^6+3t^8)}{(1-t^2)^2(1+t^2)^3}q
	+\order{q^2},\;\;
	\\
	\varphi_{\boldsymbol{28}}
	&=&
	\frac{6t}{1+t^2}
	+\order{q},
	\\
	\varphi_{\boldsymbol{35}^v}
	&=&
	2\frac{1+4t^2-t^4+4t^6-t^8+4t^{10}+t^{12}}{t^5(1+t^2)}
	+\order{q},
	\\
	\varphi_{\boldsymbol{8}^v}
	&=&
	4\frac{t^5}{(1-t^2)^2(1+t^2)^3}
	+\order{q},
	\\
	\varphi_{\boldsymbol{56}^v}
	&=&
	2\frac{1+2t^2+2t^6+t^8}{t^3(1+t^2)}
	+\order{q},
	\\
	\varphi_{\boldsymbol{2}^{\oplus 4}}
	&=&
	2\frac{(1-t^2)^2(1+t^2+t^4)}{t(1+t^2)^3}
	+\order{q}.
\eea
%
%%%%%%%%%%%%%%%%%%
\subsubsection{Frozen BPS strings on $\mathbb{C}^2/\cT$: instanton charge $\tfrac{55}{24}$}
Let us now turn to an example  for a singularity of type $\cT$ corresponding to the affine Dynkin diagram of type $E_6$. 
\begin{figure}[t]
	\centering
	\scalebox{0.8}{\begin{tikzpicture}[line width=0.4]
			\def\sh{0.7};
			\node[blue] at (2,2.9) {$\widehat{e}_{6,1}$};
			\node at (4,2.9) {$a=0,2$};
			\node at (0,2.9) {$a=1$};
			\node[emptyflavor](f00) at (3,2) {};
			\node[emptyflavor](f001) at (3,1) {};
			\node[emptyflavor](f002) at (4,0) {};
			\node[emptyflavor](f003) at (3,-1) {};
			\node[emptyflavor](f004) at (3,-2) {};
			\node at (2.8+\sh,-0.55) {${s}^{(a)}_{3,1}$} ;			
			\node[smallflavor] (f02) at (3,0) {$1$};
			\node[smallflavor,green!70!black] (f08) at (5,0) {$1$};
			\node at (5+\sh,0) {${s}^{(a)}_{0,1}$} ;
			\draw[blue] (2,2.5) -- (2,-2.5);     
			\node[smallflavor] (f10) at (0,2) {$1$};
			\node at (1.3-\sh,2) {$s^{(1)}_{1,1}$};
			\node[smallflavor](f11) at (1,0) {$1$};
			\node[smallflavor] (f13) at (0,-2) {$1$};
			\node at (2.3-\sh,0) {$s^{(1)}_{6,1}$};
			\node at (1.3-\sh,-2) {$s^{(1)}_{5,1}$};
			\node[smallgauge]   (g0)  at (-1,2) {$1$};
			\node[smallgauge]   (g1)  at (-1,1) {$1$};
			\node[smallgauge]   (g2)  at (-1,0) {$1$};
			\node[smallgauge]   (g3)  at (-1,-1) {$1$};
			\node[smallgauge]   (g4)  at (-1,-2) {$1$};
			\node[smallgauge]   (g5)  at (0,0) {$1$};
			\draw (g0) -- (g1);
			\draw (g0) -- (f10);
			\draw (g1) -- (g2);
			\draw (g2) -- (g3);
			\draw (g3) -- (g4);
			\draw (g4) -- (f13);
			\draw (g2) -- (g5);
			\draw (g5) -- (f11);
			\draw[dashed,double,double distance = 1.5pt] (g2) to[out=-30,in=-150] (f02);
	\end{tikzpicture}}
	\caption{$\cT$-dressed quiver $\cQ^{\cT,\va*{w}}_{\va*{v}}$ for a frozen BPS string of charge $\tfrac{55}{24}$ on a $\cT$ singularity.}
	\label{fig:E6quiver}
\end{figure}
We consider the frozen BPS string of Figure \ref{fig:E6quiver}, which corresponds to the following data:
\begin{equation}\label{eq:4TSU2_data}
	\va*{w}
	=
	\qty{
	\mqty(1\\0\\0\\1\\0\\0\\0),
	\mqty(0\\1\\0\\0\\0\\1\\1),
	\mqty(1\\0\\0\\1\\0\\0\\0)}
	,
	\quad
	\va*{v}
	=
	\qty{\mqty(0\\1\\1\\1\\1\\1\\1)}.
\end{equation}
and arises in the theory $\cT^{6d}_{r,4}$. The quaternionic dimension of the moduli space of instantons is 2, corresponding to $c_R=12$, and the $SU(2)_R$ level is $k_R=4$. From the choice of monodromies for $\mathfrak{f}^{6d}$ we expect $\mathfrak{g} = \mathfrak{e}_6$  to break to an $\su(3)_a\oplus\su(3)_b\oplus\su(2)\oplus\mathfrak{u}(1)$ maximal subalgebra. The elliptic genus is given by
\bea
\nn
\IE^{\cT,\va*{w}}_{\va*{v}}[\vb*{\omega}^{KK}] (\vec{\xi},\underline{\va*{s}},\epsilon_+,\tau)
=
-
\int
&&
\left(\prod_{i=1}^6 dz_i\right)
\frac{\eta^{20}\theta_1(2\epsilon_+)^6\theta_1(z_3-s^{(0)}_{3,1})\theta_1(s^{(2)}_{3,1}-z_3)}
{\prod_{l\in\{1,5,6\}}\theta_1(\epsilon_++z_l-s^{(1)}_{l,1})\theta_1(\epsilon_+-z_l+s^{(1)}_{l,1})}
\\
&&
\times
\frac{\prod_{a=1}^2\chi^{\widehat{e}_{6,1}}_{\omega^{KK,(a)}}(\vec{\xi}^{(a)})}
{\prod_{\substack{i,j=1\\i\neq j}}^{6}\theta_1(\epsilon_+ +(z_i-z_j))^{-C^{\mathfrak{e}_6}_{ij}}}
,
\eea
where the parameters $\xi^{(a)}$ are given by Equation \eqref{eq:xi-term}; explicitly:
\bea
	\xi^{(1)}_{j}
	&=&
	\xi_j+\sum_{k=0}^{\text{rk }\mathfrak{e}_6}C_{jk}^{\mathfrak{e}_6}z_k^{(a)} -S^{(1)}_j+S^{(0)}_j +\mathfrak{m},
	\\
	\xi^{(2)}_{j}
	&=&
	\xi_j-\sum_{k=0}^{\text{rk }\mathfrak{e}_6}C_{jk}^{\mathfrak{e}_6}z_k^{(a)} -S^{(2)}_j+S^{(1)}_j +\mathfrak{m},
\eea
and
\bea
\vec{S}^{(1)} &=& (0,s^{(1)}_{1,1},0,0,0,s^{(1)}_{5,1},s^{(1)}_{6,1})\\
\vec{S}^{(0)} &=& (s^{(0)}_{0,1},0,0,s^{(0)}_{3,1},0,0,0)\\
\vec{S}^{(2)} &=& (s^{(2)}_{0,1},0,0,s^{(2)}_{3,1},0,0,0).
\eea
The integral can be expressed as a sum over 21 residues evaluated at simple poles. Alternatively the computation can be performed by gluing the elliptic genera for two copies of the theory $T^{3,\vec{\boldsymbol{\mathrm{w}}}}_{\vec{\boldsymbol{\mathrm{v}}}}$ with
\bea
\vec{\boldsymbol{\mathrm{w}}}
=
\left\{
\begin{pmatrix}
	0\\
	0
\end{pmatrix}
,
\begin{pmatrix}
	1\\
	1
\end{pmatrix}
,
\begin{pmatrix}
	0\\
	0
\end{pmatrix}
\right\},
\qquad
\vec{\boldsymbol{\mathrm{v}}}=
\left\{
\begin{pmatrix}
	1\\
	1
\end{pmatrix}
\right\}.
\eea
and one copy of the theory $T^{2}_{(1)}$,
and by gauging a common $U(1)$ flavor symmetry.
As in the previous example, one obtains relatively simple expressions for the leading order terms of the elliptic genus in the unrefined limit $\vec{\underline{\boldsymbol{s}}}=0$ for the following choices of $\mathfrak{e}_6$ representations:
\bea
\IE_{\va*{v}}^{\cT,\va*{w}}[(\vb{1},\vb{1})]
&=&
-
q^{-\frac{2}{3}}\frac{2t^2(1+t+t^2)(1+t^8)}{(1-t+t^2)(1+t-t^5-t^6)^2}
({\boldsymbol{1}},{\boldsymbol{1}},\boldsymbol{1})_0
+\order{q^\frac{1}{3}}
\\
\IE_{\va*{v}}^{\cT,\va*{w}}[(\vb{1},\vb{27})]
&=&
-\frac{2t^4(1+t^2+2t^3+t^4+t^6)}{(1-t+t^2)(1+t-t^5-t^6)^2}
(\overline{\boldsymbol{3}},\overline{\boldsymbol{3}},\boldsymbol{1})_0
+
\nn
\\
&&
\frac{t^3(1+t+t^2)(-2+t-2t^3+t^5-2t^6)}{(1-t+t^2)(1+t-t^5-t^6)^2}
(({\boldsymbol{3}},{\boldsymbol{1}},\boldsymbol{2})_{-1}
+
(\boldsymbol{1},{\boldsymbol{3}},\boldsymbol{2})_1)
\nn
\\
&&
-
\frac{t^4(1+t+t^2)}{(1-t^5)^2}
(({\boldsymbol{3}},{\boldsymbol{1}},\boldsymbol{1})_{2}
+
(\boldsymbol{1},{\boldsymbol{3}},\boldsymbol{1})_{-2})
+
\order{q}.
\eea
Here, we have denoted the characters of the unbroken $\su(3)\oplus\su(3)\oplus\su(2)\oplus\mathfrak{u}(1)$ as $(\boldsymbol{R}_1,\boldsymbol{R}_2,\boldsymbol{R}_3)_{q}$, where $\boldsymbol{R}_{1,2,3}$ denote irreducible representations of the three non-abelian factors, while $q$ denotes the $\mathfrak{u}(1)$ charge. Analogous expressions for $\boldsymbol{\omega}^{KK}=(\vb*{1},\boldsymbol{\overline{27}})$ can be obtained by complex conjugation. The elliptic genus for the remaining inequivalent choices $\boldsymbol{\omega}^{KK}=(\boldsymbol{27},\boldsymbol{27})$ and $(\boldsymbol{27},\boldsymbol{\overline{27}})$ can also be expressed in a similar way although the final expressions contain a much larger number of terms.

\subsection{Integer string charge configurations}
\label{sec:full}

In this section we focus on  BPS string configurations of integer charge. We first consider the rank $r=2$ M-string SCFT $\cT^{6d}_{2,1}$ on a generic $\mathbb{C}^2/\Gamma$ orbifold in Section \ref{sec:onestring}, and provide detailed results for the elliptic genus of one instanton string in the case $\Gamma = \cQ_4$. Additional cases of instanton charge 1 for $\Gamma=\cT$, $\cO$, and $\cI$ will be considered in fuller detail in \cite{wip}. In Section \ref{sec:hr} we look at a bound state of two BPS strings for the theory $\cT^{6d}_{3,1}$ on $\mathbb{C}^2/\cQ_4$.

\subsubsection{One M-string on $\mathbb{C}^2/\Gamma$}
\label{sec:onestring}
The configurations we consider in this section correspond to the first nontrivial BPS string contribution to the partition function  \eqref{eq:PFM} of the rank $r=2$ M-string SCFT $\mathcal{T}^{6d}_{2,1}$ for arbitrary $\Gamma$.
\begin{figure}[h]
	\centering
     \scalebox{1}{\begin{tikzpicture}[line width=0.4]
        \node[smallflavor](f00) at (3,1) {$1$};
       \node[emptyflavor] (f01) at (5,1) {};
        \node[emptyflavor] (f02) at (4,0) {};
        \node[emptyflavor] (f03) at (3,-1) {};
        \node[emptyflavor] (f04) at (5,-1) {};
        \draw[blue] (2,2.5) -- (2,-2.5);     
        \node[blue] at (2,2.9) {$\widehat{\so}(8)_1$};
        \node at (0,2.9) {$a=1$};
        \node at (4,2.9) {$a=0,2$};

        \node[smallflavor] (f10) at (-1,2) {$1$};
        \node[emptyflavor](f11) at (1,2) {};
        \node[emptyflavor](f12) at (-1,0) {};
        \node[emptyflavor] (f13) at (-1,-2) {};
        \node[emptyflavor] (f14) at (1,-2) {};
        \node[smallgauge]   (g0)  at (-1,1) {$1$};
        \node[smallgauge]   (g1)  at (1,1) {$1$};
        \node[smallgauge]   (g2)  at (0,0) {$2$};
        \node[smallgauge]   (g3)  at (-1,-1) {$1$};
        \node[smallgauge]   (g4)  at (1,-1) {$1$};
        \draw (g0) -- (g2);
        \draw (g1) -- (g2);
        \draw (g3) -- (g2);
        \draw (g4) -- (g2);
        \draw (g0) -- (f10);
        \draw[dotted] (g1) -- (f11);
        \draw[dotted] (g2) -- (f12);
        \draw[dotted] (g3) -- (f13);
        \draw[dotted] (g4) -- (f14);
        \draw[dashed,double,double distance = 1.5pt] (g0)to[out=23,in=158](f00);
	\end{tikzpicture}}
    \caption{$\cQ_4$-dressed quiver $\cQ^{\cQ_4}_{(1)}$ of a single BPS string on $\IC^2/\cQ_4$. The conventions used to depict the quiver are explained in Section \ref{sec:exs}.}
	\label{quiver:1-Mstring_on_D4}
\end{figure}
The superselection sector parameters that enter the definition of the $\Gamma$-dressed quiver for this set of examples are:
\bea
\vec{\boldsymbol{w}}
=
\left\{
\begin{pmatrix}
1\\
0\\
\vdots\\
0
\end{pmatrix}
,
\begin{pmatrix}
1\\
0\\
\vdots\\
0
\end{pmatrix}
,
\begin{pmatrix}
1\\
0\\
\vdots\\
0
\end{pmatrix}
\right\},
\qquad
\vec{\boldsymbol{v}}=
\left\{
\begin{pmatrix}
a_0\\
a_1\\
\vdots\\
a_{\text{rk }\mathfrak{g}}
\end{pmatrix}
\right\}
\label{eq:wv1}
\eea
as well a pair of level-1 integrable highest weight representations $(\omega^{KK,(1)},\omega^{KK,(2)})$ of $\widehat{\mathfrak{g}}_1$. In Figure \ref{quiver:1-Mstring_on_D4} we show the 2d quiver that corresponds to the case $\Gamma = \mathcal{Q}_4$. Based on this data one finds that the moduli space has quaternionic dimension 1, so $c_R = 6$, and moreover the level with respect to $SU(2)_R$ is given according to Equation \eqref{eq:Rlevel} by 
\be
k_R = \vert\Gamma\vert-1.
\ee
In the expression for the elliptic genus \eqref{eq:ADE_elliptic_genus_loc} we can factor out the dependence of the elliptic genus on the holonomy $z_0\equiv z_{0,1}^{(1)}$ of the $U(1)$ gauge group associated to the affine node of the quiver as follows:
\begin{multline}
\label{eq:1stringEG_bare}
\mathbb{E}^{\Gamma,\vec{\boldsymbol{w}}}_{\vec{\boldsymbol{v}}}[\vb*{\omega}^{KK}]
=
\qty(\prod_{j=1}^{\text{rk }\mathfrak{g}}\frac{1}{a_j!})
\int
\qty(\prod_{j=1}^{\text{rk }\mathfrak{g}}
Z_{V_j^{(1)}}
\prod_{i=0}^{j-1}
Z_{X_{i,j}^{(1)}}^{-C^{\widehat{\mathfrak{g}}}_{ij}}){\chi}^{\widehat{\mathfrak{g}}_1}_{\omega^{KK,(1)}}\qty(\va{\xi}^{(1)},\tau){\chi}^{\widehat{\mathfrak{g}}_1}_{\omega^{KK,(2)}}\qty(\va{\xi}^{(2)},\tau)\\
    \times
   \int\frac{dz_0}{2\pi i} \frac{2\pi \eta^3 \theta_1(2\epsilon_+)}{\eta^2}\frac{\theta_1(s^{(2)}-z_{0})\theta_1(z_{0}-s_0^{(0)})}{\theta_1(\epsilon_++s_0^{(1)} -z_{0})\theta_1(\epsilon_+ +z_{0}-s_0^{(1)})},
\end{multline}
where in the first row
\be
{\xi}^{(a)}_j= \xi_j+\mu-(-1)^{a}(C^{\widehat{\mathfrak{g}}}\cdot \vec{Z}^{(1)})_j,
\ee
and the second row contains the contributions of the multiplets $V^{(1)}_0$, $W^{(1)}$, $\Sigma^{(1)}$ and $\Theta^{(1)}$. This is accomplished by performing the following shift on the gauge holonomies in Equation \eqref{eq:ADE_elliptic_genus_loc}:
\begin{equation}
    z^{(1)}_{j,k}\to z^{(1)}_{j,k}+(1-\delta_{j,0})z_0.
\end{equation}
This eliminates the $z_0$ dependence from the first row of Equation \eqref{eq:1stringEG_bare}. Performing the integral over $z_0$ we obtain:
\begin{equation}
\label{eq:factor}
\mathbb{E}^{\Gamma,\vec{\boldsymbol{w}}}_{\vec{\boldsymbol{v}}}[\vb*{\omega}^{KK}]
(\vec{\xi},\mu,\epsilon_+,\tau)
=
-
\frac{\theta_1(h^\vee_{\mathfrak{g}}\mu+\epsilon_+) \theta_1(h^\vee_{\mathfrak{g}}\mu-\epsilon_+ )}{ \eta^2}
\widetilde{\mathbb{E}}^{\Gamma,\vec{\boldsymbol{w}}}_{\vec{\boldsymbol{v}}}[\vb*{\omega}^{KK}]
(\vec{\xi},\epsilon_+,\tau),
\end{equation}
where we have used the constraint \eqref{eq:Stuckelberg_constr} to eliminate the $s_0^{(a)}$. This expression consists of the contribution of decoupled Fermi multiplets, times the elliptic genus of a frozen BPS string corresponding to the following configuration:
\bea
\vec{\boldsymbol{w}}
=
\left\{
\begin{pmatrix}
2\\
0\\
\vdots\\
0
\end{pmatrix}
,
\begin{pmatrix}
0\\
\vdots\\
1\\
\vdots\\
0
\end{pmatrix}
,
\begin{pmatrix}
2\\
0\\
\vdots\\
0
\end{pmatrix}
\right\},
\qquad
\vec{\boldsymbol{v}}=
\left\{
\begin{pmatrix}
0\\
a_1\\
\vdots\\
a_{\text{rk }\mathfrak{g}}
\end{pmatrix}
\right\}
,
\eea
where the nonzero entry of $\vec{w}^{(1)}$ corresponds to the adjoint node.
This quiver is shown in Figure \ref{fig:redq} for the case $\Gamma = \mathcal{Q}_4$. This reflects the well-known fact that for any $\Gamma$ there exists a fractional instanton of charge
\begin{equation}
    N^{(a)}=\frac{|\Gamma|-1}{|\Gamma|} ,\text{ for }a=1,\dots,r-1,
\end{equation}
whose moduli space is isomorphic to the moduli space of one instanton on $\mathbb{C}^2/\Gamma$.\\

\begin{figure}[t]
        \centering
         \scalebox{1}{\begin{tikzpicture}[line width=0.4]
        \node[blue] at (2,2.9) {$\widehat{\so}(8)_1$};
        \node at (0,2.9) {$a=1$};
        \node at (4,2.9) {$a=0,2$};
        \node[smallflavor,green!70!black](f00) at (3,1) {$2$};
       \node[emptyflavor] (f01) at (5,1) {};
        \node[emptyflavor] (f02) at (4,0) {};
        \node[emptyflavor] (f03) at (3,-1) {};
        \node[emptyflavor] (f04) at (5,-1) {};
        \draw[blue] (2,2.5) -- (2,-2.5);     
        \node[emptyflavor] (f10) at (-1,2) {};
        \node[emptyflavor](f11) at (1,2) {};
        \node[smallflavor](f12) at (-1,0) {$1$};
        \node[emptyflavor] (f13) at (-1,-2) {};
        \node[emptyflavor] (f14) at (1,-2) {};
        \node[emptygauge]   (g0)  at (-1,1) {};
        \node[smallgauge]   (g1)  at (1,1) {$1$};
        \node[smallgauge]   (g2)  at (0,0) {$2$};
        \node[smallgauge]   (g3)  at (-1,-1) {$1$};
        \node[smallgauge]   (g4)  at (1,-1) {$1$};
        \draw[dotted] (g0) -- (g2);
        \draw (g1) -- (g2);
        \draw (g3) -- (g2);
        \draw (g4) -- (g2);
        \draw[dotted] (g0) -- (f10);
        \draw[dotted] (g1) -- (f11);
        \draw (g2) -- (f12);
        \draw[dotted] (g3) -- (f13);
        \draw[dotted] (g4) -- (f14);
	\end{tikzpicture}}
    \caption{$\cQ_4$-dressed quiver for a frozen BPS string of instanton charge $\frac{7}{8}$ on $\mathbb{C}^2/\cQ_4$.}
	\label{fig:redq}
\end{figure}

Let us from now on specialize to the case $\Gamma=\mathcal{Q}_4$, corresponding to the quiver depicted in Figure~\ref{quiver:1-Mstring_on_D4}, and turn to explicit computations. A convenient way to determine the elliptic genus, in light of the discussion above, is by multiplying the contribution of a decoupled free Fermi multiplet as in Equation \eqref{eq:factor} to the elliptic genus of the theory of Figure \ref{fig:redq}, which is given by
\bea
\widetilde{\mathbb{E}}^{\cQ_4,\va*{w}}_{\va*{v}}[\boldsymbol{\omega}^{KK}]
&&
=
\sum_{\mathcal{S}[\omega^{KK,(1)}]}
\sum_{\mathcal{S}[\omega^{KK,(2)}]}
\int \frac{d z_1 d z_2}{2!}\, \eta^2
\\
&&\times \frac{\theta_1(2\epsilon_+)^2\theta_1(z_1-z_2)^2\theta_1(2\epsilon_++z_1-z_2)\theta_1(2\epsilon_++z_2-z_1)}{\prod_{k=1}^2\theta_1(\epsilon_+ +z_k)\theta_1(\epsilon_+ -z_k)}
\nonumber
\\
&&
	\times
	\prod_{I\in\{1,3,4\}}
	\IE_{(1)}^2[\vb*{\varpi}_I]
	\qty(\widehat{\xi}_I, \underline{\vb*{z}}, \epsilon_+,\tau ),
\eea
where the sets $\cS[\omega]$ are given in Equations \eqref{eq:cS1}--\eqref{eq:cS4}, and the parameters $\widehat{\xi}_I$ are given in Equation \eqref{eq:mapping}. Performing this integral requires evaluating residues at non-simple poles, which can be expressed in terms of quasi-Jacobi forms \cite{kaneko1995generalized,libgober2011elliptic}. Nevertheless, the elliptic genus obtained by summing over the residues is expected to be an ordinary (meromorphic) Jacobi form, and we have indeed been able to find a closed form for it in terms of Jacobi theta functions. The resulting expressions are however very unwieldy, and here we limit ourselves to presenting an expression in the unrefined limit where the $\mathfrak{so}(8)$ chemical potentials $\vec{\xi}$ are switched off:
\begin{equation}
\mathbb{E}^{\cQ_4,\vec{\boldsymbol{w}},unr}_{\vec{\boldsymbol{v}}}[\vb*{\omega}^{KK}]
(\mu,\epsilon_+,\tau)
=
-
\frac{\theta_1(6\mu+\epsilon_+,\tau) \theta_1(6\mu-\epsilon_+,\tau)}{ \eta(\tau)^2}
\widetilde{\mathbb{E}}^{\cQ_4,\vec{\boldsymbol{w}},unr}_{\vec{\boldsymbol{v}}}[\vb*{\omega}^{KK}]
(\epsilon_+,\tau),
\end{equation}
where
\begin{multline*}
\widetilde{\mathbb{E}}^{\cQ_4,\vec{\boldsymbol{w}},unr}_{\vec{\boldsymbol{v}}}[(\boldsymbol{1},\boldsymbol{1})]
=
\frac{1}{ 4\eta^{6}\theta_1(2\epsilon_+)}
\Bigg[
\frac{\qty(\theta_3^4+\theta_4^4)\qty(\theta_3^2\theta_3(2\epsilon_+)^2+\theta_4^2\theta_4(2\epsilon_+)^2)\theta_1(4\epsilon_+)}{\theta_1(2\epsilon_+)^2}\\-3\frac{\qty(\theta_3(2\epsilon_+)^4+\theta_4(2\epsilon_+)^4)^2}{\theta_1(4\epsilon_+)}\Bigg],
\end{multline*}
\begin{multline*}
\widetilde{\mathbb{E}}^{\cQ_4,\vec{\boldsymbol{w}},unr}_{\vec{\boldsymbol{v}}}[(\boldsymbol{1},\boldsymbol{8}^a)]
=
\frac{1}{ 4\eta^{6}\theta_1(2\epsilon_+)}
\Bigg[\frac{ \qty[\theta_3^6 \theta_3(2 \epsilon_+ )^2-\theta_4^6 \theta_4(2 \epsilon_+ )^2]\theta_1(4\epsilon_+)}{\theta_1(2 \epsilon_+ )^2 }\\-\frac{2 \theta_3(2 \epsilon_+ )^8-2 \theta_4(2 \epsilon_+ )^8+\theta_3^6 \theta_3(4 \epsilon_+ )^2-\theta_4^6 \theta_4(4 \epsilon_+ )^2}{ \theta_1(4 \epsilon_+)}\Bigg] ,
\end{multline*}
\begin{multline*}
\widetilde{\mathbb{E}}^{\cQ_4,\vec{\boldsymbol{w}},unr}_{\vec{\boldsymbol{v}}}[(\boldsymbol{8}^a,\boldsymbol{8}^b)]
=
\frac{1}{ 4\eta^{6}\theta_1(2\epsilon_+)}
\Bigg[-\frac{\theta_2^6 \theta_2(2 \epsilon_+ )^2 \theta_1(4 \epsilon_+ )}{\theta_1(2 \epsilon_+  )^2}\\+\frac{ (-1)^{1-\delta_{a,b}}2 \theta_1(2 \epsilon_+  )^8+2 \theta_2(2 \epsilon_+  )^8+\theta_2^6 \theta_2(4 \epsilon_+  )^2}{\theta_1(4 \epsilon_+ )}\Bigg] ,
\end{multline*}
for $a,b\in\{v,s,c\}$. For this class of examples the monodromy of the 6d flavor symmetry group is trivial and we expect the elliptic genus to be expressible in terms of characters of $\widehat{\mathfrak{g}}_r$. This turns out to be the case and leads to  elegant expressions for the elliptic genus which neatly encode the infrared physics of the BPS strings. These results, as well as analogous results for orbifold singularities of the exceptional type $\cT$, $\cO$, $\cI$, will appear in a separate publication \cite{wip}.

\subsubsection{A higher rank example}
\label{sec:hr}

As our last example, let us consider a BPS string configuration for the rank $r=3$ 6d M-string SCFT $\cT^{6d}_{3,1}$ on the background $T^2\times\mathbb{C}^2/\cQ_4$. We will focus on the case of a bound state of two BPS strings arising from a pair of M2 branes stretched between neighboring M5 branes, corresponding to $\boldsymbol{\kappa}=(1,1)$ or, equivalently, to the following choice of data:
\bea
\vec{\boldsymbol{w}}
=
\left\{
\begin{pmatrix}
	1\\
	0\\
	0\\
	0\\
	0
\end{pmatrix}
,
\begin{pmatrix}
	1\\
	0\\
	0\\
	0\\
	0
\end{pmatrix}
,
\begin{pmatrix}
	1\\
	0\\
	0\\
	0\\
	0
\end{pmatrix}
,
\begin{pmatrix}
	1\\
	0\\
	0\\
	0\\
	0
\end{pmatrix}
\right\},
\qquad
\vec{\boldsymbol{v}}=
\left\{
\begin{pmatrix}
	1\\
	1\\
	2\\
	1\\
	1
\end{pmatrix},
\begin{pmatrix}
	1\\
	1\\
	2\\
	1\\
	1
\end{pmatrix}
\right\}.
\label{eq:wv2}
\eea
The $\cQ_4$-dressed quiver for this configuration is shown in Figure~\ref{quiver:r3onQ4}. The model has $SU(2)_R$ level $k_R = 6$.
\begin{figure}[h]
	\centering
	\centering
	\scalebox{1}{\begin{tikzpicture}[line width=0.4]
			\def\sh{4};
			%%%%%%%%%%%%%%%%%
			%a=0
			%%%%%%%%%%%%%%
			\node at (-4,2.9) {$a=0$};
			\node[smallflavor](f00) at (-5,1) {$1$};
			\node at (-5.6,1) {$s^{(0)}$};
			\node[emptyflavor] (f01) at (-3,1) {};
			\node[emptyflavor] (f02) at (-4,0) {};
			\node[emptyflavor] (f03) at (-3,-1) {};
			\node[emptyflavor] (f04) at (-5,-1) {};
			%%%%%%%%%%%%%
			% a=1
			%%%%%%%%%%%%
			\node at (0,2.9) {$a=1$};
			\draw[blue] (-2,2.5) -- (-2,-2.5);     
			\node[blue] at (-2,2.9) {$\widehat{\so}(8)_1$};
			\node[smallflavor] (f10) at (-1,2) {$1$};
			\node at (-1.6,2) {$s^{(1)}$};
			\node[emptyflavor](f11) at (1,2) {};
			\node[emptyflavor](f12) at (-1,0) {};
			\node[emptyflavor] (f13) at (-1,-2) {};
			\node[emptyflavor] (f14) at (1,-2) {};
			\node[smallgauge]   (g10)  at (-1,1) {$1$};
			\node[smallgauge]   (g11)  at (1,1) {$1$};
			\node[smallgauge]   (g12)  at (0,0) {$2$};
			\node[smallgauge]   (g13)  at (-1,-1) {$1$};
			\node[smallgauge]   (g14)  at (1,-1) {$1$};
			\draw (g10) -- (g12);
			\draw (g11) -- (g12);
			\draw (g13) -- (g12);
			\draw (g14) -- (g12);
			\draw (g10) -- (f10);
			\draw[dotted] (g11) -- (f11);
			\draw[dotted] (g12) -- (f12);
			\draw[dotted] (g13) -- (f13);
			\draw[dotted] (g14) -- (f14);
			%%%%%%%%%%%%%
			% a=2
			%%%%%%%%%%%%
			\node at (4,2.9) {$a=2$};
			\draw[blue] (2,2.5) -- (2,-2.5);     
			\node[blue] at (2,2.9) {$\widehat{\so}(8)_1$};
			\node[smallflavor] (f20) at (-1+\sh,2) {$1$};
			\node at (-0.4+\sh,2) {$s^{(2)}$};
			\node[emptyflavor](f21) at (1+\sh,2) {};
			\node[emptyflavor](f22) at (-1+\sh,0) {};
			\node[emptyflavor] (f23) at (-1+\sh,-2) {};
			\node[emptyflavor] (f24) at (1+\sh,-2) {};
			\node[smallgauge]   (g20)  at (-1+\sh,1) {$1$};
			\node[smallgauge]   (g21)  at (1+\sh,1) {$1$};
			\node[smallgauge]   (g22)  at (0+\sh,0) {$2$};
			\node[smallgauge]   (g23)  at (-1+\sh,-1) {$1$};
			\node[smallgauge]   (g24)  at (1+\sh,-1) {$1$};
			\draw (g20) -- (g22);
			\draw (g21) -- (g22);
			\draw (g23) -- (g22);
			\draw (g24) -- (g22);
			\draw (g20) -- (f20);
			\draw[dotted] (g21) -- (f21);
			\draw[dotted] (g22) -- (f22);
			\draw[dotted] (g23) -- (f23);
			\draw[dotted] (g24) -- (f24);
			%%%%%%%%%%%%%
			% a=3
			%%%%%%%%%%%%
			\node at (8,2.9) {$a=3$};
			\draw[blue] (6,2.5) -- (6,-2.5);     
			\node[blue] at (6,2.9) {$\widehat{\so}(8)_1$};
			\node[smallflavor](f30) at (-5+3*\sh,1) {$1$};
			\node at (-0.4+2*\sh,1) {$s^{(3)}$};
			\node[emptyflavor] (f31) at (-3+3*\sh,1) {};
			\node[emptyflavor] (f32) at (-4+3*\sh,0) {};
			\node[emptyflavor] (f33) at (-3+3*\sh,-1) {};
			\node[emptyflavor] (f34) at (-5+3*\sh,-1) {};
			%%%%%%%%%%%%%
			% intra quivers nodes
			%%%%%%%%%%%%%
			%  Fermi gauge-flavor
			\draw[dashed] (g10)to[out=150,in=30](f00);
			\draw[dashed] (g10)--(f20);
			\draw[dashed] (g20)--(f10);
			\draw[dashed] (f30)to[out=150,in=30](g20);
			%  Purple hypers
			\draw[purple] (g10)to[out=-20,in=-160](g20);
			\draw[purple] (g11)to[out=-20,in=-160](g21);
			\draw[purple] (g12)--(g22);
			\draw[purple] (g13)to[out=-20,in=-160](g23);
			\draw[purple] (g14)to[out=-20,in=-160](g24);
			% Green Fermi
			\draw[green!70!black, dashed] (g10)--(g22);
			\draw[green!70!black, dashed] (g11)--(g22);
			\draw[green!70!black, dashed] (g13)--(g22);
			\draw[green!70!black, dashed] (g14)--(g22);
			\draw[green!70!black, dashed] (g20)--(g12);
			\draw[green!70!black, dashed] (g21)--(g12);
			\draw[green!70!black, dashed] (g23)--(g12);
			\draw[green!70!black, dashed] (g24)--(g12);
	\end{tikzpicture}}
	\caption{$\cQ_4$-dressed quiver for the $\vb*{\kappa}=(1,1)$ BPS string on $\IC^2/\cQ_4$.}
	\label{quiver:r3onQ4}
\end{figure}

In this example it is convenient to construct the theory by combining three copies of theory $(T_{(1,1)}^2)_I$, $I=1,3,4$ and one of theory $(\widetilde{T}^2_{(1,1)})_0$. The $\cC_2$-dressed quivers for the two type of building blocks are displayed respectively in Figures \ref{quiver:T112} and \ref{quiver:T112tilde}. They are glued by coupling to the following degrees of freedom
\begin{equation*}
	\begin{tikzpicture}[line width=0.4]
		\node (q) at (-0.5,0) 	{$\cQ^{glue}:$};
		\node[smallgauge] (g1) at (1,0) {$2$};
		\node at (1,-0.6) {$\underline{z}_c^{(1)}$};
		\node[smallgauge] (g2) at (2,0) {$2$};
		\node at (2,-0.6) {$\underline{z}_c^{(2)}$};
		\draw [purple] (g1)--(g2);
	\end{tikzpicture}
\end{equation*}
with the following identification of chemical potentials:
\be
(\underline{x}^{(a)})_I = \underline{z}_c^{(a)}
\ee
and
\be
\widehat{\xi}_I = \pi_I(\vec{\xi}) + \mu
\ee
with $\pi_I$ as in Equation \ref{eq:pi}.
%
%%%%%%%%%%%%%%%%%%
%% T_{11}^2
%%%%%%%%%%%%%%%%
\begin{figure}[t]
	\centering
	\scalebox{1}{\begin{tikzpicture}[line width=0.4]
			\def\sh{2.5};
			\def\ytop{3.5}
			\def\ybot{1}
			\def\YSH{1}
			%%%%%%%%%%%%%%%%%
			%a=0
			%%%%%%%%%%%%%%
			%%%%%%%%%%%%%
			% a=1
			%%%%%%%%%%%%
			\draw[blue] (\sh/2,\ytop-\YSH) -- (\sh/2,\ybot-\YSH);     
			\node[blue] at (\sh/2,\ytop+0.4-\YSH) {$\widehat{\su}(2)_1$};
			\node[smallgauge]   (g11)  at (\sh,\ytop-1.5) {$1$};
			\node at (\sh-0.6,\ytop-1.5 ) {$z^{(1)}$};
			\node[smallflavor]   (f11)  at (\sh,\ytop-2.5) {$2$};
			\node at (\sh,\ytop-3.1) {$\underline{x}^{(1)}$};
			\draw (g11)--(f11);
			%%%%%%%%%%%%%
			% a=2
			%%%%%%%%%%%%
			\draw[blue] (3*\sh/2,\ytop-\YSH) -- (3*\sh/2,\ybot-\YSH);     
			\node[blue] at (3*\sh/2,\ytop+0.4-\YSH) {$\widehat{\su}(2)_1$};
			\node[smallgauge]   (g21)  at (2*\sh,\ytop-1.5) {$1$};
			\node at (2*\sh+0.6,\ytop-1.5 ) {$z^{(2)}$};
			\node[smallflavor]   (f21)  at (2*\sh,\ytop-2.5) {$2$};
			\node at (2*\sh,\ytop-3.1) {$\underline{x}^{(2)}$};
			\draw (g21)--(f21);
			%%%%%%%%%%%%%%%%%
			%a=3
			%%%%%%%%%%%%%%
			\node[blue] at (5*\sh/2,\ytop+0.4-\YSH) {$\widehat{\su}(2)_1$};
			\draw[blue] (5*\sh/2,\ytop-\YSH) -- (5*\sh/2,\ybot-\YSH);     
			%%%%%%%%%
			% intra quiver lines
			%%%%%%%%%%
			\draw[dashed] (g11)--(f21);
			\draw[dashed] (g21)--(f11);
			\draw[purple] (g11)--(g21) ;
	\end{tikzpicture}}
	\caption{$\cC_2$-dressed quiver for $T_{(1,1)}^2$.}
	\label{quiver:T112}
\end{figure}
%
%
%%%%%%%%%%%%%%%%%%
%% T_{11}^2 TILDE
%%%%%%%%%%%%%%%%
\begin{figure}[t]
	\centering
	\centering
	\scalebox{1}{\begin{tikzpicture}[line width=0.4]
			\def\sh{2.5};
			\def\ytop{3.5}
			\def\ybot{0}
			%%%%%%%%%%%%%%%%%
			%a=0
			%%%%%%%%%%%%%%
			\node[smallflavor](f00) at (0,\ytop-0.5) {$1$};
			\node at (-0.6,\ytop-0.5) {$s^{(0)}$};
			%%%%%%%%%%%%%
			% a=1
			%%%%%%%%%%%%
			\draw[blue] (\sh/2,\ytop) -- (\sh/2,\ybot);     
			\node[blue] at (\sh/2,\ytop+0.4) {$\widehat{\su}(2)_1$};
			\node[smallgauge]   (g11)  at (\sh,\ytop-1.5) {$1$};
			\node at (\sh-0.6,\ytop-1.5-0.2 ) {$z^{(1)}$};
			\node[smallflavor]   (f11)  at (\sh,\ytop-2.5) {$2$};
			\node at (\sh,\ytop-3.1) {$\underline{x}^{(1)}$};
			\node[smallflavor](f10) at (\sh,\ytop-0.5) {$1$};
			\node at (\sh-0.6,\ytop-0.5) {$s^{(1)}$};
			\draw (g11)--(f11);
			\draw  (g11)--(f10);
			%%%%%%%%%%%%%
			% a=2
			%%%%%%%%%%%%
			\draw[blue] (3*\sh/2,\ytop) -- (3*\sh/2,\ybot);     
			\node[blue] at (3*\sh/2,\ytop+0.4) {$\widehat{\su}(2)_1$};
			\node[smallgauge]   (g21)  at (2*\sh,\ytop-1.5) {$1$};
			\node at (2*\sh+0.6,\ytop-1.5-0.1 ) {$z^{(2)}$};
			\node[smallflavor]   (f21)  at (2*\sh,\ytop-2.5) {$2$};
			\node at (2*\sh,\ytop-3.1) {$\underline{x}^{(2)}$};
			\node[smallflavor](f20) at (2*\sh,\ytop-0.5) {$1$};
			\node at (2*\sh+0.6,\ytop-0.5) {$s^{(2)}$};
			\draw (g21)--(f21);
			\draw (g21)--(f20);
			%%%%%%%%%%%%%%%%%
			%a=3
			%%%%%%%%%%%%%%
			\node[blue] at (5*\sh/2,\ytop+0.4) {$\widehat{\su}(2)_1$};
			\draw[blue] (5*\sh/2,\ytop) -- (5*\sh/2,\ybot);     
			\node[smallflavor](f30) at (3*\sh,\ytop-0.5) {$1$};
			\node at (3*\sh+0.6,\ytop-0.5) {$s^{(3)}$};
			%%%%%%%%%
			% intra quiver lines
			%%%%%%%%%%
			\draw[dashed] (g11)--(f21);
			\draw[dashed] (g21)--(f11);
			\draw[dashed] (g11)--(f00);
			\draw[dashed] (g21)--(f10);
			\draw[dashed] (g21)--(f30);
			\draw[dashed] (g11)--(f20);
			\draw[purple] (g11)--(g21) ;
	\end{tikzpicture}}
	\caption{$\cC_2$-dressed quiver for $\widetilde{T}^2_{(1,1)}=T^{2,(1,1+2,1+2,1)}_{(1,1)}$.}
	\label{quiver:T112tilde}
\end{figure}
The elliptic genus is given by:
\begin{multline}
	\IE_{\va*{v}}^{\cQ_4,\va*{w}}[\vb*{\omega}^{KK}](\vec{\xi},\mu,\epsilon_+,\tau)=\sum_{\cS[\omega^{(1)}]} \sum_{\cS[\omega^{(2)}]}  \sum_{\cS[\omega^{(3}]} 
	\int Z^{glue}
	\prod_{I={1,3,4}}\IE_{(1,1)}^{2}[\vb*{\varpi}_I]\qty(\widehat{{\xi}}_I,\underline{\vb*{z}}_c,\epsilon_+,\tau)
	\\
	\times \widetilde{\IE}_{(1,1)}^{2}[\vb*{\varpi}_0]\qty(\widehat{{\xi}}_0,\underline{\vb*{z}}_c,\boldsymbol{s},\epsilon_+,\tau)
\end{multline}
where $\mathcal{S}[\omega]$ are as in Equations \eqref{eq:cS1}--\eqref{eq:cS4}, the central node quiver factor is given by
\begin{equation}
	Z^{glue}= \frac{\eta^4 \theta_1(2\epsilon_+)^4}{(2!)^2}  	\frac{\prod_{a=1}^2 dz^{(a)}_{c,1}dz^{(a)}_{c,2} \theta_1(z^{(a)}_{c,1}-z^{(a)}_{c,2})^2 \theta_1(2\epsilon+z^{(a)}_{c,1}-z^{(a)}_{c,2})\theta_1(2\epsilon+z^{(a)}_{c,2}-z^{(a)}_{c,1})  }{\prod_{k,l=1}^2 \theta_1(\epsilon_+ + z^{(1)}_{c,k}-z^{(2)}_{c,l}) \theta_1(\epsilon_+ + z^{(2)}_{c,l}-z^{(1)}_{c,k})}, 
\end{equation}
the elliptic genus for theory $T^{2}_{(1,1)}$ is given by
\bea
&&
\IE_{(1,1)}^{2}[\vb*{\varpi}]\qty(\widehat{\xi},\underline{\vb*{x}},\epsilon_+,\tau)
=
\nn
\\
&&
\int
\eta^4
dz^{(1)}
dz^{(2)}
\frac{
\theta_1(2\epsilon_+)^2
\prod_{a=1}^2
\prod_{k=1}^2
\theta_1(z^{(a)}-x^{(3-a)}_k)
}
{\prod_{s=\pm}
\theta_1(\epsilon_+-s(z^{(1)}-z^{(2)}))
\prod_{a=1}^2
\prod_{k=1}^2
\theta_1(\epsilon_+-s(z^{(a)}-x^{(a)}_k))
}
\nn\\
&&
\nn
\times
\chi^{\widehat{\su}(2)_1}_{\varpi_1}
(\widehat{\xi}+2z^{(1)}-x^{(1)}_1-x^{(1)}_2)
\\
\nn
&&
\times
\chi^{\widehat{\su}(2)_1}_{\varpi_2}
(\widehat{\xi}+2z^{(2)}-2z^{(1)}+x^{(1)}_1+x^{(1)}_2-x^{(2)}_1-x^{(2)}_2)
\\
&&
\times
\chi^{\widehat{\su}(2)_1}_{\varpi_3}
(\widehat{\xi}+2z^{(2)}+x^{(2)}_1+x^{(2)}_2),
\label{eq:et112}
\eea
the one for theory $\widetilde{T}^{2}_{(1,1)}$ is given by
\bea
&&
\IE^{2}_{(1,1)}[\vb*{\varpi}]\qty(\widehat{\xi},\underline{\vb*{x}},\vb*{s},\epsilon_+,\tau)
=
\nn
\\
&&
\int
\eta^4
dz^{(1)}
dz^{(2)}
\frac{
\theta_1(2\epsilon_+)^2
\prod_{a=1}^2
\prod_{k=1}^2
\theta_1(z^{(a)}-x^{(3-a)}_k)
}
{\prod_{s=\pm}
\theta_1(\epsilon_+-s(z^{(1)}-z^{(2)}))
\prod_{a=1}^2
\prod_{k=1}^2
\theta_1(\epsilon_+-s(z^{(a)}-x^{(a)}_k))
}
\nn
\\
\nn
&&
\times
\prod_{a=1}^{2}
\frac{\theta_1(z^{(a)}-s^{(a+1)})\theta_1(z^{(a)}-s^{(a-1)})}{\prod_{s=\pm}\theta_1(\epsilon_++s(z^{(a)}-s^{(a)}))}
\\
&&
\nn
\times
\chi^{\widehat{\su}(2)_1}_{\varpi_1}
(\widehat{\xi}+2z^{(1)}-x^{(1)}_1-x^{(1)}_2+s^{(0)}-s^{(1)})
\\
\nn
&&
\times
\chi^{\widehat{\su}(2)_1}_{\varpi_2}
(\widehat{\xi}+2z^{(2)}-2z^{(1)}+x^{(1)}_1+x^{(1)}_2-x^{(2)}_1-x^{(2)}_2+s^{(1)}-s^{(2)})
\\
&&
\times
\chi^{\widehat{\su}(2)_1}_{\varpi_3}
(\widehat{\xi}+2z^{(2)}+x^{(2)}_1+x^{(2)}_2+s^{(2)}-s^{(3)}),
\label{eq:et112}
\eea
and
\be
s^{(a+1)}= s^{(a)}+6\mu.
\ee
The computation of the elliptic genus of the tails only involves residues at simple poles and can be performed straightforwardly; on the other hand, integration over central node holonomies involves higher-order residues and leads to  complicated expressions. In this paper we content ourselves with presenting the elliptic genus in integral form, while in \cite{wip} we will provide further details on the infrared physics of this BPS string configuration.

\section{Conclusions}
\label{sec:concl}
In this paper we have constructed a partition function \eqref{eq:PF} for the 6d SCFTs $\cT^{6d}_{r,W}$ on the equivariant background $T^2\ltimes \mathbb{C}^2/\Gamma$, where $\Gamma$ is an arbitrary choice of discrete subgroup of $SU(2)$. These 6d theories are the worldvolume theories for stacks of $r$ M5 branes probing a transverse $TN_W$ space. The partition function is built out of familiar ingredients: on the one hand, contributions of BPS particles and BPS strings, and on the other hand current algebras associated to the McKay dual $\widehat{\mathfrak{g}}$ of $\Gamma$, which also famously contribute to the Vafa--Witten partition function of $\mathcal{N}=4$ SYM in four dimensions on $\mathbb{C}^2/\Gamma$ \cite{Vafa:1994tf}.

The BPS particle factor in the partition function is the simplest to formulate: it is a product of plethystic exponentials coming from the contributions of (infinitely many) free BPS particles in 5d arising from the KK modes of the 6d fields. The main novelty we observe is that, whereas in the more familiar case of equivariant partition functions on $T^2\ltimes \mathbb{C}^2$ the single-particle contributions are encoded in terms of the Hilbert series which counts holomorphic functions of $\mathbb{C}^2$, in our setting the single-particle contributions correspond to holomorphic sections of nontrivial vector bundles over the orbifold space and are encoded by $\Gamma$-covariant Hilbert series which we introduced in Section \ref{sec:gammaH} and compute explicitly in Appendix \ref{app:Hilbert}. This is due to the coupling of BPS particles to gauge connections that possess nontrivial monodromy at asymptotic infinity.

The remaining contributions to the partition function have a more intricate structure, due to the fact that the current algebras and BPS strings interact nontrivially with each other. This is required to ensure the cancelation of 2d gauge anomalies and ultimately is a manifestation of the fact that BPS strings are charged under the two-form fields of the 6d SCFT. The combined system is captured by  2d $\mathcal{N}=(0,4)$ relative QFTs which possess conformal blocks labeled by choices of integrable highest weight representations for $\widehat{\mathfrak{g}}$. We have obtained a UV description for them in terms of $\Gamma$-dressed quivers, which are built out of collections of Kronheimer--Nakajima quivers interacting at interfaces, and have developed techniques to compute their elliptic genera which enter the 6d partition function. While in this paper we have largely focused on the UV description of the BPS string QFTs, in a separate publication we plan to give a fuller account of the CFTs which capture the infrared behavior of the BPS strings and are described by nonlinear sigma models on instanton moduli spaces on $\mathbb{C}^2/\Gamma$ coupled nontrivially to the current algebras.

With suitable modifications, the technology we have developed should make it possible to determine the partition functions on general ADE singularities for various classes of quantum field theories:
\begin{itemize}
	\item[-] M-string orbifold SCFTs in the presence of defect strings, as in \cite{Agarwal:2018tso};
	\item[-] Six-dimensional SCFTs arising from M5 branes probing transverse singularities other than $\mathbb{C}^2/\mathbb{Z}_W$, along the lines of \cite{Gadde:2015tra};
	\item[-] Six-dimensional $\cN=(1,0)$ little string theories \cite{Hohenegger:2016eqy,Bastian:2017ary,Filoche:2024vne};
	\item[-] Five- and four-dimensional QFTs with eight supercharges obtained by compactification of the theories $\cT^{6d}_{r,W}$, including the case of twisted circle compactifications to 5d \cite{Bhardwaj:2019fzv,Lee:2022uiq}.
\end{itemize}

From a mathematical point of view, the partition functions on orbifold singularities should have an interpretation as generating functions of higher rank Donaldson--Thomas invariants counting sheaves of ADE type on elliptic Calabi--Yau threefolds. It would be very interesting to clarify the connection with the proposal of \cite{Nekrasov:2014nea} and in particular to understand how to adapt their framework to include the contributions from current algebras and corresponding monodromy data which play a crucial role in the partition functions we constructed.

\section*{Acknowledgements}
We would like to thank Yann Proto and Francesco Sala for helpful discussions. The work of DC and GL has received funding from the European Research Council (ERC) under the Horizon Europe (grant agreement No. 101078365) research and innovation program.

\appendix
\section{The $\Gamma$-covariant Hilbert series of ALE spaces}
\label{app:Hilbert}

In this appendix we determine explicit expressions for the $\Gamma$-covariant Hilbert series of $\mathbb{C}^2$ defined in Section \ref{sec:gammaH}. We remark that analogous expressions have previously appeared elsewhere in the literature, see \emph{e.g.}\ \cite{kostant2006coxeter} and references therein.

\subsection{$\Gamma=\mathcal{C}_N$}
The discrete subgroup $\mathcal{C}_N$ of $SU(2)_L$ is the only one which leaves the Cartan subgroup $U(1)_L$ unbroken. In particular, we can identify the generator of $\mathcal{C}_N$ with a rotation $ e^{\frac{2\pi i }{N}}\in U(1)_x$ of the Cartan, so that the fundamental representation of $SU(2)$ splits as:
\be\label{eq:SU2toZn_projection}
(\boldsymbol{2})
\to
\underline{1}^{(1)}\oplus\underline{1}^{(N-1)}.
\ee
The branching rules for arbitrary representations are determined by taking tensor products
\be
(\boldsymbol{2})\otimes(\boldsymbol{n})
=
(\boldsymbol{n-1})\oplus(\boldsymbol{n+1})
\ee
for $n\geq 1$, and using the relation
\be\label{eq:Zn-tensor}
\underline{1}^{(j)}
\otimes
\underline{1}^{(k)}
=
\underline{1}^{(j+k\text{ mod }N)}.
\ee
It is straightforward to determine the refined $\mathcal{C}_N$-covariant Hilbert series
\be
\mathcal{H}^{\mathcal{C}_N}_{\underline{1}^{(j)}}(t,x)
=
\frac{1}{(1-t^2)}
\left(
\frac{x^jt^j}{1-t^N x^N}
+
\frac{x^{j-N}t^{N-j}}{1-t^N x^{-N}}
\right)
\ee
which in the unrefined limit reduce to
\be
\mathcal{H}^{\mathcal{C}_N}_{\underline{1}^{(j)}}(t)
=
\frac{t^j+t^{N-j}}{(1-t^2)(1-t^N)}
\ee
and satisfy
\be\label{eq:equiv_hs_sum}
\sum_{j=0}^{N-1}
\mathcal{H}^{\mathcal{C}_N}_{\underline{1}^{(j)}}(t,x)
=
\mathcal{H}(t,x).
\ee
\subsection{$\Gamma=\mathcal{Q}_N$}
Next let us consider the case where $\Gamma$ is the binary dihedral group $\mathcal{Q}_N$ of order $2N$, corresponding to the affine Dynkin diagram for the Lie algebra $\mathfrak{so}(N+4)$ shown in Figure \ref{fig:ADEc}. $\mathcal{Q}_N$ possesses four one-dimensional irreducible representations $\underline{1}$, $\underline{1}^v$, $\underline{1}^s$, $\underline{1}^c$ and $N/2-1$ two-dimensional irreps $\underline{2}^{(1)}$,  $\dots$, $\underline{2}^{(N/2-1)}$. Branching rules from $SU(2)$ to $\Gamma$ have been determined in \cite{ishimori2012introduction} for arbitrary rank, and from those one can readily compute the $\mathcal{Q}_N$-covariant Hilbert series. These are given by:
\bea
\mathcal{H}^{\mathcal{Q}_N}_{\underline{1}}
&= &
\frac{1+t^{N+2}}{(1-t^4)(1-t^{N})},
\\
\mathcal{H}^{\mathcal{Q}_N}_{\underline{1}^v}
&= &
\frac{t^{2}+t^{N}}{(1-t^4)(1-t^{N})},
\\
\mathcal{H}^{\mathcal{Q}_N}_{\underline{1}^s}
=
\mathcal{H}^{\mathcal{Q}_N}_{\underline{1}^c}
&= &
\frac{t^{N/2}}{(1-t^2)(1-t^{N})},
\\
\mathcal{H}^{\mathcal{Q}_N}_{\underline{2}^{(a)}}
&= &
\frac{t^a+t^{N-a}}{(1-t^2)(1-t^{N})}.
\eea

\subsection{$\Gamma=\mathcal{T}$}
The irreps of $\mathcal{T}$ are denoted by $\underline{1}$, $\underline{1}'$, $\underline{1}''$, $\underline{2}$, $\underline{2}'$, $\underline{2}''$, $\underline{3}$. The decomposition of irreps of $SU(2)$ into irreps of the binary tetrahedral group $\mathcal{T}$ has been performed in \cite{Fallbacher:2015pga}.  From this it is straightforward to determine the $\mathcal{T}$-covariant Hilbert series:
\bea
\mathcal{H}^{\mathcal{T}}_{\underline{1}}
&=& 
\frac{1-t^{24}}{(1-t^6)(1-t^{8})(1-t^{12})},
\\
\mathcal{H}^{\mathcal{T}}_{\underline{1}'}
=
\mathcal{H}^{\mathcal{T}}_{\underline{1}''}
&=& 
\frac{t^{4}}{1-t^4-t^{6}+t^{10}},
\\
\mathcal{H}^{\mathcal{T}}_{\underline{2}}
&=& 
\frac{t-t^3+t^5}{1-t^2-t^6+t^8},
\\
\mathcal{H}^{\mathcal{T}}_{\underline{2}'}
=
\mathcal{H}^{\mathcal{T}}_{\underline{2}''}
&=& 
\frac{t^3}{1-t^2-t^6+t^8},
\\
\mathcal{H}^{\mathcal{T}}_{\underline{3}}
&=& 
\frac{t^2}{(1-t^2)(1-t^4)}.
\eea

\subsection{$\Gamma=\mathcal{O}$}
The irreps of $\mathcal{O}$ are denoted by $\underline{1}$, $\underline{1}'$, $\underline{2}$, $\underline{2}'$, $\underline{2}''$, $\underline{3}$, $\underline{3}'$, $\underline{4}$. While the decomposition of $SU(2)$ irreps into $\mathcal{O}$ irreps has not been carried out in the literature, it has been carried out for the symmetric group $\mathcal{S}_4$, which is an index-two subgroup of $\mathcal{O}$. This makes it possible to determine the $\mathcal{O}$-covariant Hilbert series for representations $\underline{1}$, $\underline{1}'$, $\underline{2}''$, $\underline{3}$, and $\underline{3}'$:
\bea
\mathcal{H}^{\mathcal{O}}_{\underline{1}}
&=& 
\frac{1-t^{36}}{(1-t^8)(1-t^{12})(1-t^{18})},
\\
\mathcal{H}^{\mathcal{O}}_{\underline{1}'}
&=&
\frac{t^{6}}{1-t^6-t^{8}+t^{14}},
\\
\mathcal{H}^{\mathcal{O}}_{\underline{2}''}
&=& 
\frac{t^4}{1-t^4-t^6+t^{10}},
\\
\mathcal{H}^{\mathcal{O}}_{\underline{3}}
&=& 
\frac{t^2-t^4+t^6}{1-t^2-t^8+t^{10}},
\\
\mathcal{H}^{\mathcal{O}}_{\underline{3}'}
&=& 
\frac{t^4}{1-t^2-t^8+t^{10}}.
\eea
To determine the remaining Hilbert series, we can exploit the fact that
\be
\underline{2}\otimes \rho_j = \sum_{k\neq j} C^{\widehat \Gamma}_{jk} \rho_k
\label{eq:repr}
\ee
to work out the branching of even-dimensional representations of $SU(2)$ into irreps of $\mathcal{O}$: specifically, one obtains the relation
\be
\left(\frac{1}{t}+t\right)
\mathcal{H}^{\mathcal{O}}_{\rho_j}
=
\sum_{k\neq j}
C^{\widehat \Gamma}_{jk} \mathcal{H}^{\mathcal{O}}_{\rho_j},
\ee
from which we can read off
\bea
\mathcal{H}^{\mathcal{O}}_{\underline{2}}
&=& 
\frac{t-t^3+t^{5}-t^7+t^9}{(1-t^2)^2(1+t^4)(1+t^2+t^4)},
\\
\mathcal{H}^{\mathcal{O}}_{\underline{2}'}
&=& 
\frac{t^{5}}{(1-t^2)^2(1+t^4)(1+t^2+t^4)},
\\
\mathcal{H}^{\mathcal{O}}_{\underline{4}}
&=& 
\frac{t^{3}}{1-t^2-t^6+t^8}.
\eea
As expected, one can easily check that
\be
\mathcal{H}^{\mathcal{O}}_{\underline{1}}
+
\mathcal{H}^{\mathcal{O}}_{\underline{1}'}
+
2\mathcal{H}^{\mathcal{O}}_{\underline{2}}
+
2\mathcal{H}^{\mathcal{O}}_{\underline{2}'}
+
2\mathcal{H}^{\mathcal{O}}_{\underline{2}''}
+
3\mathcal{H}^{\mathcal{O}}_{\underline{3}'}
+
3\mathcal{H}^{\mathcal{O}}_{\underline{3}}
+
4\mathcal{H}^{\mathcal{O}}_{\underline{4}}
=
\mathcal{H}.
\ee

\subsection{$\Gamma=\mathcal{I}$}
The irreps of $\mathcal{I}$ are denoted by $\underline{1}$, $\underline{2}$, $\underline{2}'$, $\underline{3}$, $\underline{3}'$, $\underline{4}$, $\underline{4}'$, $\underline{5}$, $\underline{6}$. As in the previous example, the decomposition of $SU(2)$ irreps into $\mathcal{I}$ irreps has not been carried out, but it has been carried out for the alternating group $\mathcal{A}_5$ which is an index-two subgroup of $\mathcal{I}$. From this we can determine the $\mathcal{I}$-covariant Hilbert series for representations $\underline{1}$, $\underline{3}$, $\underline{3}'$, $\underline{4}$, and $\underline{5}$, while by using the relation \eqref{eq:repr} we can determine the remaining Hilbert series. All in all, we find:
\bea
\mathcal{H}^{\mathcal{I}}_{\underline{1}}
&=& 
\frac{1-t^{60}}{(1-t^{12})(1-t^{20})(1-t^{30})},
\\
\mathcal{H}^{\mathcal{I}}_{\underline{2}}
&=&
\frac{t-t^{7}+t^{13}}{1-t^6-t^{10}+t^{16}},
\\
\mathcal{H}^{\mathcal{I}}_{\underline{2}'}
&=&
\frac{t^7}{1-t^6-t^{10}+t^{16}},
\\
\mathcal{H}^{\mathcal{I}}_{\underline{3}}
&=&
\frac{t^2-t^{6}+t^{10}}{1-t^4-t^{10}+t^{14}},
\\
\mathcal{H}^{\mathcal{I}}_{\underline{3}'}
&=& 
\frac{t^{6}}{1-t^4-t^{10}+t^{14}},
\\
\mathcal{H}^{\mathcal{I}}_{\underline{4}}
&=& 
\frac{t^{3}+t^{11}}{1-t^6-t^{10}+t^{16}},
\\
\mathcal{H}^{\mathcal{I}}_{\underline{4}'}
&=& 
\frac{t^{6}+t^8}{1-t^6-t^{10}+t^{16}},
\\
\mathcal{H}^{\mathcal{I}}_{\underline{5}}
&=& 
\frac{t^4}{1-t^4-t^6+t^{10}},
\\
\mathcal{H}^{\mathcal{I}}_{\underline{6}}
&=& 
\frac{t^5}{1-t^2-t^{10}+t^{12}},
\eea
which satisfy the relation
\be
\mathcal{H}^{\mathcal{I}}_{\underline{1}}
+
2\mathcal{H}^{\mathcal{I}}_{\underline{2}}
+
2\mathcal{H}^{\mathcal{I}}_{\underline{2}'}
+
3\mathcal{H}^{\mathcal{I}}_{\underline{3}}
+
3\mathcal{H}^{\mathcal{I}}_{\underline{3}'}
+
4\mathcal{H}^{\mathcal{I}}_{\underline{4}}
+
4\mathcal{H}^{\mathcal{I}}_{\underline{4}'}
+
5\mathcal{H}^{\mathcal{I}}_{\underline{5}}
+
6\mathcal{H}^{\mathcal{I}}_{\underline{6}}
=
\mathcal{H}.
\ee

\section{Cancelation of abelian anomalies}
\label{app:anomaly-canc}

In this appendix we check that the quadratic polynomial $f_{\va*{v}}^{\Gamma,\va*{w}}(\va*{\xi},\vec{\underline{\boldsymbol{s}}},\epsilon_+)$ appearing in Equation \eqref{eq:Str}, which encodes the anomalies of the 2d theory $\cQ_{\va*{v}}^{\Gamma,\va*{w}}$, 
 does not depend on the gauge holonomies $z_{j,k}^{(a)}$, indicating that the theory $\cQ^{\Gamma,\va*{w}}_{\va*{v}}$ is free from gauge anomalies.  Each of the 1-loop contributions $Z_\Phi$ from $\mathcal{N}=(0,4)$ multiplets listed in \eqref{eq:Z_V}--\eqref{eq:Z1loop_Theta} transforms as 
\be
Z_\Phi\to Z_\Phi e^{\frac{2\pi i}{\tau} L_\Phi}
\ee
modulo an overall constant phase. Using the modular transformations of the Jacobi theta function and the Dedekind eta function given in Equations \eqref{eq:dedm} and \eqref{eq:thm}, we can determine the exponent $L$ for each of these 1-loop factors:
\begin{align*}
	L_{V_j^{(a)}}&=2 \qty(v_{j}^{(a)})^2\epsilon_+^2-2\qty(Z_j^{(a)})^2 +2 v_{j}^{(a)} \mathscr{Z}^{(a)}_j;\\
	L_{Y_j^{(a)}}&=-v_j^{(a)}v_j^{(a+1)}\epsilon_+^2-v_j^{(a)}\mathscr{Z}^{(a+1)}_j -v_j^{(a+1)}\mathscr{Z}^{(a)}_j+2 Z_j^{(a)}Z_j^{(a+1)}; \\
	L_{X_{i,j}^{(a)}}&=-v_{i}^{(a)}v_{j}^{(a)}\epsilon_+^2-v_{i}^{(a)}\mathscr{Z}^{(a)}_j  -v_{j}^{(a)} \mathscr{Z}^{(a)}_i+2Z_i^{(a)}Z_{j}^{(a)};\\
	L_{\Psi_{i,j}^{(a)}}&=\frac{1}{2} v_i^{(a)} \mathscr{Z}^{(a+1)}_j+\frac{1}{2}v_{j}^{(a+1)} \mathscr{Z}^{(a)}_i-Z_i^{(a)}Z_{j}^{(a+1)};\\
	L_{\widetilde{\Psi}_{i,j}^{(a)}}&=\frac{1}{2} v_j^{(a)} \mathscr{Z}^{(a+1)}_i+\frac{1}{2}v_{i}^{(a+1)} \mathscr{Z}^{(a)}_j-Z_j^{(a)}Z_{i}^{(a+1)};\\
	L_{W_j^{(a)}}&=-v_j^{(a)}w_j^{(a)}\epsilon_+^2-v_j^{(a)}\mathscr{S}^{(a)}_j -w_j^{(a)}\mathscr{Z}^{(a)}_j+2 Z_j^{(a)}S_j^{(a)}; \\
	L_{\Sigma_j^{(a)}}&=\frac{v_j^{(a)}}{2} \mathscr{S}^{(a+1)}_j+\frac{w_j^{(a+1)}}{2}\mathscr{Z}^{(a)}_j-S_j^{(a+1)}Z_j^{(a)};\\
	L_{\Theta_j^{(a)}}&=\frac{v_j^{(a)}}{2}\mathscr{S}^{(a-1)}_j +\frac{w_j^{(a-1)}}{2}\mathscr{Z}^{(a)}_j-S_j^{(a-1)}Z_j^{(a)}.
\end{align*}
The parameters $Z_{j}^{(a)}$ and $S_j^{(a)}$ were introduced in \eqref{eq:Zj_and_Sj}, and we define
\begin{equation}
\mathscr{Z}^{(a)}_j=\sum_{l=1}^{v_j^{(a)}}\qty(z_{j,l}^{(a)})^2\quad,\quad \mathscr{S}^{(a)}_j=\sum_{l=1}^{w_j^{(a)}}\qty(s_{j,l}^{(a)})^2.
\end{equation}
The factors $Z_{\widehat{\mathfrak{g}}_1^{(a)}}^{\omega^{(a)}}$ are 
characters of  i.h.w.r.'s of $\widehat{\mathfrak{g}}_1$, which, under a modular S-transformation transform as
\begin{equation}
	{\chi}^{\widehat{\mathfrak{g}}_1}_{\omega}\qty(\frac{\va{\xi}}{\tau},-\frac{1}{\tau})=e^{\frac{1}{2}\frac{2\pi i}{\tau}\va{\xi}\cdot (C^{\mathfrak{g}})^{-1}\cdot\va{\xi}} \sum_{\upsilon} \mathcal{S}_{\omega,\upsilon} {\chi}^{\widehat{\mathfrak{g}}_1}_{\upsilon}\qty(\va{\xi},\tau) ,
\end{equation}
where $\mathcal{S}_{\omega,\upsilon}$ is the modular S-matrix of $\widehat{\mathfrak{g}}_1$ and the sum is over all the level-1 i.h.w.r.'s. Adding up all contributions, we find that
\begin{align*}
	f_{\va*{v}}^{\Gamma,\va*{w}}&
	=
	\sum_{j=0}^{\text{rk }\mathfrak{g}}
	\qty[\sum_{a=1}^{r-1} \qty(L_{V_j^{(a)}}+L_{W_j^{(a)}}+L_{\Sigma_j^{(a)}}+L_{\Theta_j^{(a)}})+\sum_{a=1}^{r-2}L_{Y_j^{(a)}}]
	\\
    &-\sum_{j=1}^{\text{rk }\mathfrak{g}}
    \sum_{i=0}^{j-1}C^{\widehat{\mathfrak{g}}}_{ij}\qty[\sum_{a=1}^{r-1}L_{X_{i,j}^{(a)}}+\sum_{a=1}^{r-2}\qty(L_{\Psi^{(a)}_{i,j}}+L_{\widetilde{\Psi}^{(a)}_{i,j}})] \\
	&+\sum_{a=1}^{r}   \frac{1}{2}\va{\xi}^{(a)}\cdot (C^{\mathfrak{g}})^{-1}\cdot \va{\xi}^{(a)}  ,
\end{align*}
where the expression for the $\xi_j^{(a)}$'s is given in Equation \eqref{eq:xi-term}. Let us introduce the notation
\begin{equation}
    \Delta Q^{(a)} = Q^{(a)}-Q^{(a-1)} ,
\end{equation}
for any quantity $Q^{(a)}$ that depends on the index $(a)$. After some manipulation, we can rewrite $f_{\va*{v}}^{\Gamma,\va*{w}}$ as:
\begin{equation}\label{eq:L_intermediate}
	\begin{aligned}
		f_{\va*{v}}^{\Gamma,\va*{w}}=\frac{1}{2}\sum_{a=1}^r\Bigg\{&\epsilon_+^2\qty[\qty(\Delta \vec{v}^{(a)})^2+ \vec{v}^{(a)}\cdot C^{\widehat{\mathfrak{g}}}\cdot\vec{v}^{(a)}-2\vec{v}^{(a)}\cdot \vec{w}^{(a)}]-\Delta \vec{v}^{(a)}\cdot \Delta\vec{\mathscr{S}}^{(a)}\\
        &+\Delta\vec{\mathscr{Z}}^{(a)}\cdot \qty( C^{\widehat{\mathfrak{g}}}\cdot\Delta\vec{v}^{(a)}-\Delta \vec{w}^{(a)})\\
		&-\Delta \vec{Z}^{(a)}\cdot C^{\widehat{\mathfrak{g}}}\cdot\Delta \vec{Z}^{(a)}+\Delta \vec{Z}^{(a)}\cdot\Delta \vec{S}^{(a)}   \\
		&+\va{\xi}^{(a)}\cdot (C^{\mathfrak{g}})^{-1}\cdot \va{\xi}^{(a)}\Bigg\} .
	\end{aligned}
\end{equation}
The second row vanishes by virtue of Equation \eqref{eq:c2} while the remaining rows depend on the gauge fugacities $z_{j,k}^{(a)}$. In particular, the third row arises from the abelian factors of the gauge nodes of the quiver $\cQ^{\Gamma,\va*{w}}_{\va*{v}}$ and contributes to the abelian gauge anomaly. We will now see how these terms cancel out with an opposite contribution from the term in the fourth row, which originates from the shift \eqref{eq:xi-term} of the chemical potentials $\xi_j$'s.

Let us focus on the fourth row of \eqref{eq:L_intermediate} and expand the $\vec{\xi}^{(a)}$'s using \eqref{eq:xi-term}, keeping in mind that $\vec{\xi}$ and $\vec{\xi}^{(a)}$ have $\text{rk }\mathfrak{g}$ components with index $j=1,\dots,\text{rk }\mathfrak{g}$, while the index for other vector quantities $\vec{{Q}}$ have ranges from $0$ to $\text{rk }\mathfrak{g}$. We find:

\begin{equation}\label{eq:xi-expansion}
\begin{aligned}
\frac{1}{2}\sum_{a=1}^r\va{\xi}^{(a)}\cdot (C^{\mathfrak{g}})^{-1}\cdot \va{\xi}^{(a)}
=
\sum_{a=1}^r\sum_{j,k=1}^{\text{rk }\mathfrak{g}}\Bigg\{ &\frac{1}{2}(\xi_j+\mathfrak{m}-\Delta S_j^{(a)}) \qty(C^{\mathfrak{g}})^{-1}_{jk} (\xi_k+\mathfrak{m}-\Delta S_k^{(a)})\\ 
    &+(\xi_j+\mathfrak{m})\qty(C^{\mathfrak{g}})^{-1}_{jk}  \qty(C^{\widehat{\mathfrak{g}}}\cdot\Delta \vec{Z}^{(a)})_k  \\
	&+ \frac{1}{2}\qty(C^{\widehat{\mathfrak{g}}}\cdot\Delta \vec{Z}^{(a)})_{j} \qty(C^{\mathfrak{g}})^{-1}_{jk}\qty(C^{\widehat{\mathfrak{g}}}\cdot\Delta \vec{Z}^{(a)})_{k} \\
    &-\qty(C^{\widehat{\mathfrak{g}}}\cdot\Delta \vec{Z}^{(a)})_{j} \qty(C^{\mathfrak{g}})^{-1}_{jk}\Delta S_k^{(a)} \Bigg\}.
\end{aligned}
\end{equation}
The second row vanishes in the sum over $a=1,\dots,r$ since
\begin{equation}
    \sum_{a=1}^r \Delta Q^{(a)}=Q^{(r)}-Q^{(0)}
\end{equation}
for any $(a)$-indexed quantity $Q^{(a)}$ and
\begin{equation}
Z_j^{(0)}=Z_j^{(r)}=0,\quad\forall\, j=0,\dots,n ,
\end{equation}
which follows from the definition \eqref{eq:Zj_and_Sj} and the fact that $v_j^{(0)}=v_{j}^{(r)}=0$ for every $j=0,\dots,\mathrm{rk}\,\mathfrak{g}$. We can further manipulate the terms in the third and fourth row of Equation \eqref{eq:xi-expansion}, by noticing that
\begin{equation}
    \sum_{j,k=1}^{\text{rk }\mathfrak{g}}
    \qty(C^{\widehat{\mathfrak{g}}}\cdot\Delta \vec{Z}^{(a)})_{j} \qty(C^{\mathfrak{g}})^{-1}_{jk}\qty(C^{\widehat{\mathfrak{g}}}\cdot\Delta \vec{Z}^{(a)})_{k}=\Delta \vec{Z}^{(a)}\cdot C^{\widehat{\mathfrak{g}}} \cdot \Delta \vec{Z}^{(a)} ,
\end{equation}
and
\begin{equation}
    \sum_{a=1}^r\sum_{j,k=1}^{\text{rk }\mathfrak{g}}
    \qty(C^{\widehat{\mathfrak{g}}}\cdot\Delta \vec{Z}^{(a)})_{j} \qty(C^{\mathfrak{g}})^{-1}_{jk}\Delta S_k^{(a)}=\sum_{a=1}^r\Delta\vec{Z}^{(a)}\cdot \Delta\vec{S}^{(a)} ,
\end{equation}
where in the latter we have also used Equation \eqref{eq:Stuckelberg_constr} and $\sum_{a=1}^r\Delta Z_0^{(a)}=0$. This leads to the following expression:
\begin{multline}
\label{eq:xi-expansion2}
    \frac{1}{2}\sum_{a=1}^r\va{\xi}^{(a)}\cdot (C^{\mathfrak{g}})^{-1}\cdot \va{\xi}^{(a)}=\sum_{a=1}^r\Bigg\{\frac{1}{2}\vec{\xi}^{(a)}_{\mathfrak{m},\underline{\va*{s}}}\cdot (C^{\mathfrak{g}})^{-1}\cdot \vec{\xi}^{(a)}_{\mathfrak{m},\underline{\va*{s}}}\\+ \Delta \vec{Z}^{(a)}\cdot C^{\widehat{\mathfrak{g}}} \cdot \Delta \vec{Z}^{(a)} - \Delta\vec{Z}^{(a)}\cdot \Delta\vec{S}^{(a)}\Bigg\}, 
\end{multline}
where
\begin{equation}
    (\vec{\xi}^{(a)}_{\mathfrak{m},\underline{\va*{s}}})_j=\xi_j+\mathfrak{m}-\Delta S_j^{(a)}.
\end{equation}
The two terms in the third row of Equation \eqref{eq:xi-expansion2} come respectively from the couplings \eqref{eq:currents-gauge-coupling} and \eqref{eq:mixed-anomaly-coupling} and are exactly what we need to cancel the second row of Equation \eqref{eq:L_intermediate}. Indeed, upon inserting \eqref{eq:xi-expansion2} into \eqref{eq:L_intermediate}, we obtain:
\begin{multline}\label{eq:mod_anom_raw}
		f_{\va*{v}}^{\Gamma,\va*{w}}=\frac{1}{2}\sum_{a=1}^r\Bigg\{\epsilon_+^2\qty[\qty(\Delta \vec{v}^{(a)})^2+ \vec{v}^{(a)}\cdot C^{\widehat{\mathfrak{g}}}\cdot\vec{v}^{(a)}-2\vec{v}^{(a)}\cdot \vec{w}^{(a)}]\\
        -\Delta \vec{v}^{(a)}\cdot \Delta\vec{\mathscr{S}}^{(a)}+\vec{\xi}^{(a)}_{\mathfrak{m},\underline{\va*{s}}}\cdot (C^{\mathfrak{g}})^{-1}\cdot \vec{\xi}^{(a)}_{\mathfrak{m},\underline{\va*{s}}}\Bigg\}  ,
\end{multline}
which is independent of gauge holonomies $z_{j,k}^{(a)}$.
Equation \eqref{eq:mod_anom_raw} can be rewritten as:
\begin{equation}
    f_{\va*{v}}^{\Gamma,\va*{w}}(\va*{\xi},\underline{\va*{s}},\mathfrak{m},\epsilon_+)=\frac{1}{2}\sum_{a=1}^{r}\vec{\xi}^{(a)}_{\mathfrak{m},\underline{\va*{s}}}\cdot (C^{\mathfrak{g}})^{-1}\cdot \vec{\xi}^{(a)}_{\mathfrak{m},\underline{\va*{s}}}+ k_R \epsilon_+^2 +\sum_{a=0}^r \sum_{j=0}^n \frac{k_{\mathfrak{u}(w_j^{(a)})}}{2}\sum_{K=1}^{w_j^{(a)}} \qty(s_{j,K}^{(a)})^2 ,
\end{equation}
where the levels for $SU(2)_t$ and $\mathfrak{u}(w_j^{(a)})$ are given respectively by:
\begin{align}
    k_R&=-\sum_{a=1}^{r-1}\dim_{\IH}\cM_{\vec{w}^{(a)},\vec{v}^{(a)}}^\Gamma+\frac{1}{2}\sum_{j=0}^{\text{rk }\mathfrak{g}} \vb*{v}_j\cdot C^{\su(r)}\cdot \vb*{v}_j
    \\
    \label{eq:kt2}
    &=-\frac{c_R}{6}+\frac{1}{2}\sum_{j=0}^{\text{rk }\mathfrak{g}} \vb*{v}_j\cdot C^{\su(r)}\cdot \vb*{v}_j,\\
    k_{\mathfrak{u}(w_j^{(a)})}&=\begin{cases}
    -\qty(C^{\su(r)}\cdot \vb*{v}_j)^{(a)} & \text{ for } a=1,\dots,r-1\\
    v_{j}^{(1)} & \text{ for } a=0\\
    v_{j}^{(r-1)} & \text{ for } a=r\\
    \end{cases}.
\end{align}
Here, $\dim_{\IH}\cM_{\vec{w}^{(a)},\vec{v}^{(a)}}^\Gamma$ is given in Equation \eqref{eq:mod_space_dim}, and we have used Equation \eqref{eq:Rcentral_charge} to obtain Equation \eqref{eq:kt2}.\\

\section{$T^{l,\vec{\boldsymbol{w}}}_{\vec{\boldsymbol{v}}}$  from frozen BPS strings on $\mathbb{C}^2/\cC_l$}
\label{app:TVW}
\begin{figure}[t]
	\centering
		\scalebox{1}{\begin{tikzpicture}[line width=0.8]
				\def\xsh{3.5};
				\def\gap{2};
				\draw[blue] (-2.5,-5.7)--(-2.5,5.7);
				\node[blue] at (-2.5,6) {$\widehat{\mathfrak{su}}(l)^{(1)}_1$};
				\draw[blue] (1,-5.7)--(1,5.7);		
				\node[blue] at (1,6) {$\widehat{\mathfrak{su}}(l)^{(2)}_1$};
				\draw[blue] (1+\xsh,-5.7)--(1+\xsh,5.7);		
				\node[blue] at (1+\xsh,6) {$\widehat{\mathfrak{su}}(l)^{(3)}_1$};
				\draw[blue] (1+\gap+\xsh,-5.7)--(1+\gap+\xsh,5.7);		
				\node[blue] at (1+\gap+\xsh,6) {$\widehat{\mathfrak{su}}(l)^{(r-1)}_1$};
				\draw[blue] (1+\gap+2*\xsh,-5.7)--(1+\gap+2*\xsh,5.7);		
				\node[blue] at (1+\gap+2*\xsh,6) {$\widehat{\mathfrak{su}}(l)^{(r)}_1$};
				%%%%%    (0)		
				%%%%%%%%%
				\node[flavor] (f01) at (-\xsh,4) {$w_1^{(0)}$};
				\node[flavor] (f02) at (-\xsh,2) {$w_2^{(0)}$};
				\node   (f0i) at (-\xsh,0) {$\vdots$};
				\node[flavor] (f0M) at (-\xsh,-2) {$w_{l-1}^{(0)}$};
				\node[flavor] (f00) at (-\xsh,-4) {$w_{0}^{(0)}$};
				%%%%%%%%%
				%%%%%    (1)		
				%%%%%%%%%
				\node[gauge]   (g11) at (0,4) {$v_1^{(1)}$};
				\node[flavor] (f11) at (-1.5,4) {$w_1^{(1)}$};
				\node[gauge]   (g12) at (0,2) {$v_2^{(1)}$};
				\node[flavor] (f12) at (-1.5,2) {$w_2^{(1)}$};
				\node  (g1i) at (0,0) {$\vdots$};
				\node[gauge]   (g1M) at (0,-2) {$v_{l-1}^{(1)}$};
				\node[flavor] (f1M) at (-1.5,-2) {$w_{l-1}^{(1)}$};
				\draw (g11) -- (g12);
				\draw (g11) -- (f11);
				\draw (g12) -- (f12);
				\draw (g12) -- (g1i);
				\draw (g1i) -- (g1M);
				\draw (g1M) -- (f1M);
				\node[gauge,dotted]  (g10) at (0,-4) {$0$};
				\node[flavor] (f10) at (-1.5,-4) {$w_{0}^{(1)}$};
				\draw[dotted] (g1M) -- (g10);
				\node at (0,5.5) {$=$};
				\node at (0,-5.5) {$=$};
				\draw[dotted] (g10) -- (f10);
				\draw[dotted] (g11)--(0,5.5);
				\draw[dotted] (g10)--(0,-5.4);
				%%%%%%%%%
				%%%%%    (2)		
				%%%%%%%%%
				\node[gauge]   (g21) at (0+\xsh,4) {$v_1^{(2)}$};
				\node[flavor] (f21) at (-1.5+\xsh,4) {$w_1^{(2)}$};
				\node[gauge]   (g22) at (0+\xsh,2) {$v_2^{(2)}$};
				\node[flavor] (f22) at (-1.5+\xsh,2) {$w_2^{(2)}$};
				\node   (g2i) at (0+\xsh,0) {$\vdots$};
				\node[gauge]   (g2M) at (0+\xsh,-2) {$v_{l-1}^{(2)}$};
				\node[flavor] (f2M) at (-1.5+\xsh,-2) {$w_{l-1}^{(2)}$};
				\draw (g21) -- (g22);
				\draw (g21) -- (f21);
				\draw (g22) -- (f22);
				\draw (g22) -- (g2i);
				\draw (g2i) -- (g2M);
				\draw (g2M) -- (f2M);
				\node[gauge,dotted]  (g20) at (0+\xsh,-4) {$0$};
				\node[flavor] (f20) at (-1.5+\xsh,-4) {$w_{0}^{(2)}$};
				\draw[dotted] (g2M) -- (g20);
				\node at (\xsh,5.5) {$=$};
				\node at (\xsh,-5.5) {$=$};
				\draw[dotted] (g20) -- (f20);
				\draw[dotted] (g21)--(\xsh,5.5);
				\draw[dotted] (g20)--(\xsh,-5.4);
				%%%%%%%%%%(dots)%%%%%%%%%%%%
				\node  (d1) at (1+\xsh+\gap/2,4) {\Large$\cdots$};
				\node  (d2) at (1+\xsh+\gap/2,2) {\Large$\cdots$};
				\node  (di)  at (1+\xsh+\gap/2,0) {\Large$\cdots$};
				\node (dM) at (1+\xsh+\gap/2,-2) {\Large$\cdots$};
				\node (d0) at (1+\xsh+\gap/2,-4) {\Large$\cdots$};
				%%%%%%%%%
				%%%%%    (r-1)		
				%%%%%%%%%
				\node[gauge]   (gr-11) at (0+\gap+2*\xsh,4) {\tiny$v_1^{(r-1)}$};
				\node[flavor] (fr-11) at (-1.5+\gap+2*\xsh,4) {\tiny$w_1^{(r-1)}$};
				\node[gauge]   (gr-12) at (0+\gap+2*\xsh,2) {\tiny$v_2^{(r-1)}$};
				\node[flavor] (fr-12) at (-1.5+\gap+2*\xsh,2) {\tiny$w_2^{(r-1)}$};
				\node   (gr-1i) at (0+\gap+2*\xsh,0) {$\vdots$};
				\node[gauge]   (gr-1M) at (0+\gap+2*\xsh,-2) {\tiny$v_{l-1}^{(r-1)}$};
				\node[flavor] (fr-1M) at (-1.5+\gap+2*\xsh,-2) {\tiny$w_{l-1}^{(r-1)}$};
				\draw (gr-11) -- (gr-12);
				\draw (gr-11) -- (fr-11);
				\draw (gr-12) -- (fr-12);
				\draw (gr-12) -- (gr-1i);
				\draw (gr-1i) -- (gr-1M);
				\draw (gr-1M) -- (fr-1M);
						%%%%add 0 nodes
				\node[gauge,dotted]  (gr-10) at (\gap+2*\xsh,-4) {$0$};
				\node[flavor] (fr-10) at (-1.5+\gap+2*\xsh,-4) {\tiny $w_{0}^{(r-1)}$};
				\draw[dotted] (gr-1M) -- (gr-10);
				\node at (\gap+2*\xsh,5.5) {$=$};
				\node at (0+\gap+2*\xsh,-5.5) {$=$};
				\draw[dotted] (gr-10) -- (fr-10);
				\draw[dotted] (gr-11)--(0+\gap+2*\xsh,5.5);
				\draw[dotted] (gr-10)--(0+\gap+2*\xsh,-5.4);
				%%%%%%%%%
				%%%%%    (r)		
				%%%%%%%%%
				\node[flavor] (fr1) at (-1.5+\gap+3*\xsh,4) {$w_1^{(r)}$};
				\node[flavor] (fr2) at (-1.5+\gap+3*\xsh,2) {$w_2^{(r)}$};
				\node   (fri) at (-1.5+\gap+3*\xsh,0) {$\vdots$};
				\node[flavor] (frM) at (-1.5+\gap+3*\xsh,-2) {$w_{l-1}^{(r)}$};
				\node[flavor] (fr0) at (-1.5+\gap+3*\xsh,-4) {$w_{0}^{(r)}$};
				%%%%%%%%%%%%%%%
				% (0) -- (1) 
				%%%%%%%%%%%%%%%
				\draw[dashed] (f01)to[out=-30,in=-150](g11);
				\draw[dashed] (f02)to[out=-30,in=-150](g12);
				\draw[dashed] (f0M)to[out=-30,in=-150](g1M);
				%%%%%%%%%%%%%%%
				% (1)--(2)
				%%%%%%%%%%%%%%%
				\draw[purple] (g11)to[out=-30,in=-150](g21);
				\draw[dashed] (g11)--(f21);
				\draw[dashed] (f11)to[out=30,in=150](g21);
				\draw[green!70!black,dashed] (g11)--(g22);
				\draw[green!70!black,dashed] (g12)--(g21);
				%%%%%%%%%%%%%%%%%
				\draw[purple] (g12)to[out=-30,in=-150](g22);
				\draw[dashed] (g12)--(f22);
				\draw[dashed] (f12)to[out=30,in=150](g22);
				\draw[green!70!black,dashed] (g12)--(g2i);
				\draw[green!70!black,dashed] (g1i)--(g22);
				%%%%%%%%%%%%%%%%%%%%
				\draw[purple] (g1M)to[out=-30,in=-150](g2M);
				\draw[dashed] (g1M)--(f2M);
				\draw[dashed] (f1M)to[out=30,in=150](g2M);
				\draw[green!70!black,dashed] (g1M)--(g2i);
				\draw[green!70!black,dashed] (g1i)--(g2M);
				%%%%%%%%%%%%%%%%%%%%
				% (2)--(dots)
				%%%%%%%%%%%%%%%
				\draw[purple] (g21)to[out=-30,in=-150](d1);
				\draw[dashed] (g21)--(d1);
				\draw[dashed] (f21)to[out=30,in=150](d1);
				\draw[green!70!black,dashed] (g21)--(d2);
				\draw[green!70!black,dashed] (g22)--(d1);
				%%%%%%%%%%%%%%%%%
				\draw[purple] (g22)to[out=-30,in=-150](d2);
				\draw[dashed] (g22)--(d2);
				\draw[dashed] (f22)to[out=30,in=150](d2);
				\draw[green!70!black,dashed] (g22)--(di);
				%%%%%%%%%%%%%%%%%%%%
				\draw[purple] (g2M)to[out=-30,in=-150](dM);
				\draw[dashed] (g2M)--(dM);
				\draw[dashed] (f2M)to[out=30,in=150](dM);
				\draw[green!70!black,dashed] (g2M)--(di);
				%%%%%%%%%%%%%%%%%%%%
				% (dots)--(r-1)
				%%%%%%%%%%%%%%%
				\draw[purple] (d1)to[out=-30,in=-150](gr-11);
				\draw[dashed] (d1)--(fr-11);
				\draw[dashed] (d1)to[out=30,in=150](gr-11);
				\draw[green!70!black,dashed] (d1)--(gr-12);
				\draw[green!70!black,dashed] (d2)--(gr-11);
				%%%%%%%%%%%%%%%%%
				\draw[purple] (d2)to[out=-30,in=-150](gr-12);
				\draw[dashed] (d2)--(fr-12);
				\draw[dashed] (d2)to[out=30,in=150](gr-12);
				\draw[green!70!black,dashed] (di)--(gr-12);
				%%%%%%%%%%%%%%%%%%%%
				\draw[purple] (dM)to[out=-30,in=-150](gr-1M);
				\draw[dashed] (dM)--(fr-1M);
				\draw[dashed] (dM)to[out=30,in=150](gr-1M);
				\draw[green!70!black,dashed] (di)--(gr-1M);
				%%%%%%%%%%%%%%%
				% (r-1) -- (r) 
				%%%%%%%%%%%%%%%
				\draw[dashed] (gr-11)--(fr1);
				\draw[dashed] (gr-12)--(fr2);
				\draw[dashed] (gr-1M)--(frM);
		\end{tikzpicture}}
    \caption{This figure depicts a frozen BPS string configuration into which the  $\mathcal{N}=(0,4)$ $\cC_l$-dressed quiver theory $T^{l,\vec{\boldsymbol{w}}}_{\vec{\boldsymbol{v}}}$ can be embedded.}
    \label{fig:Clembed}
\end{figure}
In this appendix we show that the theory $T^{l,\vec{\boldsymbol{w}}}_{\vec{\boldsymbol{v}}}$ displayed in Figure \ref{fig:gammaquiv}, for an arbitrary choice of selection sector $\boldsymbol{\varpi}$, can always be embedded into the frozen BPS string configuration of Figure \ref{fig:Clembed} for a suitable 6d theory $\cT^{6d}_{r,W}$ on a $\mathbb{C}^2/\cC_{l}$ singularity, which is obtained by adding an empty gauge nodes with rank $v^{(a)}_0=0$ as well as a decoupled flavor symmetry nodes with rank $w^{(a)}_0$. The field content of the two quivers is identical, and the only thing to check is that it is possible to embed theory $T^{l,\vec{\boldsymbol{w}}}_{\vec{\boldsymbol{v}}}$ into the $\mathcal{C}_l$-dressed quiver theory for arbitrary values of the chemical potentials. Specifically, we need to ensure that the St\"{u}ckelberg constraint \eqref{eq:Stuckelberg_constr} on the $\mathcal{C}_l$-dressed quiver does not restrict the possible assignments of chemical potentials for quiver $T^{l,\vec{\boldsymbol{w}}}_{\vec{\boldsymbol{v}}}$. Let us report the constraint here for convenience:
\be
    \sum_{j=0}^{\text{rk }\mathfrak{g}} a_j\qty(S^{(a)}_j-S^{(a-1)}_j) = h^{\vee}_{\Gamma}\mathfrak{m}.
    \label{eq:stu}
\ee
This constraint relates the chemical potentials associated to the abelian factors of the $U(w^{(a)}_j)$ global symmetries of the $\mathcal{C}_l$ quiver. However, we can always choose arbitrary values of the $S_j^{(a)}$ for $j >0$ provided that the global symmetry ranks $w^{(a)}_0$ on the affine nodes are all positive, in which case we can use them to trivialize the constraint \eqref{eq:stu} on the $S^{(a)}_j$ for $j>0$. This in particular can also be accomplished if we set the parameter $\mathfrak{m}, $ which does not appear in theory $T^{l,\vec{\boldsymbol{w}}}_{\vec{\boldsymbol{v}}}$, to zero. Thus the only condition we find on the $\mathcal{C}_l$ quiver theory is that for $a=0,\dots,r$
\be
w_0^{(a)}> 0 \rightarrow w^{(0)}_0 \geq w^{(0),min}_0 = \max_{a=1.\dots,r-1}(v^{(a)}_{1}+v^{(a)}_{l-1})+1,
\ee
which we have obtained by using Equation \eqref{eq:c1diag} and \eqref{eq:uw0}. If we take the minimal allowed value, we find that we can embed the $T^{l,\vec{\boldsymbol{w}}}_{\vec{\boldsymbol{v}}}$ theory in a frozen BPS string configuration in the theory $\mathcal{T}^{6d}_{r,W}$ with 6d gauge algebra rank given by
\be
W = \sum_{j=1}^{l-1} w^{(a^\star)}_j+1,
\ee
where $a^\star$ is the value of the index $a$ which maximizes $v^{(a)}_1+v^{(a)}_{l-1}$. The discrete parameters of the corresponding quiver are given by 
\begin{equation}
	\vec{\boldsymbol{\mathrm{ v}}} = \mqty(\boldsymbol{0}\\ \boldsymbol{v}_1\\\vdots\\\boldsymbol{v}_{l-1}),\quad
	\vec{\boldsymbol{\mathrm{ w}}} = \mqty(W-\sum_{j=1}^{l-1}\boldsymbol{w}_j\\\boldsymbol{w}_1\\\vdots\\\boldsymbol{w}_{l-1}).
\end{equation}
and
\be
{\boldsymbol{\omega}}^{KK} = \boldsymbol{\varpi},
\ee
and we have the following identity between the elliptic genus $\IE^{l,\va*{w}}_{\va*{v}}[\mathcal{\vb*{\varpi}}]$ of theory $T^{l,\va*{w}}_{\va*{v}}$ and the one of the frozen BPS string (which is given by Equation \eqref{eq:ADE_elliptic_genus_loc}):
\be
\IE^{l,\va*{w}}_{\va*{v}}
[\vb*{\varpi}]
(\vec{\xi},\vec{\underline{\boldsymbol{s}}},\epsilon_+,\tau)
=
\IE^{\cC_{l},\vec{\boldsymbol{\mathrm{ w}}}}_{\vec{\boldsymbol{\mathrm{ v}}}}
[\vb*{\omega}^{KK}]
(\vec{\xi},\underline{\va{s}},\epsilon_+,\tau),
\ee
where the $U(w^{(a)}_j)$ chemical potentials that enter on the right hand side are given by $\mathrm{s}^{(a)}_{j,K}= s^{(a)}_{j,K}$ for $j>0$, while
\be
\mathrm{s}^{(a)}_{0,1}
=
\begin{cases}
0 & a=0
\\
\mathrm{s}^{(a-1)}_{0,1} + \sum_{j=1}^{\text{rk }\mathfrak{g}} a_j\qty(S^{(a-1)}_j -S^{(a)}_j) & a=1,\dots,r
\end{cases}
.
\ee

\bibliography{untitled}
\bibliographystyle{utphys}

\end{document}